\documentclass[12pt]{article}
\usepackage{epsfig, epsf, graphicx, subfigure, color}
\usepackage{pstricks, pst-node, psfrag}
\usepackage{amssymb,amsmath}
\usepackage{verbatim,enumerate}
\usepackage{rotating, lscape}
\usepackage{setspace}
\usepackage{bbm}
\usepackage{natbib}
\usepackage{amsthm}
\usepackage{url}
\usepackage{appendix}
\usepackage{multirow}
\usepackage{caption}
\usepackage{hyperref}
\usepackage{arydshln}
\hypersetup{
  colorlinks   = true, 
  urlcolor     = blue, 
  linkcolor    = blue, 
  citecolor   =  blue  
}

\theoremstyle{plain}

\newtheorem{Theorem}{Theorem}

\newtheorem{Proposition}{Proposition}

\newcommand{\bi}{\begin{itemize}}
\newcommand{\ei}{\end{itemize}}

\begin{document}
	\thispagestyle{empty} \baselineskip=28pt \vskip 5mm
	\begin{center} {\Large{\bf Semiparametric Estimation of Cross-covariance Functions for Multivariate Random Fields}}
	\end{center}
	
	\baselineskip=12pt \vskip 5mm
	
	\begin{center}\large
		Ghulam A. Qadir\footnote[1]{
			\baselineskip=10pt CEMSE Division, King Abdullah University of Science and Technology, Thuwal 23955-6900, Saudi Arabia.
			E-mail:  ghulam.qadir@kaust.edu.sa; ying.sun@kaust.edu.sa
		} and Ying Sun$^1$ \end{center}
	
	\baselineskip=16pt \vskip 1mm \centerline{\today} \vskip 8mm
	
	\begin{center}
		{\large{\bf Abstract}}\\
	\end{center}
    The prevalence of spatially referenced multivariate data has impelled researchers to develop a procedure for the joint modeling of multiple spatial processes.  This ordinarily involves modeling marginal and cross-process dependence for any arbitrary pair of locations using a multivariate spatial covariance function. However, building a flexible multivariate spatial covariance function that is nonnegative definite is challenging. Here, we propose a semiparametric approach for multivariate spatial covariance function estimation with approximate Mat{\'e}rn marginals and highly flexible cross-covariance functions via their spectral representations. The flexibility in our cross-covariance function arises due to B-spline based specification of the underlying coherence functions, which in turn allows us to capture non-trivial cross-spectral features. We then develop a likelihood-based estimation procedure and perform multiple simulation studies to demonstrate the performance of our method, especially on the coherence function estimation. Finally, we analyze particulate matter concentrations ($\text{PM}_{2.5}$) and wind speed data over the North-Eastern region of the United States, where we illustrate that our proposed method outperforms the commonly used full bivariate Mat{\'e}rn model and the linear model of coregionalization for spatial prediction.	\baselineskip=17pt

	\begin{doublespace}
		
		\par\vfill\noindent
		{\bf Some key words}: Coherence, co-kriging, Mat{\'e}rn covariance, nonnegative definite, multivariate spatial data.
	\par\medskip\noindent
		 
	\end{doublespace}
	
	\clearpage\pagebreak\newpage \pagenumbering{arabic}
	\baselineskip=26.5pt
	

\section{Introduction}\label{sec:intro}
Recent technological advances have led to the exposition of spatially indexed multivariate data in a wide range of applications, such as, for instance, in geophysical, environmental and atmospheric sciences, to name but a few \citep{sain2011,greasby2011}. This has motivated and facilitated researchers to jointly model multiple spatial processes for gaining scientific insights into the dynamics within each variable and between distinct variables.  Modeling spatial data conventionally involves quantifying spatial dependence through valid covariance functions, which call for marginal and cross-covariance functions in the case of multivariate spatial data. Let $\textbf{X}(\textbf{s})=\big(X_1(\textbf{s}),\dots,X_p(\textbf{s})\big)^\text{T}$ be a $p$-variate zero mean Gaussian random field defined on a spatial domain $\mathcal{D}\subset\mathbb{R}^d,d\geq1$. Under the assumption of second-order stationarity, the covariance functions associated with $\textbf{X}(\textbf{s})$ are defined as:\[\textrm{C}_{ij}(\textbf{s}_1-\textbf{s}_2)=\mathbb{E}[X_i(\textbf{s}_1)X_j(\textbf{s}_2)], \;i,j=1,\dots,p,\; \textbf{s}_1,\textbf{s}_2\in\mathcal{D},\]where $\text{C}_{ii}(\cdot),\;i=1,\dots,p$ are the marginal covariance functions that describe the spatial dependence of the $i^{th}$ process component $\{X_i(\textbf{s}):\textbf{s}\in \mathcal{D}\}$, whereas $\text{C}_{ij}(\cdot),\;1\leq i\neq j\leq p$, often termed as the cross-covariance function, describes the spatial dependence between $i^{th}$ and $j^{th}$ process components. If the covariance function depends on the spatial lag $\textbf{s}_1-\textbf{s}_2$ only through its Euclidean norm, i.e., $\|\textbf{s}_1-\textbf{s}_2\|$, then the random field $\textbf{X}$ is said to be isotropic. The assumptions of stationarity and isotropy state that the covariances are invariant under rigid transformations of the coordinates, and hence may seem unrealistic for many applications. However, this class of models is important, as they form the basic ingredients for more complex and sophisticated non-stationary and anisotropic models. Construction of a valid and flexible model for multivariate covariances entails the difficulty of guaranteeing the nonnegative definiteness, or the nonnegative definite covariance matrix $\mathbf{\Sigma}$ for the random vector $\big(\textbf{X}(\textbf{s}_1)^\text{T},\dots,\textbf{X}(\textbf{s}_n)^\text{T}\big)^\text{T}\in\mathbb{R}^{np}$. Specifically, the main challenge is to build a flexible model for $\text{C}_{ij}(\cdot)$ that yields {$\mathbf{\Sigma}$}, and ensures $\textbf{c}^\text{T}\mathbf{\Sigma}\textbf{c}\geq 0$ for any nonzero vector $\textbf{c}\in\mathbb{R}^{np}$, any set of spatial coordinates $\textbf{s}_1,\dots,\textbf{s}_n$, and any positive integer $n$.

The growing interest in building models for multivariate spatial fields has led to the development of a fairly rich literature in the last few decades, and a comprehensive summary of the existing approaches can be found in the review paper \cite{genton2015}. Many of these models have their genesis in combining univariate covariance functions. Perhaps the most rudimentary modeling approach is to introduce separability by setting $\text{C}_{ij}(\textbf{s}_1-\textbf{s}_2)=\textbf{A}\text{C}(\textbf{s}_1-\textbf{s}_2)$, where $\textbf{A}$ is a $p\times p$ nonnegative definite matrix, and $\text{C}(\cdot)$ is any valid univariate covariance function \citep{mardia1993,Helterbrand1994,article}. Such a specification enforces the same shape of covariance function for all the marginal and cross components, which inhibits its use for modeling complex dependencies. The linear model of coregionalization (LMC) is another univariate covariance function based model, which decomposes the multivariate random field as a linear combination of independent univariate random fields \citep{Goulard1992,doi:10.1029/2002JD002905,wackernagel2010multivariate,doi:10.1002/env.807}. The roughest underlying univariate field in the LMC governs the smoothness of all the components of a multivariate random field, making it inflexible for modeling distinct smoothness in components. \cite{doi:10.1093/biomet/asp078} introduced an approach that can produce flexible multivariate models with distinct smoothnesses in each component while controlling nonseparability. However, this approach involves representing a multivariate random field as a univariate random field in a higher dimensional Euclidean space, which in turn requires the estimation of latent dimensions for each component. Moreover, kernel convolution \citep{ver1998constructing,doi:10.1198/1061860043498} and covariance convolution \citep{doi:10.1002/qj.49712555417,doi:10.1256/qj.05.08,Majumdar2007} methods are other popular univariate covariance function based approaches for building valid cross-covariance functions. 

In the context of univariate random fields, the Mat{\'e}rn class \citep{matern,guttorp} has become a preferred choice for modeling covariances, primarily due to its smoothness controlling parameter that governs the correlations at small distances. \cite{bimat} extended this class for multivariate random fields and introduced a matrix-valued covariance function such that both marginal and cross-covariances are of the Mat{\'e}rn type. For the bivariate case ($p=2$), these authors provided full characterization of the parameter values that lead to a valid \textit{full bivariate Mat{\'e}rn} model, whereas for $p>2$, they specified a \textit{parsimonious multivariate Mat{\'e}rn} model that admits only common spatial scale parameters and constrained smoothness parameters. Further generalization of this idea in \cite{multimat} provided sufficient validity conditions on the parameter space for any $p>1$ and introduced the \textit{flexible multivariate Mat{\'e}rn} model.

Recently \cite{cohkl} analyzed the spectral properties of a number of existing multivariate spatial models, and pointed out that many of them are not sufficiently flexible to capture non-trivial coherence between components. For instance, separable, kernel convolution and the  parsimonious multivariate Mat{\'e}rn model impose constant coherence between components. The full bivariate Mat{\'e}rn model although is quite flexible as its parameters can control the decay rate of coherence at high frequency, as well as  supervise the frequency of the greatest coherence, its flexibility is limited to its parametric form of coherence function that can capture only certain shapes of coherence and not beyond that. For example, the full bivariate Mat{\'e}rn model cannot comprehend a multivariate process with an underlying coherence function that shows oscillations or multiple peaks. In this article, we propose a semiparametric multivariate spatial covariance model with highly flexible underlying coherence functions. The proposed model specifies an approximate Mat{\'e}rn marginal for each component and highly flexible cross-covariances for every pair of components. We specify the coherence functions as a linear combination of cubic splines (B-splines of order 4). Such a specification enables our coherence functions to represent a wide range of smooth curves and allows us to model non-trivial coherence between every pair of process components. The flexibility of our coherence functions is also reflected in the corresponding cross-covariances in the space domain. Additionally, we enact the exact likelihood based inference method jointly for both the parametric marginal and nonparametric coherence function in the proposed model, for both the regularly and irregularly spaced multivariate spatial data.

The rest of our paper is organized as follows. In Section~\ref{sec:method}, we describe the construction of our model and its properties. We also provide sufficient conditions on B-spline coefficients to ensure the validity of our model. We perform multiple simulation studies to explore the performance of our model in Section~\ref{sec:simulation}. In particular, we estimate the coherence of the processes generated from the full bivariate Mat{\'e}rn model and the LMC, using our model with maximum likelihood estimation (MLE). In Section~\ref{sec:app}, we illustrate the application of our proposed model on a bivariate dataset of particulate matter concentrations ($\text{PM}_{2.5}$) and wind speed over the North-Eastern region of the United States. We compare our model with the full bivariate Mat{\'e}rn model and the LMC on the basis of commonly used prediction scores. We conclude in Section~\ref{sec:disc} with a discussion and potential future extension. 

\section{Multivariate Spatial Model}\label{sec:method}
In this section, we introduce our proposed semiparametric model through its origin in the spectral domain, and provide sufficient conditions to ensure its validity. We revisit some notions and concepts of spectral domain in Section~\ref{subsec:spec} that are crucial to our model construction in Section~\ref{subsec:semimod}. 
\subsection{Spectral Representation}\label{subsec:spec}
Let $\textbf{X}(\textbf{s})=\big(X_1(\textbf{s}),\dots,X_p(\textbf{s})\big)^\text{T}$ be a $p$-variate weakly stationary random field defined on a spatial domain $\mathcal{D}\subset\mathbb{R}^d,\;d\geq1$, and $\textbf{C}(\textbf{h})=\{\text{C}_{ij}(\textbf{h})\}_{i,j=1}^p$ be a matrix valued covariance function for $\textbf{X}$ such that $\text{C}_{ij}(\textbf{h})=\text{Cov}\big(X_i(\textbf{s}),X_j(\textbf{s}+\textbf{h})\big)$. The validity of $\textbf{C}(\cdot)$ is generally ensured by using the Cram{\'e}r's Theorem \citep{cramer} in its spectral density version (\citealp[p.~215]{wackernagel2010multivariate}; \citealp{cohkl}) which states that: \\\emph{The necessary and sufficient condition for the matrix valued function} $\textbf{C}:\mathbb{R}^d\rightarrow\mathbb{C}^{p\times p}$, $\textbf{C}(\textbf{h})=\{\text{C}_{ij}(\textbf{h})\}_{i,j=1}^p$ \emph{to be nonnegative definite is its representation as} 
\begin{equation}\label{eq1}
    \text{C}_{ij}(\textbf{h})=\int_{\mathbb{R}^d}\text{exp}(\imath\textbf{u}^{\text{T}}\textbf{h})g_{ij}(\textbf{u})\text{d}\textbf{u},\;\;\;(\imath=\sqrt{-1}),
\end{equation}\emph{for} $i,j=1,\dots,p$ \emph{such that the matrix} $\textbf{g}(\textbf{u})=\{g_{ij}(\textbf{u})\}_{i,j=1}^p$ \emph{is nonnegative definite for all} $\textbf{u}\in\mathbb{R}^d.$\\Here the functions $g_{ij}:\mathbb{R}^d\rightarrow\mathbb{C}$, such that $g_{ij}(\textbf{u})=\overline{g_{ji}(\textbf{u})}$, are the spectral densities for marginal and cross-covariance functions, that admit the $d$-dimensional frequencies $\textbf{u}$ as an argument and return a complex or real value. Under the assumption of isotropy,  $g_{ij}(\textbf{u}_1)=g_{ij}(\textbf{u}_2)\; \forall\; i,j=1,\dots,p$  whenever $\|\textbf{u}_1\|=\|\textbf{u}_2\|$ and therefore (\ref{eq1}) can be reduced to a one dimensional integral \cite[p.~42-44]{stein2012}:\begin{equation}\label{eq2}
    \text{C}_{ij}(\textbf{h})=\int_0^\infty\|\textbf{h}\|\Bigg(\frac{2\pi\omega}{\|\textbf{h}\|}\Bigg)^{\kappa+1} J_\kappa(\omega\|\textbf{h}\|)f_{ij}(\omega)\text{d}\omega
,\end{equation}where $\omega=\|\textbf{u}\|\geq0$, $\kappa=\frac{d}{2}-1$, $J_\kappa(\cdot)$ is a Bessel function of the first kind of order $\kappa$ \citep{watson1995treatise} and $f_{ij}:\mathbb{R}\rightarrow\mathbb{C}$ are the isotropic spectral densities such that $g_{ij}(\textbf{u})=f_{ij}(\|\textbf{u}\|),\;\forall\textbf{u}\in\mathbb{R}^d,\;i,j=1,\dots,p$.

For given spectral densities $\{g_{ij}(\cdot),\;i,j=1,\dots,p\}$, the coherence between the $i^{th}$ and $j^{th}$ components of the process $\textbf{X}$ at a particular frequency $\textbf{u}$ is defined as:\begin{equation}\label{eq3}
    \gamma_{ij}(\textbf{u})=\frac{g_{ij}(\textbf{u})}{\sqrt{g_{ii}(\textbf{u})g_{jj}(\textbf{u})}}\:\forall\;1\leq i\neq j \leq p.
\end{equation}Coherence functions in general can be complex-valued depending on the codomain of the spectral densities $\{g_{ij}(\cdot),\;i,j=1,\dots,p\}$, and therefore absolute coherence functions $|\gamma_{ij}(\cdot)|$ are examined in practice. The isotropic version of the coherence function can be obtained trivially by replacing the argument $\textbf{u}$ by $\omega$ and functions $g_{ij}$ by $f_{ij}$ in (\ref{eq3}).
For a more detailed account on coherence functions in spatial case, we refer readers to \cite{cohkl}. In the subsequent sections, we develop our semiparametric multivariate covariance functions using the above-mentioned notions.
\subsection{Semiparametric Multivariate Spatial Model}\label{subsec:semimod}
We consider the isotropic spectral densities $\{f_{ij}(\cdot),\;i,j=1,\dots,p\}$ up to a certain sufficiently large threshold frequency $\omega_t$. We choose the marginal spectral densities $\{f_{ii}(\cdot),i=1,\dots,p\}$ to be of Mat{\'e}rn type \cite[A.1]{bimat}, truncated for frequencies greater than $\omega_t$, i.e,\begin{equation}\label{eq4}
    f_{ii}(\omega|\sigma_i,\nu_i,a_i)=\sigma_i^2\frac{\Gamma(\nu_i+d/2)a_i^{2\nu_i}}{\Gamma(\nu_i)\pi^{d/2}(a_i^2+\omega^2)^{\nu_i+d/2}}, \: 0\leq\omega\leq\omega_t,\: \sigma_i,\nu_i,a_i>0.
\end{equation} 
The untruncated version of (\ref{eq4}) corresponds to the spectral density of the isotropic Mat{\'e}rn covariance function \citep{matern,guttorp} :\[\text{M}(\textbf{h}|\sigma,\nu,a)=\sigma^2\frac{2^{1-\nu}}{\Gamma(\nu)}(a\|\textbf{h}\|)^\nu K_\nu(a\|\textbf{h}\|),\]where $\sigma>0$ is the marginal standard deviation, $a>0$ represents a spatial scale parameter, $\nu>0$ is a smoothness parameter and $K_\nu$ is a modified Bessel function of the second kind of order $\nu$.

For given marginal spectral densities in (\ref{eq4}), we specify the cross-spectral densities $\{f_{ij}(\cdot), 1\leq i< j \leq p\}$ using the linear combination of B-splines as follows:
\begin{equation}\label{eq5}
   f_{ij}(\omega|f_{ii},f_{jj},\textbf{S}_{ij},K)=\sum_{k=-3}^Kb_k^{(ij)}B_k(\omega)\sqrt{f_{ii}(\omega)f_{jj}(\omega)}, \:0\leq\omega\leq\omega_t, 
\end{equation}where $B_k$'s are the cubic splines (B-splines of order 4) (\citealp[chapter~IX]{de2001}; \citealp*{techreport}), for a sequence of uniform knots $\big(-3\Delta,\dots,0,\Delta,2\Delta,\dots,(K+1)\Delta\big)$ such that $\omega_t\in\big(K\Delta,(K+1)\Delta]$, and $\{b_k^{(ij)},\;k=-3,\dots,K,\; 1\leq i< j \leq p\}$ are the B-spline coefficients. We begin the B-splines combinations from $k=-3$ to $k=K$ in order to include all the B-splines that have support on the interval $[0,\omega_t]$. \cite{semipzhu} used a similar B-spline representation for defining the nonparametric part of their univariate semiparametric spectral density. Here $K$ supervises the number of knots, $\Delta$ represents its uniform spacing and $\textbf{S}_{ij}$ constitutes the set of coefficients required to fully specify the B-spline part of (\ref{eq5}), i.e., $\textbf{S}_{ij}=\{b_k^{(ij)},\;k=-3,\dots,K\},\;1\leq i< j \leq p$. Note that the cross-spectral densities specified in (\ref{eq5}) are real valued, therefore $f_{ij}(\cdot)=f_{ji}(\cdot),\;\forall\; 1\leq i <j \leq p$, and consequently $\textbf{S}_{ij}=\textbf{S}_{ji},\;\forall\; 1\leq i <j \leq p$.  We choose the B-spline of order 4, however, a higher order B-spline can also be incorporated in (\ref{eq5}) with only slight modifications.

Following the definition in (\ref{eq3}), the coherence between  $i^{th}$ and $j^{th}$ process components at frequency $\omega$ for the spectral densities specified in (\ref{eq4}) and (\ref{eq5}) is given as:\begin{equation}\label{eq:coh}\gamma_{ij}(\omega)=\sum_{k=-3}^Kb_k^{(ij)}B_k(\omega),\:0\leq\omega\leq\omega_t.\end{equation} Here, our specified spectral densities lead to fully nonparametric  coherence functions based on the linear combination of B-splines that can accommodate a wide range of smooth functions, and therefore induces a great deal of flexibility in our proposed coherence model that can be controlled by the value of $\Delta$. The smaller values of $\Delta$ produce more flexible coherence functions, however, it makes the estimation computationally challenging due to a large number of B-spline coefficients, whereas the large values of $\Delta$ generate relatively less flexible coherence functions, but the estimation is computationally more feasible due to a smaller number of B-spline coefficients. For an appropriate choice of $\Delta$, our proposed approach can model coherence functions that are beyond the comprehension of existing multivariate models.

\begin{figure}[!t]
\centering     
\subfigure[]{\label{fig:1a}\includegraphics[width=60mm]{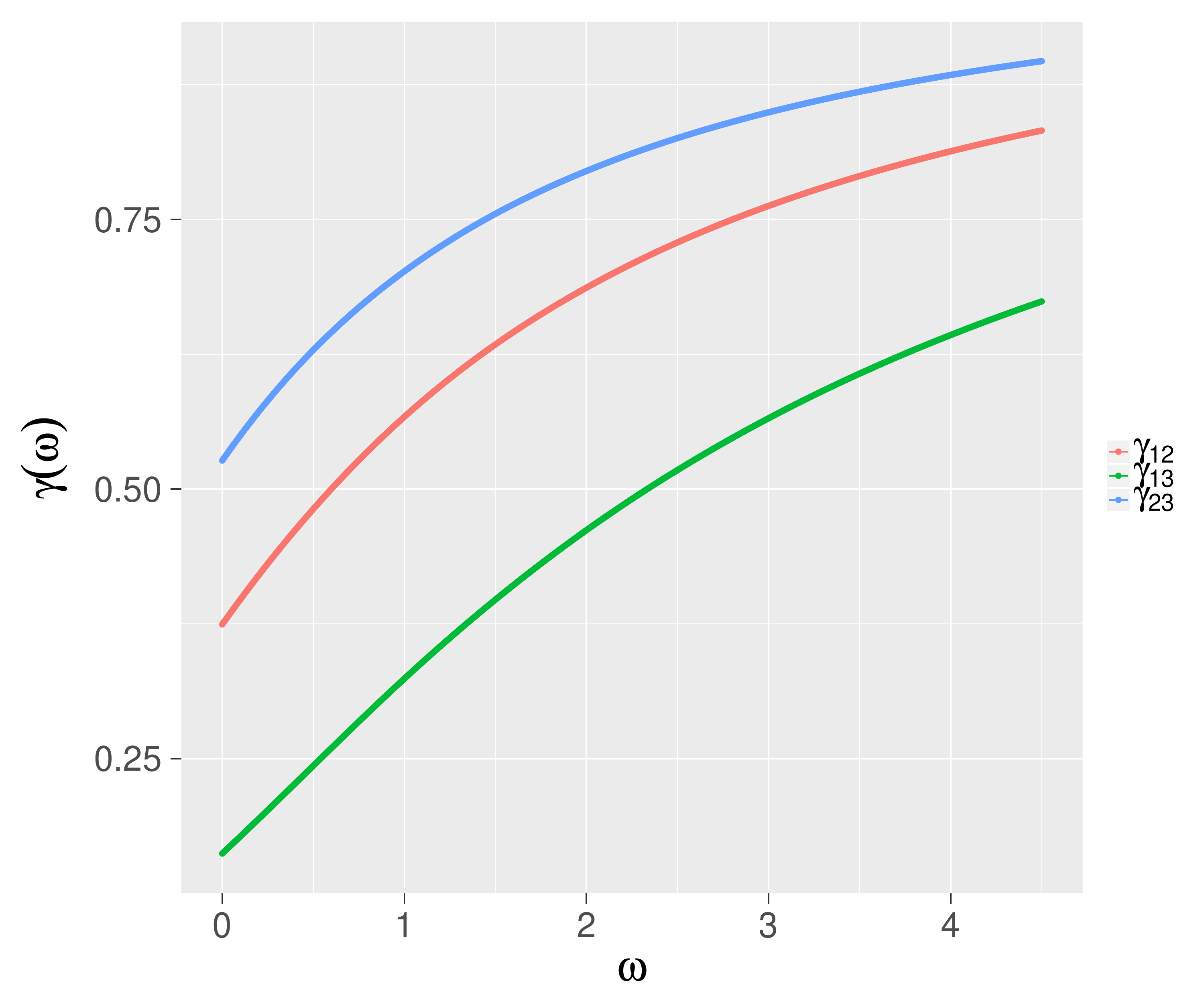}}
\subfigure[]{\label{fig:1b}\includegraphics[width=60mm]{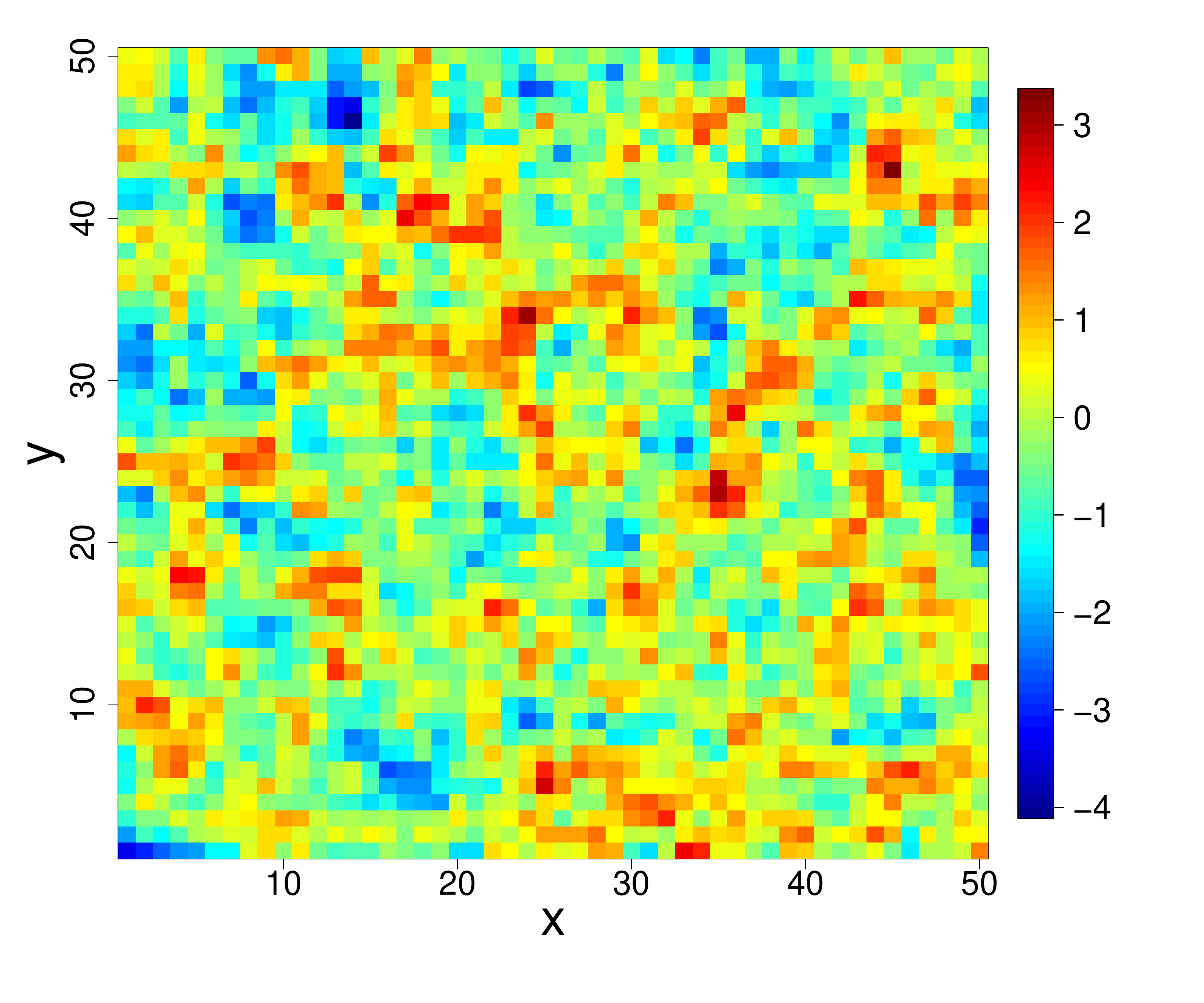}}
\subfigure[]{\label{fig:1c}\includegraphics[width=60mm]{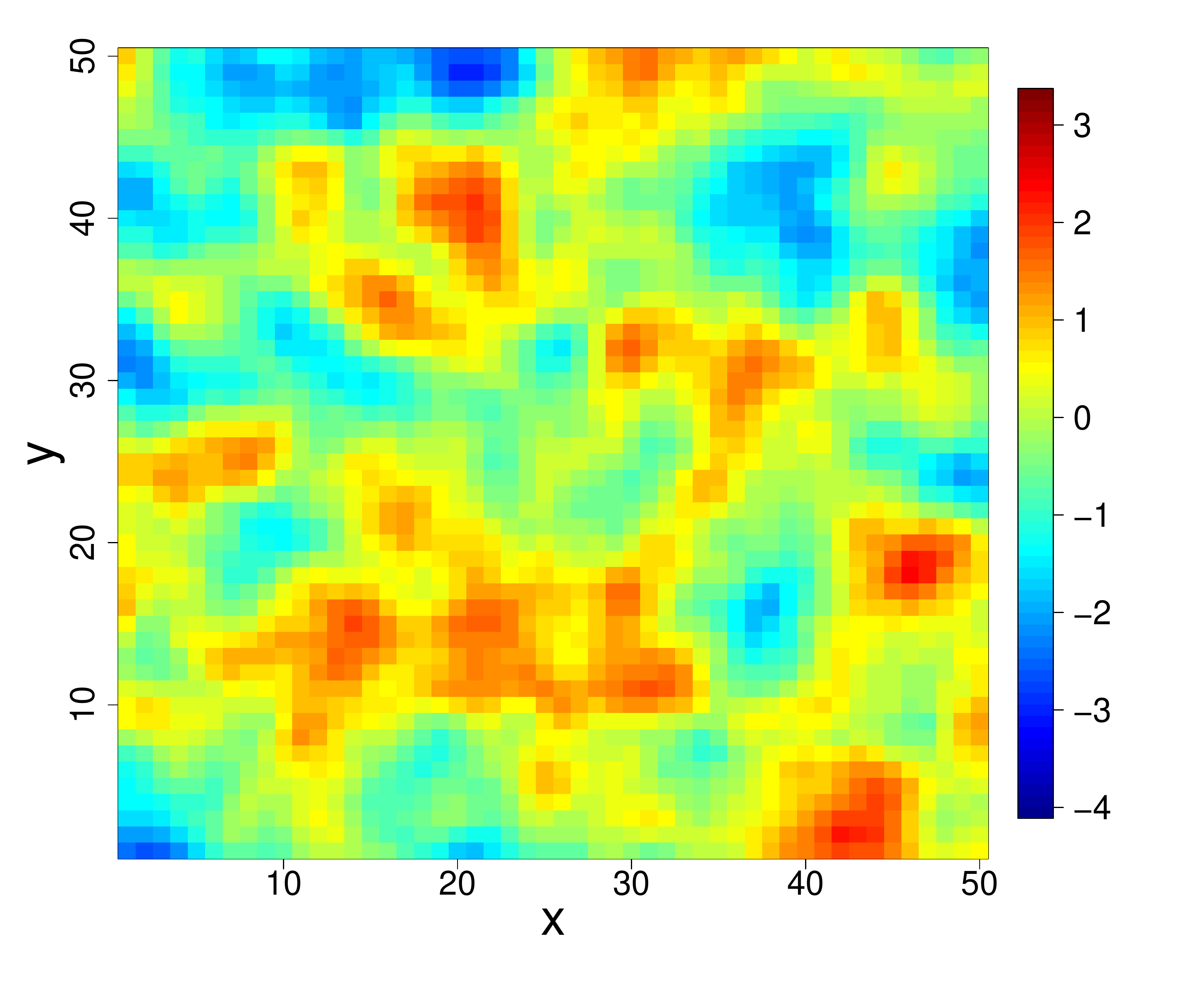}}
\subfigure[]{\label{fig:1d}\includegraphics[width=60mm]{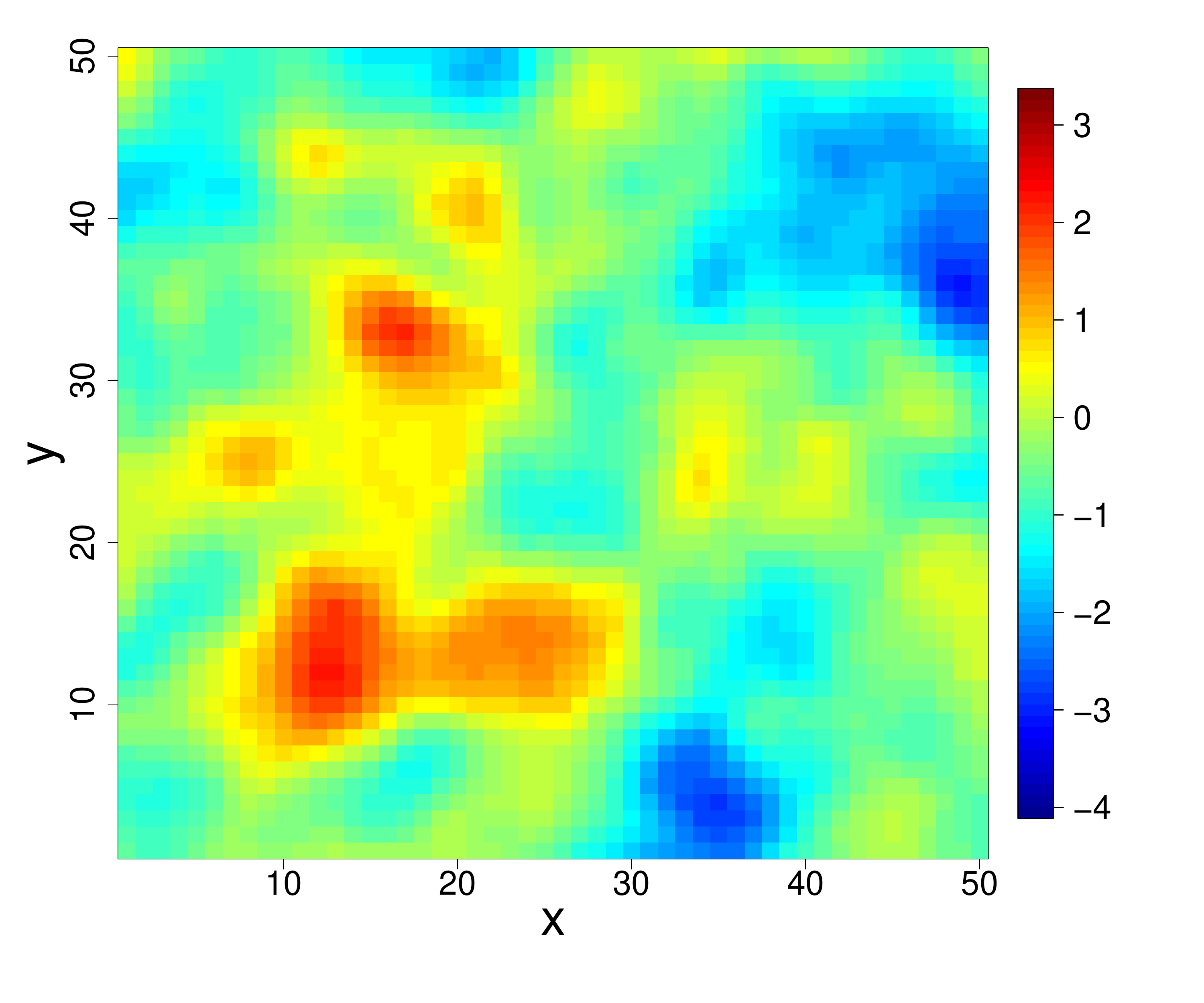}}
\caption{(a) Coherence functions for each pair of variables. (b) Simulated realization for $X_1$ $(\sigma_1=1,\nu_1=1,a_1=1)$. (c) Simulated realization for $X_2$ $(\sigma_2=1,\nu_2=2,a_2=0.5)$. (d) Simulated realization for $X_3$ $(\sigma_3=1,\nu_3=2.5,a_3=0.4)$.} 
\label{fig:1}
\end{figure}

\begin{figure}[!t]
\centering     
\subfigure[]{\label{fig:dec1a}\includegraphics[width=50mm]{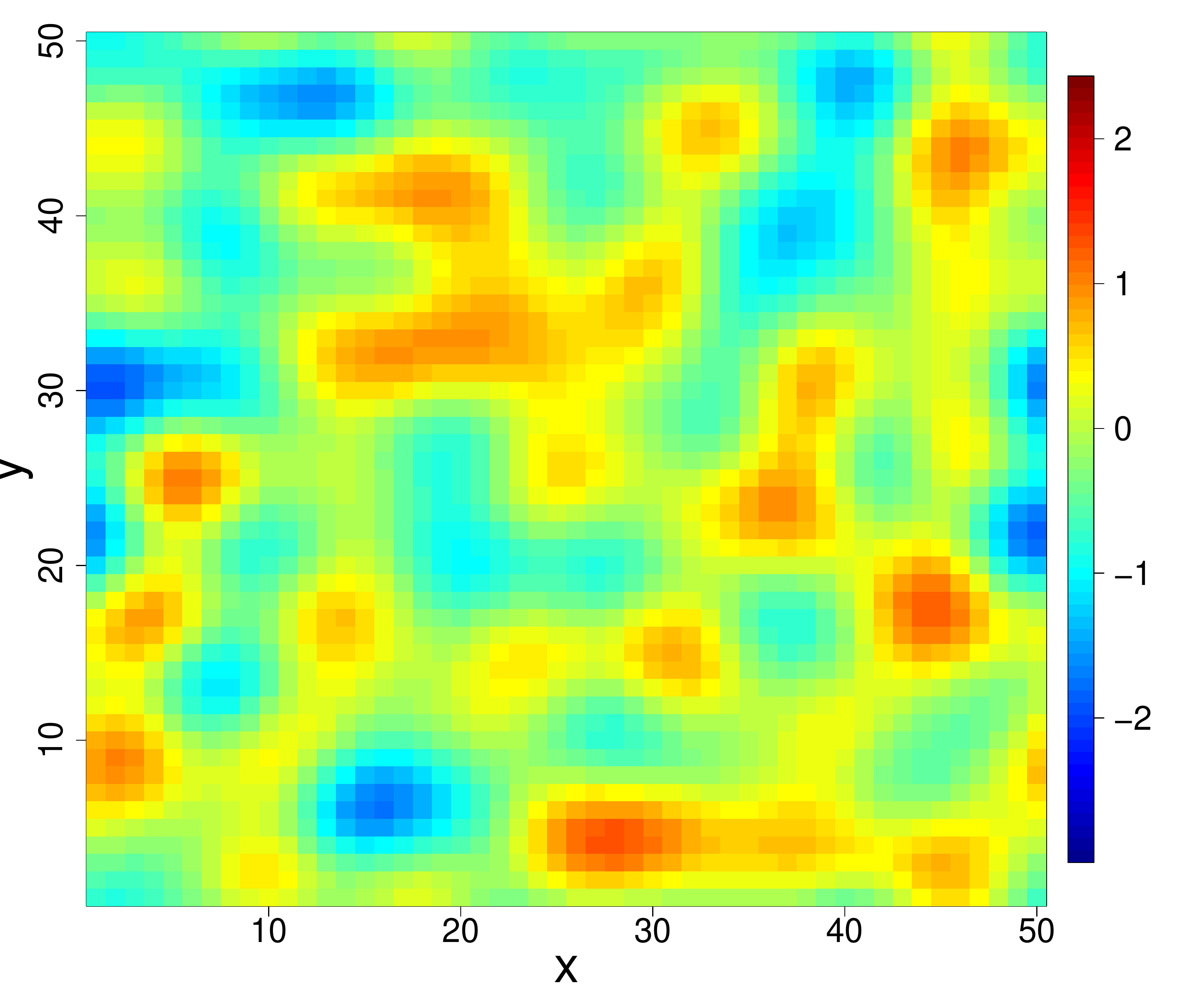}}
\subfigure[]{\label{fig:dec1b}\includegraphics[width=50mm]{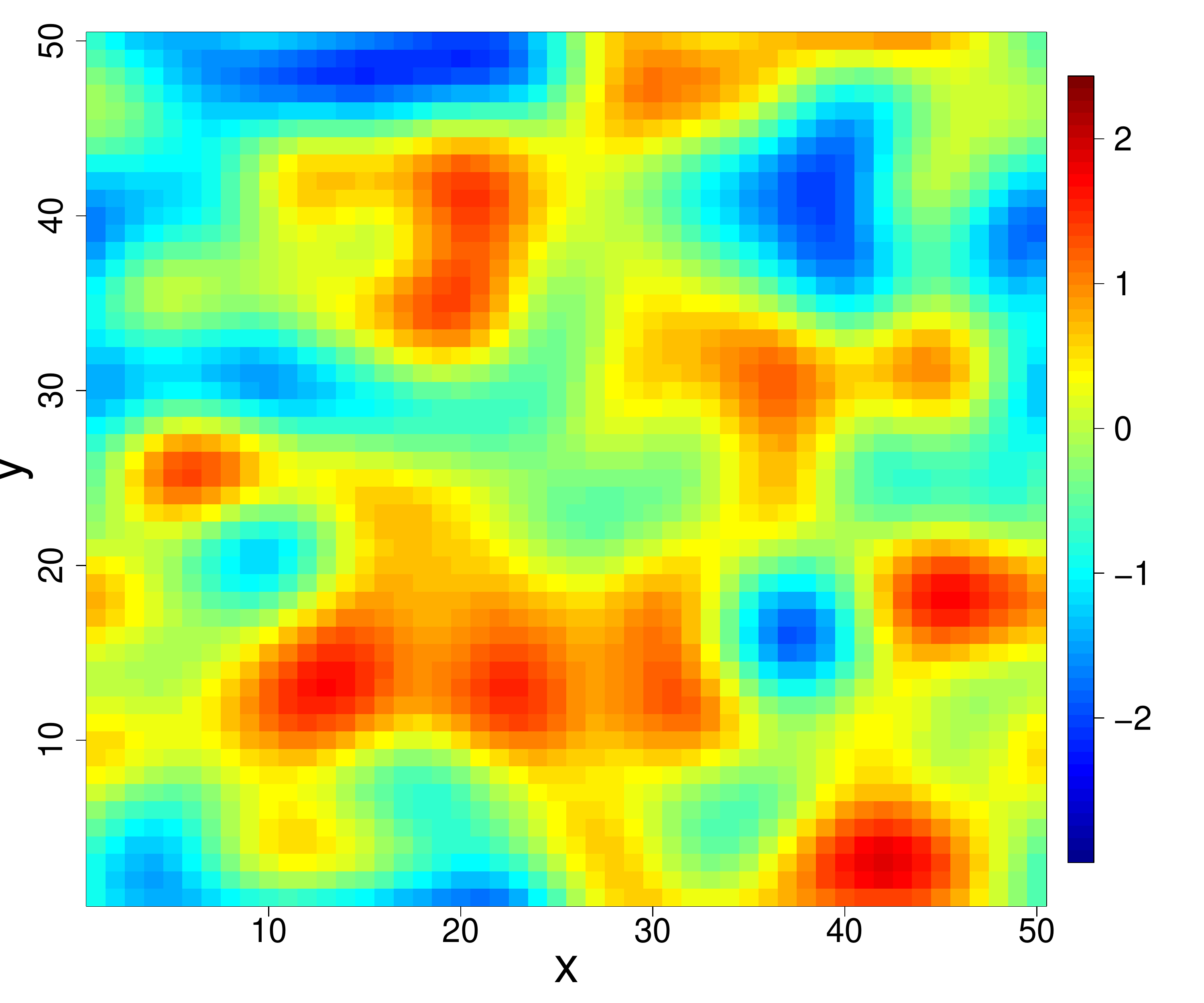}}
\subfigure[]{\label{fig:dec1c}\includegraphics[width=50mm]{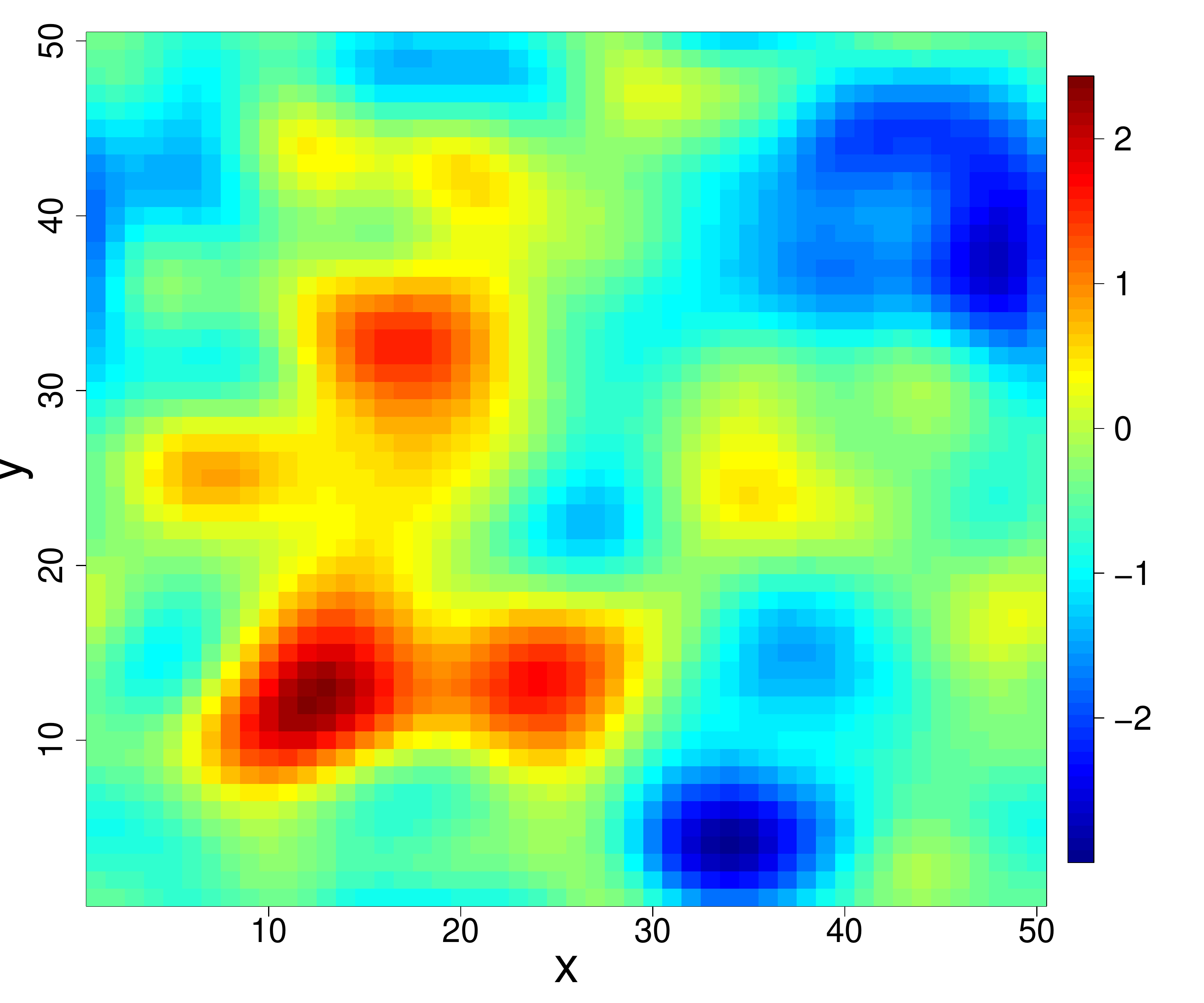}}\\
\subfigure[]{\label{fig:dec1d}\includegraphics[width=50mm]{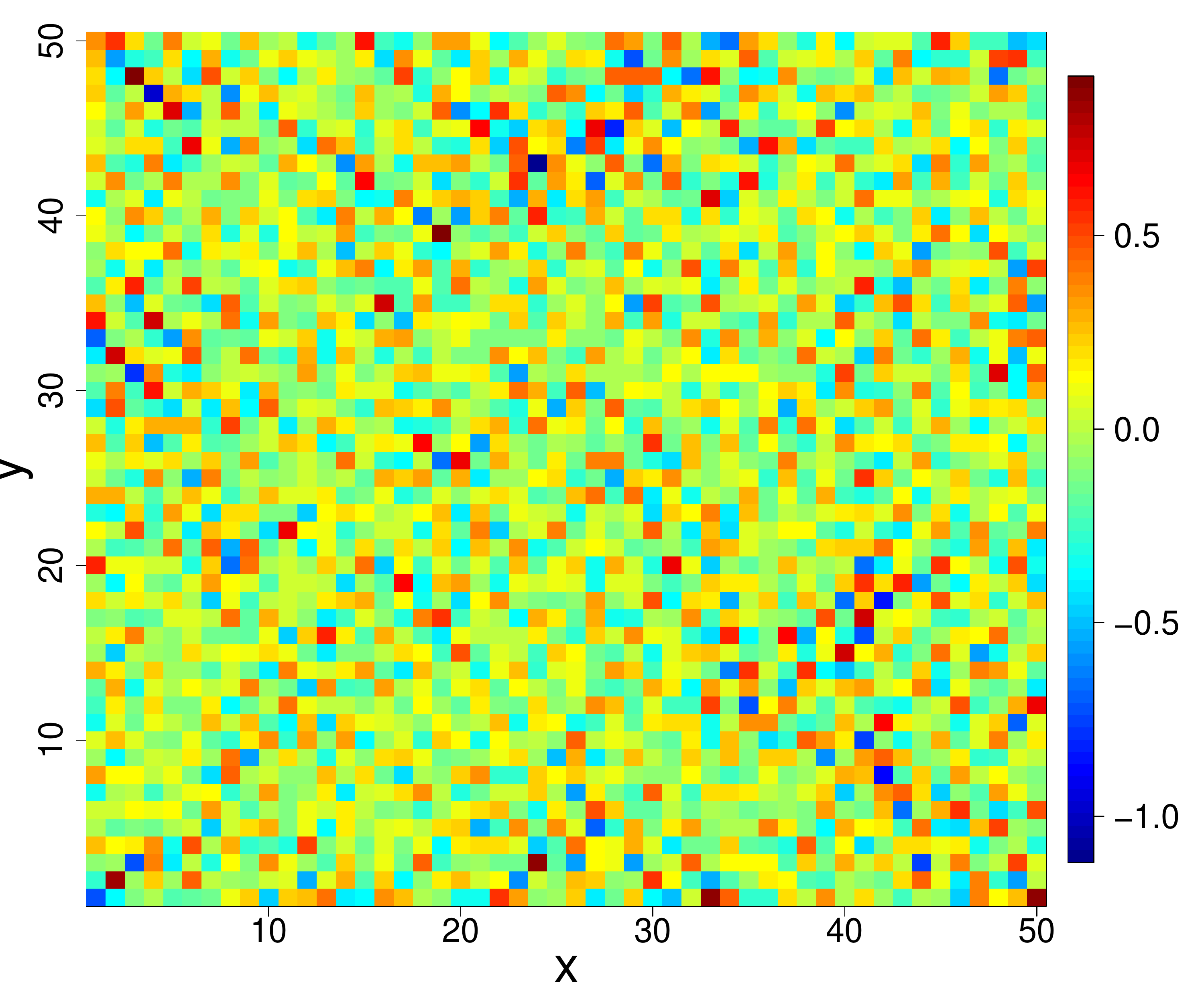}}
\subfigure[]{\label{fig:dec1e}\includegraphics[width=50mm]{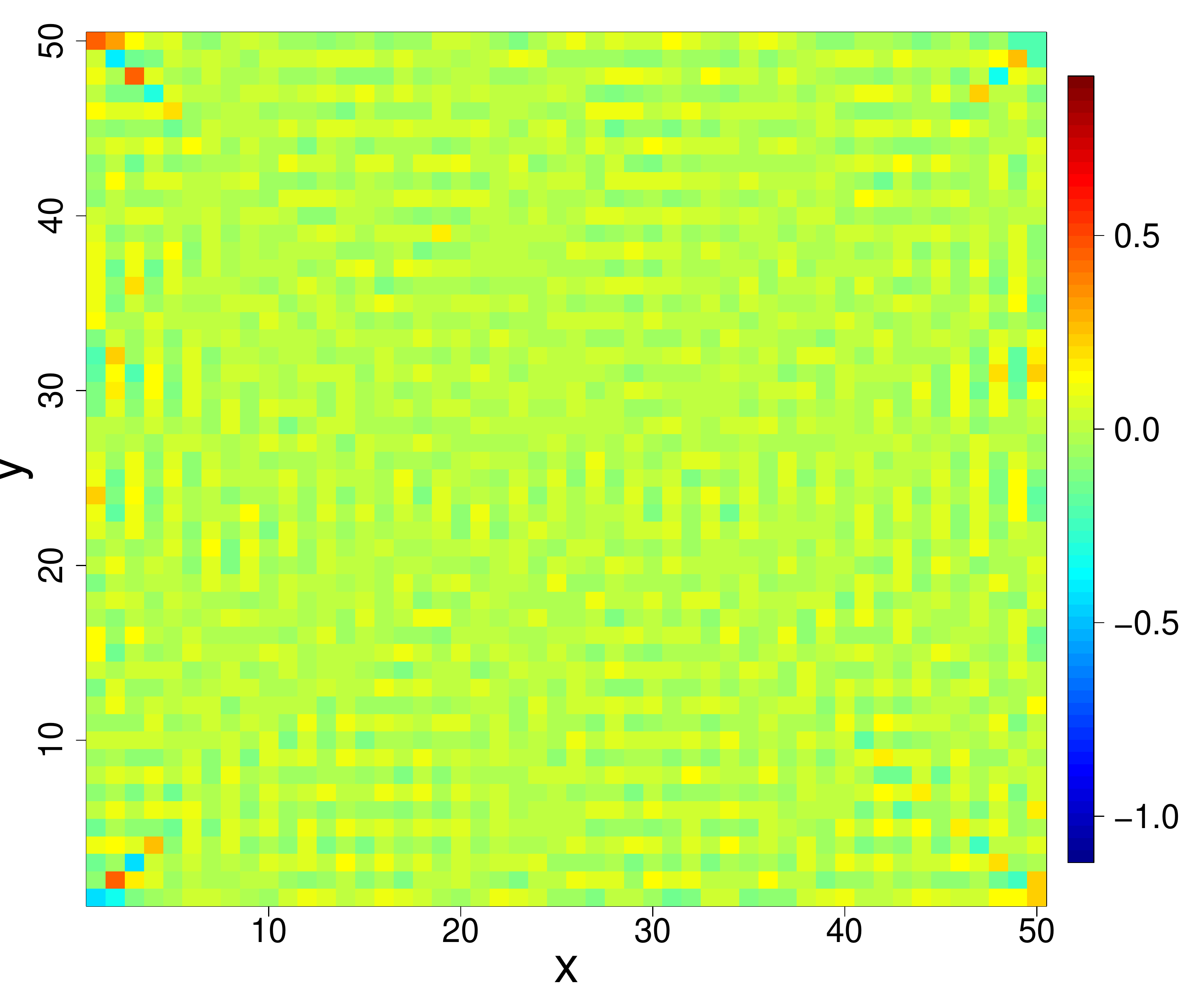}}
\subfigure[]{\label{fig:dec1f}\includegraphics[width=50mm]{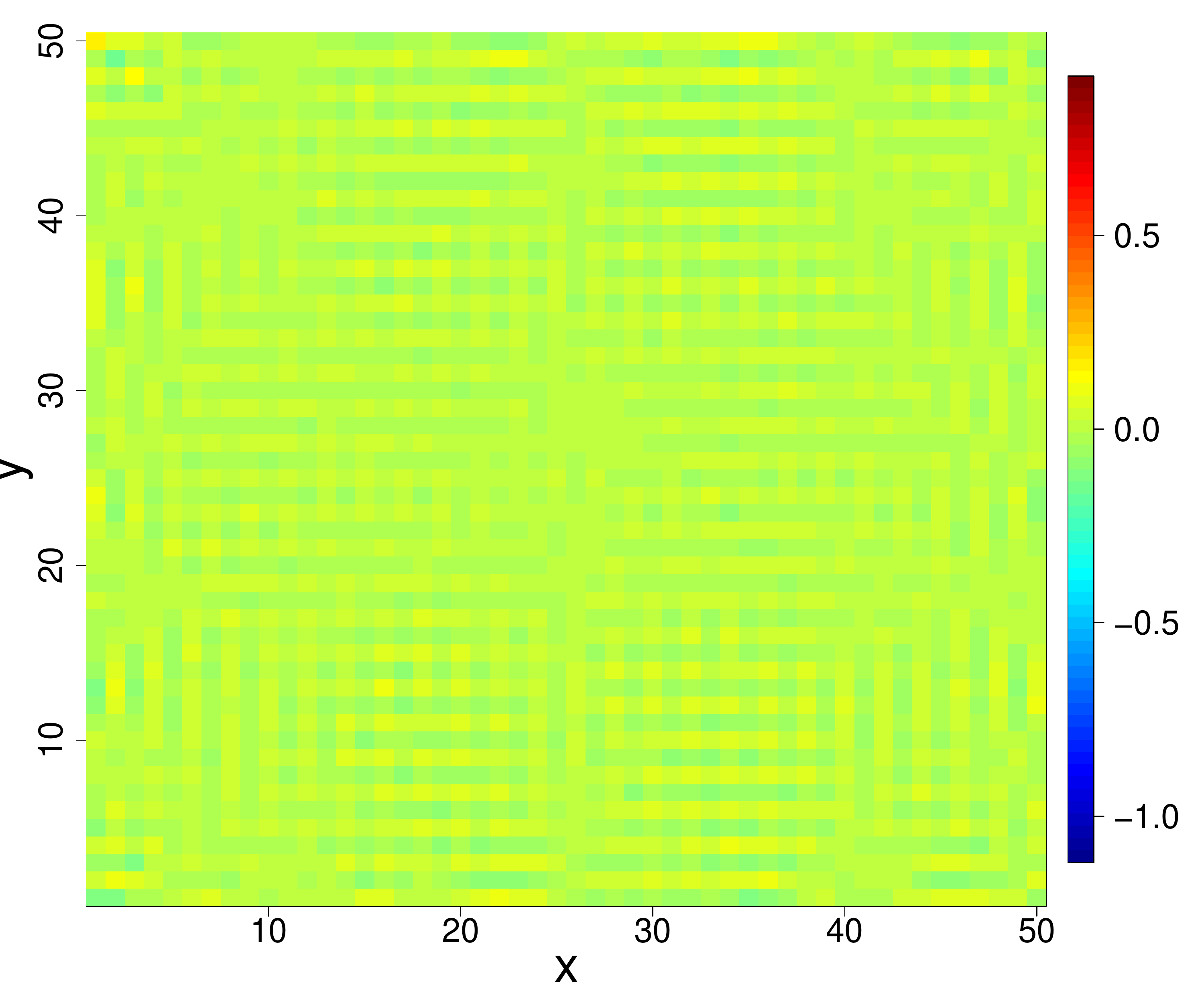}}\caption{(a) $\check{X_1}^{lf}$ $(X_1$ filtered at $0\leq\omega\leq 1)$. (b) $\check{X_2}^{lf}$ ($X_2$ filtered at $0\leq\omega\leq 1)$. (c) $\check{X_3}^{lf}$ $(X_3$ filtered at $0\leq\omega\leq 1)$. (d) $\check{X_1}^{hf}$ $(X_1$ filtered at $3.25\leq\omega\leq 4.25).$ (e) $\check{X_2}^{hf}$ $(X_2$ filtered at $3.25\leq\omega\leq 4.25)$. (f) $\check{X_3}^{hf}$ $(X_3$ filtered at $3.25\leq\omega\leq 4.25).$ } 
\label{fig:dec1}
\end{figure}

In order to obtain the multivariate covariance functions from any given isotropic marginal and cross spectral densities, we resort to the integral (\ref{eq2}), also known as the Hankel transform of the order $\kappa$. However, in our proposed framework, integral (\ref{eq2}) cannot be computed for the spectral densities defined in (\ref{eq4}) and (\ref{eq5}) because of their truncation to $\omega_t$ and unknown closed form solutions. Consequently, we choose a small value of $\delta$ to define a discrete set of frequencies $\mathcal{F}=\{\delta,\dots,m\delta\}$ such that $m\delta=\omega_t$, and then we compute the following finite sum approximation of (\ref{eq2}) to obtain the multivariate spatial covariance function:\begin{equation}\label{eq6}
    \begin{cases}\text{C}_{ii}(\textbf{h})=\sum\limits_{\omega\in\mathcal{F}}\frac{(2\pi\omega)^{\kappa+1}}{\|\textbf{h}\|^\kappa} J_\kappa(\omega\|\textbf{h}\|)f_{ii}(\omega|\sigma_i,\nu_i,a_i)\delta,\quad  i=1,\dots,p\\
  \text{C}_{ij}(\textbf{h})=\sum\limits_{\omega\in\mathcal{F}}\frac{(2\pi\omega)^{\kappa+1}}{\|\textbf{h}\|^\kappa} J_\kappa(\omega\|\textbf{h}\|)f_{ij}(\omega|f_{ii},f_{jj},\textbf{S}_{ij},K)\delta,\quad  1\leq i\neq j\leq p,
  \end{cases}
 \end{equation} where $\{f_{ij}(\cdot),\;i,j=1,\dots,p\}$ corresponds to the spectral densities defined in (\ref{eq4}) and (\ref{eq5}). The finite sum based approach has been commonly used to propose nonparametric univariate covariance functions \citep{botha1991,gor2002,Gor2004}, however, its extension to a multivariate setting is not very popular yet. For a reasonably small value of $\delta$ (or large value of $m$), a large value of $\omega_t$ and an appropriate normalization of finite sums, the marginal covariance functions $\text{C}_{ii}(\cdot)$ in (\ref{eq6}) are numerically equivalent to the corresponding exact Mat{\'e}rn covariance functions, and hence the parameters $(\sigma_i,\nu_i,a_i,\;i=1,\dots,p)$ retain their interpretations of the exact Mat{\'e}rn. In order to ensure the validity of the cross-covariances $\text{C}_{ij}(\cdot)$ in (\ref{eq6}), we need to impose certain constraints on the set of B-spline coefficients $\textbf{S}_{ij},\; 1\leq i\neq j \leq p$. In Theorem \ref{th1},  we provide sufficient conditions for the validity of our proposed multivariate covariance function $\textbf{C}(\textbf{h})=\{\text{C}_{ij}(\textbf{h})\}_{i,j=1}^p$ in (\ref{eq6}):
\begin{Theorem}\label{th1}
Let $\boldsymbol{\beta}_k=\{b_k^{(ij)}\}_{i,j=1}^p$ $,k=-3,-2,\dots,K$ be the $p\times p$ symmetric matrices with diagonal elements $\{b_k^{(ii)}=1$ $\forall \;i=1,2,\dots,p,\;\;k=-3,-2,\dots,K\}$, then the matrix-valued covariance function $\textbf{C}(\textbf{h})=\{\text{C}_{ij}(\textbf{h})\}_{i,j=1}^p$ in (\ref{eq6}) is valid if the matrices $\{\boldsymbol{\beta}_k,\;k=-3,\dots,K\}$ are nonnegative definite.
\end{Theorem}Figure \ref{fig:1} shows a realization of a trivariate zero mean Gaussian random field $\textbf{X}$, simulated from our proposed model (\ref{eq6}) with threshold frequency $\omega_t=4.5$, and $m=990$ for discretization of frequencies. The coherence functions (shown in Figure \ref{fig:1a}) are generated from suitably selected  $\textbf{S}_{ij},\;1\leq i\neq j \leq p$, such that $X_3$ has the highest coherence with $X_2$ and lowest coherence with $X_1$, at all frequencies. The marginal parameters $(\sigma_1=\sigma_2=\sigma_3=1,\nu_1=1,a_1=1, \nu_2=2,a_2=0.5,\nu_3=2.5,a_3=0.4)$ induce distinct features in the process components, varying from lowest smoothness $(\nu_1)$ and correlation range $(1/a_1)$ in $X_1$ (shown in Figure \ref{fig:1b}), moderate in $X_2$ (shown in Figure \ref{fig:1c}) to the highest smoothness $(\nu_3)$ and correlation range $(1/a_3)$ in $X_3$ (shown in Figure \ref{fig:1d}). The interpretation of the coherence functions become clearer when we look at the filtered signal $\check{\textbf{X}}^{fb}$ of the simulated trivariate dataset $\textbf{X}$ at a frequency band ${fb}$. We apply a low-pass and a high-pass filter to obtain the filtered signals at low frequency $(lf)$ and high frequency $(hf)$ bands. In particular, we consider $lf=0\leq\omega\leq 1$ and $hf=3.25\leq\omega\leq4.25$ to asses the  signal behavior in low frequencies and high frequencies, respectively. Figures \ref{fig:dec1a}-\ref{fig:dec1f} show the filtered signals for the chosen frequency bands. The empirical correlation between filtered signal pairs $(\check{X_1}^{lf},\check{X_2}^{lf})$, $(\check{X_2}^{lf},\check{X_3}^{lf})$ and $(\check{X_1}^{lf},\check{X_3}^{lf})$ are $0.46, 0.55$ and $0.14$, respectively, and for the pairs $(\check{X_1}^{hf},\check{X_2}^{hf})$, $(\check{X_2}^{hf},\check{X_3}^{hf})$ and $(\check{X_1}^{hf},\check{X_3}^{hf})$ the correlations are $0.54, 0.65$ and $0.17$, respectively. The empirical correlations mimic the underlying coherence function as the pair $(\check{X_2}^{fb},\check{X_3}^{fb})$ exhibits the highest correlation and the pair $(\check{X_1}^{fb},\check{X_3}^{fb})$ shows the weakest correlation, at both frequency bands $fb=\{lf, hf\}$. Moreover, similar to the underlying coherence function, all the pairwise correlations  at  $hf$ are stronger than those at $lf$.

The sufficient conditions stated in Theorem \ref{th1} can be corroborated during model estimation by further parameterizing $\boldsymbol{\beta}_k's$, such that $\{\boldsymbol{\beta}_k=\boldsymbol{\Lambda_{\boldsymbol{\theta}_k}},\;k=-3,\dots,K\}$ where $\boldsymbol{\Lambda}_{\boldsymbol{\theta}_k}'s$ essentially are the correlation matrices of size $p \times p$ that allows for both the negative and nonnegative off-diagonal entries that can be derived from any valid correlation function that depends on the set of parameters $\boldsymbol{\theta}_k$. For example, let $\boldsymbol{\theta}_k=\{t_{ij,k}\in\mathbb{R},\;t_{ii,k}=1,\;i=1,\dots,p,\;1\leq i< j\leq p\}$, then  $\boldsymbol{\Lambda_{\boldsymbol{\theta}_k}}=\{\frac{\sum_{l=j}^pt_{il,k}t_{jl,k}}{\sqrt{\sum_{u=i}^p(t_{iu,k})^2}\sqrt{\sum_{v=j}^p(t_{jv,k})^2}}\}_{i,j=1}^p$ is one valid and flexible parameterization that requires the total $(K+4) {{p}\choose{2}}$ parameters to define $\{\boldsymbol{\beta}_k,\;k=-3,\dots,K\}$. Alternatively, we can consider a smaller set $\boldsymbol{\theta}_k=\{t_{i,k}\in\mathbb{R},\;i=1,\dots,p\}$ and define the parameterization as $\boldsymbol{\Lambda_{\boldsymbol{\theta}_k}}=\{\text{exp}(-|t_{i,k}-t_{j,k}|)\}_{i,j=1}^p$, in which case the total number of parameters required to define $\{\boldsymbol{\beta}_k,\;k=-3,\dots,K\}$ is $(K+4)p$, which is much less than $(K+4) {{p}\choose{2}}$. However, this is a relatively less flexible parameterization as it will lead to only positive values of spline coefficients that will produce only positive coherence functions and positive cross-covariance functions, and therefore, should be considered only when the coherence functions are known to be positive for all frequencies. In the case $p=2$, a bivariate  random field, the sufficient conditions are \begin{equation}\label{eq7}
    -1\leq b_k^{(12)}\leq 1,\;k=-3,\dots,K. 
\end{equation}
Thus, the B-spline coefficients should lie between $-1$ to 1 in a bivariate case to ensure that the absolute coherence never exceeds unity at any frequency band.

\begin{figure}[!t]
\centering     
\subfigure[]{\label{fig:2a}\includegraphics[width=45mm]{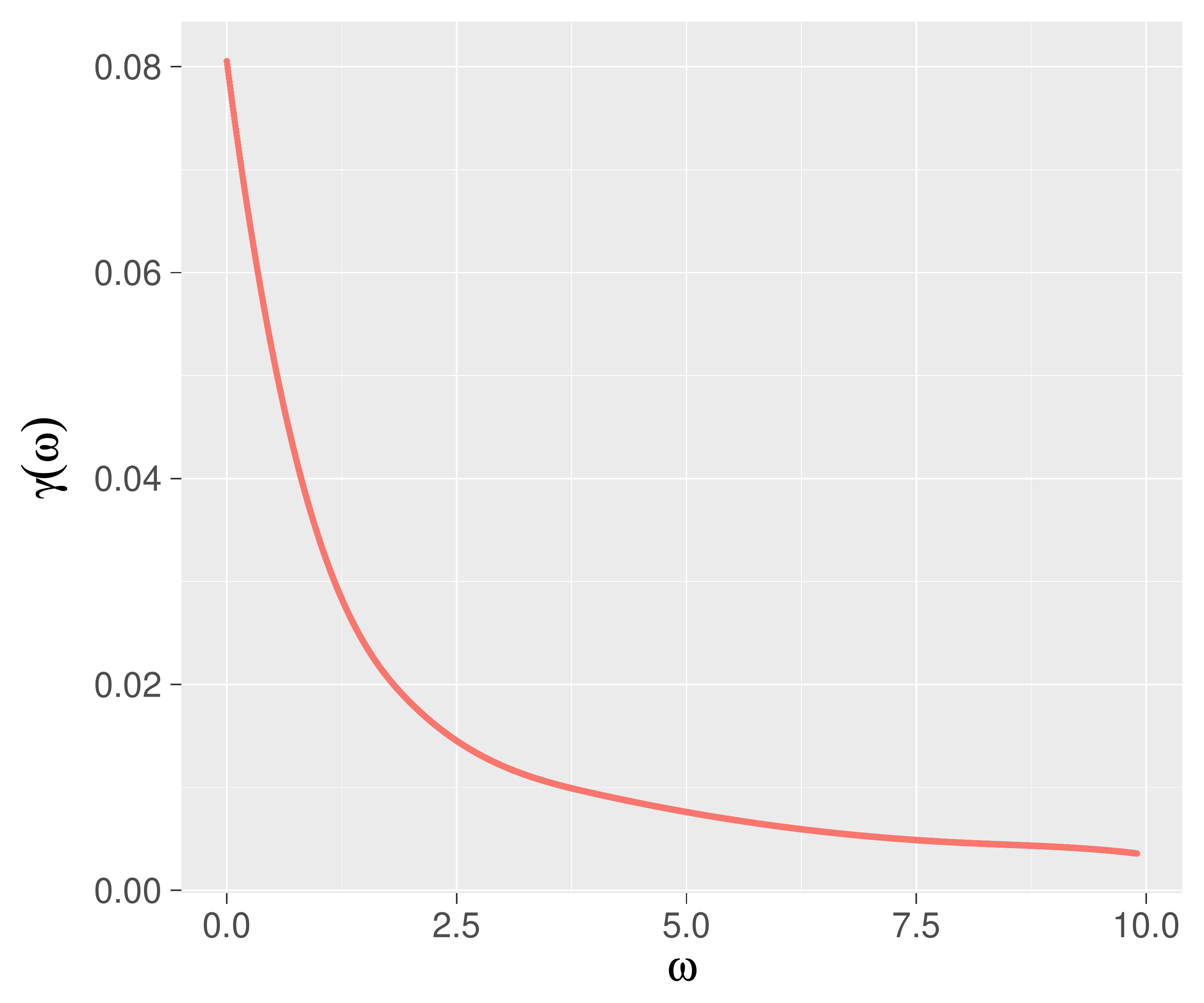}}
\subfigure[]{\label{fig:2b}\includegraphics[width=45mm]{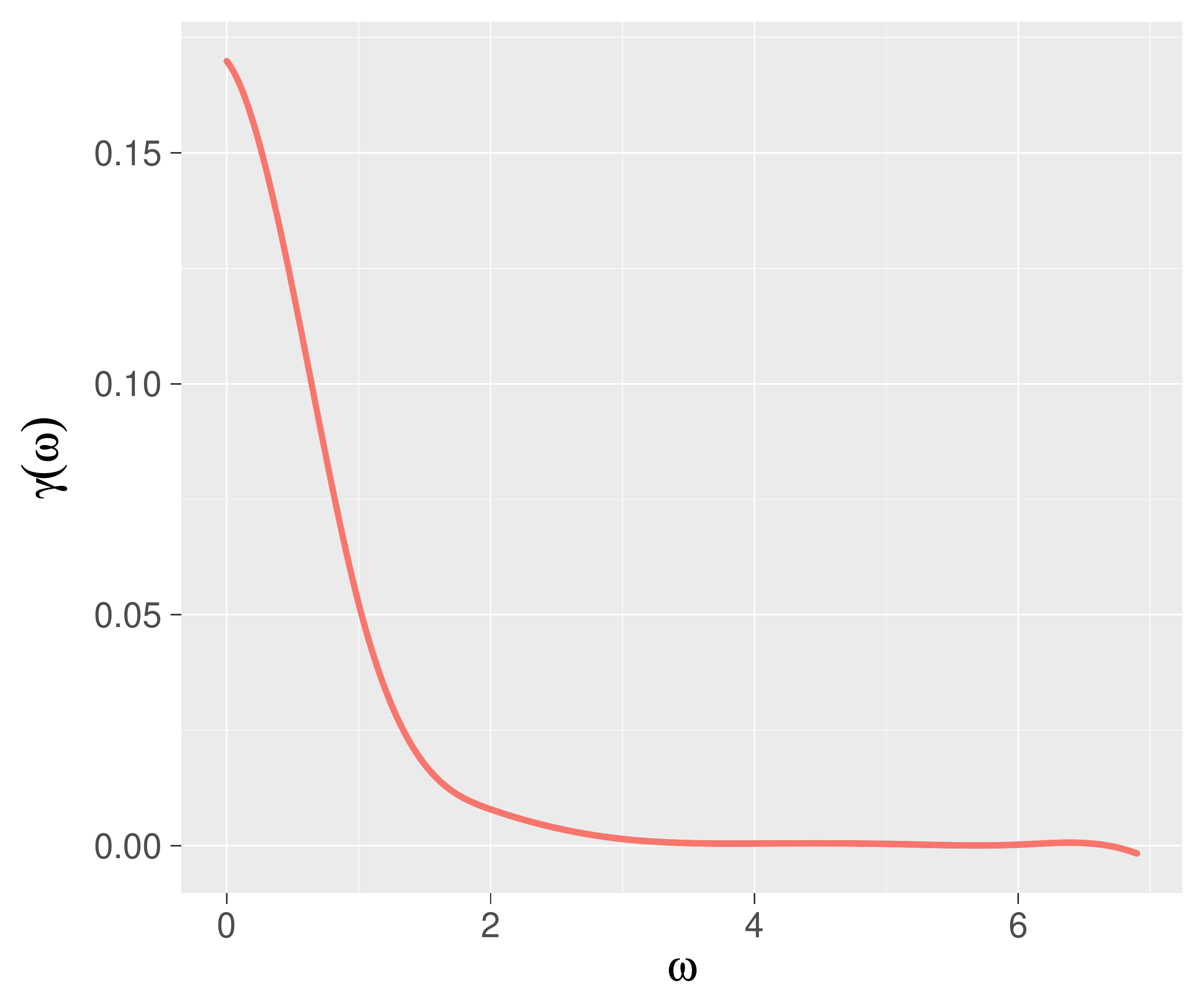}}
\subfigure[]{\label{fig:2c}\includegraphics[width=45mm]{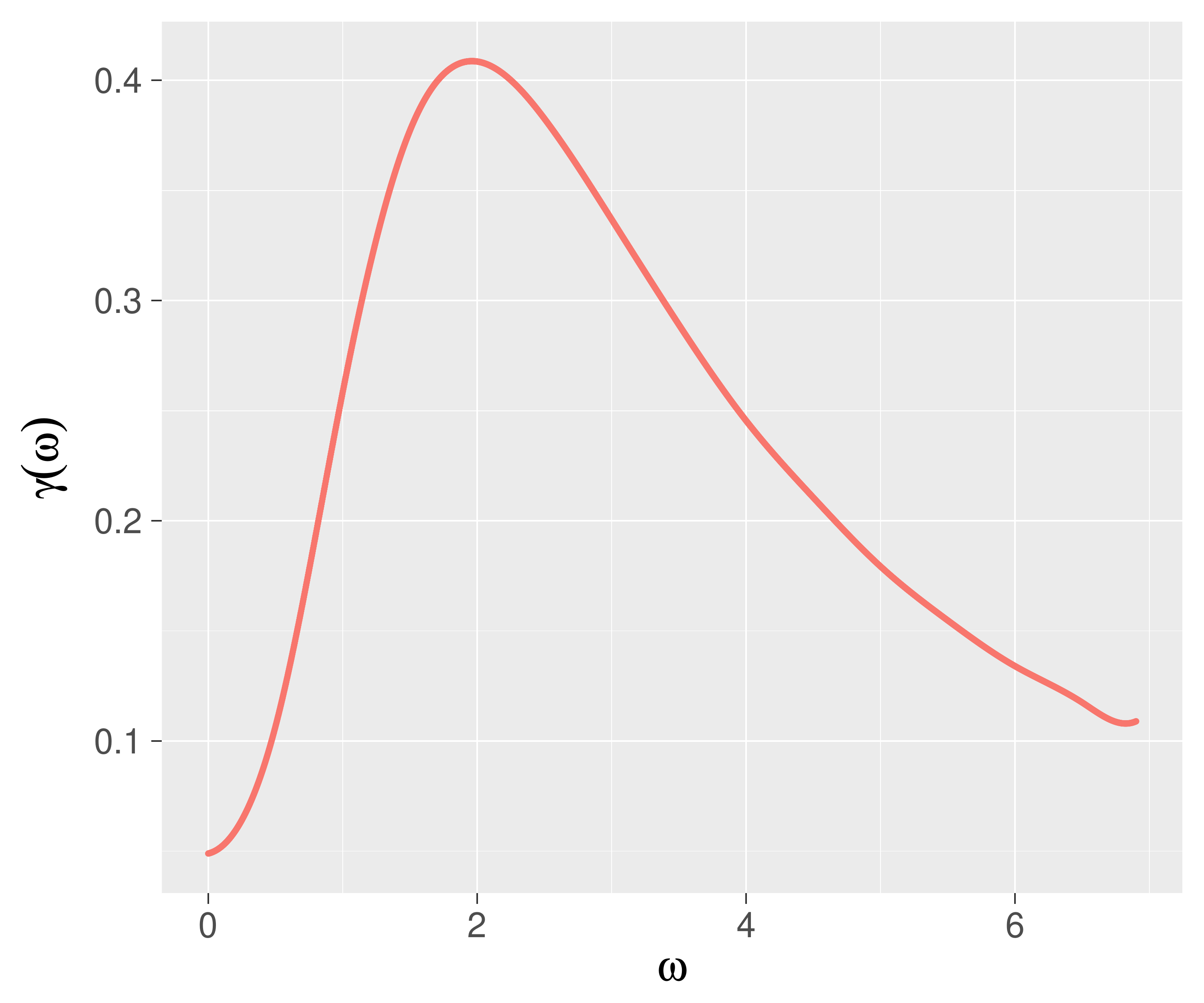}}\\
\subfigure[]{\label{fig:2d}\includegraphics[width=45mm]{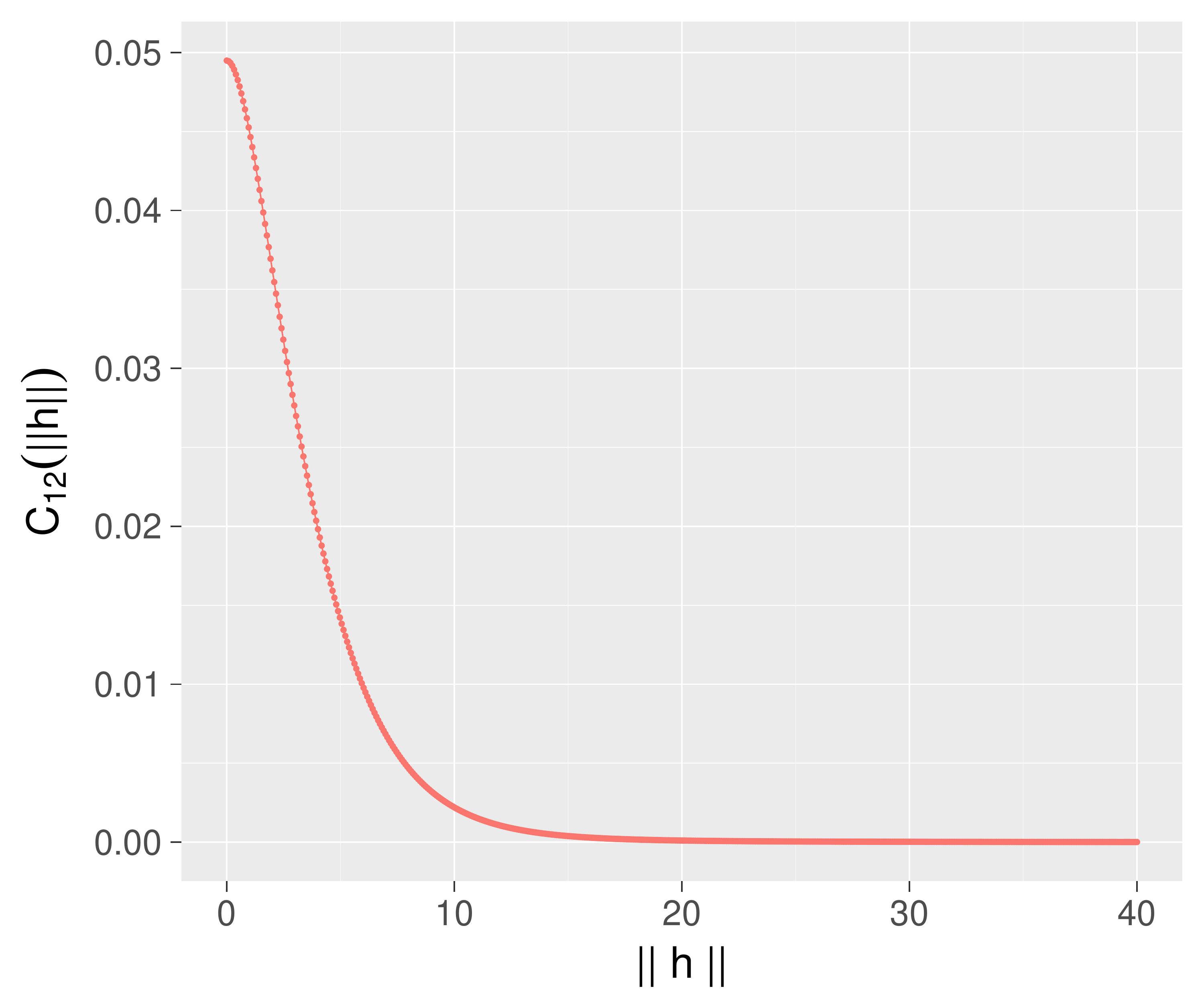}}
\subfigure[]{\label{fig:2e}\includegraphics[width=45mm]{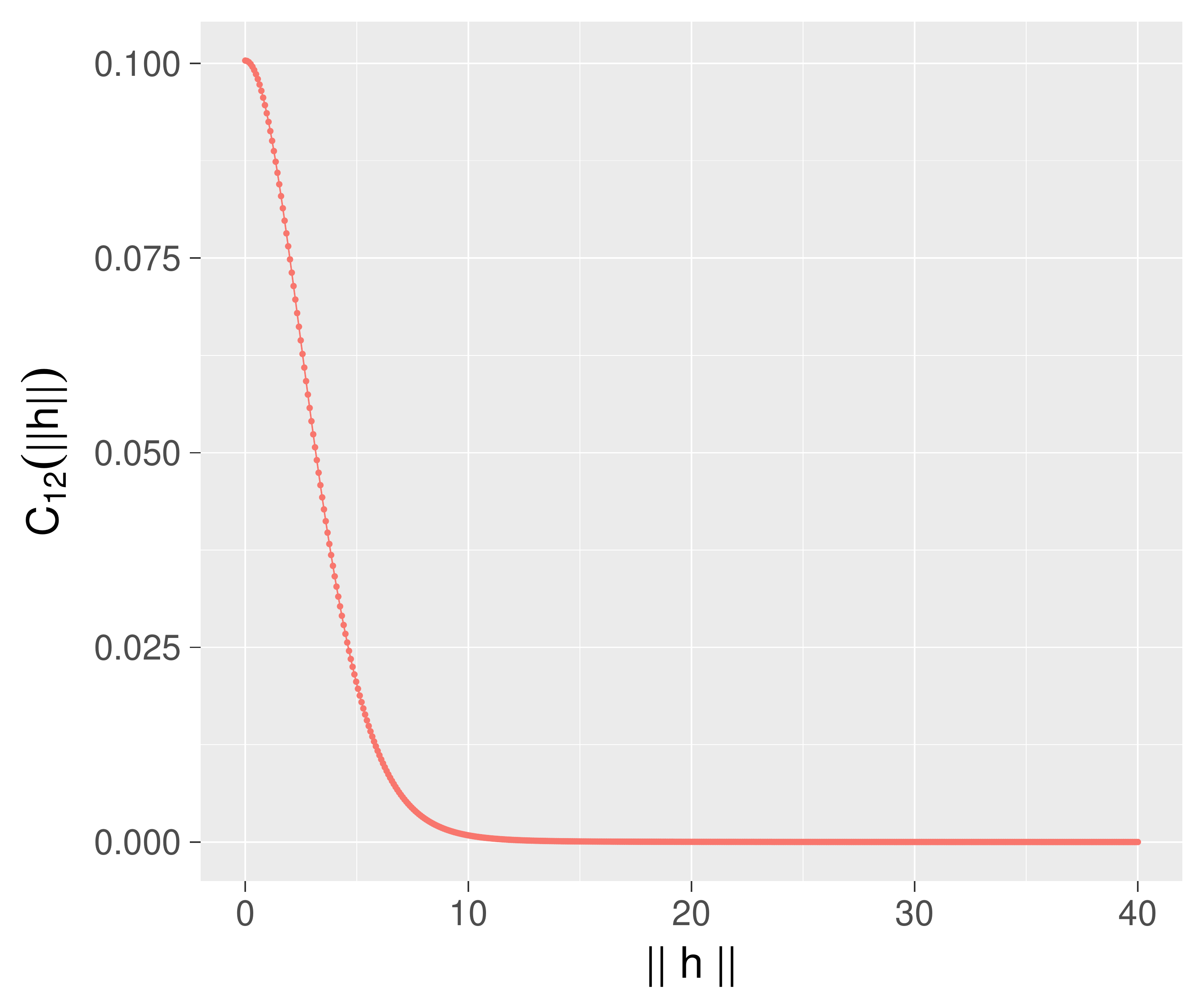}}
\subfigure[]{\label{fig:2f}\includegraphics[width=45mm]{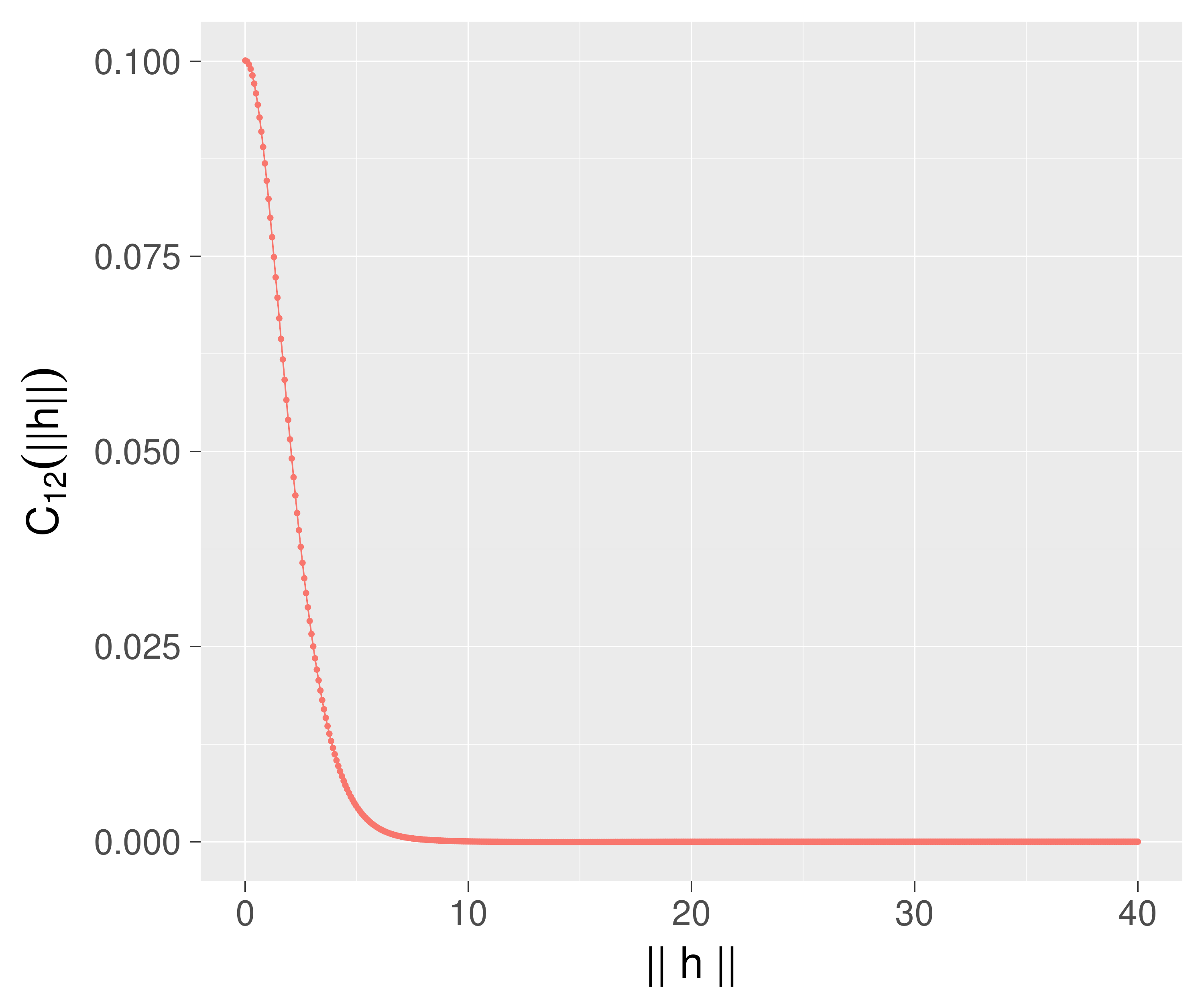}}

\caption{Example of bivariate coherence function for case-1 (a), case-2 (b) and case-3 (c). Corresponding cross-covariance function for case-1 (d), case-2 (e) and case-3 (f).}
\label{fig:2}
\end{figure}

The advantage of B-spline based specification (\ref{eq:coh}) of the coherence functions is that our proposed model (\ref{eq6}) approximately accommodates many existing classes of cross-covariance models that are constructed from the Mat{\'e}rn family, e.g., Multivariate Mat{\'e}rn,  Separable models with Mat{\'e}rn components, etc. For a sufficiently large value of $\omega_t$ and $m$, and appropriately specified B-splines, our proposed method can almost exactly reproduce those multivariate cross-covariances. For instance, the three examples of coherence functions shown in Figures \ref{fig:2a}, \ref{fig:2b} and \ref{fig:2c} are generated from our coherence model (\ref{eq:coh}) for suitably selected spline coefficients $\textbf{S}_{12}$. They closely match with the coherence functions of the full bivariate Mat{\'e}rn model for three settings listed as Model 1-3 in Table \ref{tab1}. Figures \ref{fig:2d}, \ref{fig:2e}, and \ref{fig:2f} show the computed cross-covariances from our model (\ref{eq6}) corresponding to the coherence functions in Figures \ref{fig:2a}, \ref{fig:2b} and \ref{fig:2c} and the marginal parameter values of Model 1-3 from Table \ref{tab1}, respectively. The computed cross-covariances  from our model are numerically equivalent to the corresponding full bivariate Mat{\'e}rn cross-covariances, thus exemplifying the generality of our proposed model. Furthermore, for a specific setting of parameters, the so-called parsimonious multivariate Mat{\'e}rn model is a special case in our proposed construction: 

\begin{Proposition}\label{prop1}
For a common spatial scale parameter $a_i=a,\;i=1,\dots,p$, $K\rightarrow\infty$, $\omega_t\rightarrow\infty$, and common spline coefficients $b_k^{(ij)}=\tau_{ij},\;k=-3,\dots,K,\;1\leq i \neq j \leq p$ (or equivalently constant coherence function $\gamma_{ij}(\omega)=\tau_{ij},\;\forall\omega\geq 0$) satisfying the sufficient conditions of Theorem \ref{th1}, the closed form solution of the integral (\ref{eq2}) for the spectral densities in (\ref{eq4}) and (\ref{eq5}) exists, and is equal to the parsimonious multivariate Mat{\'e}rn model.
\end{Proposition}

\begin{table}[h!]
\centering
\caption{Three parameter settings of full bivariate Mat{\'e}rn} \label{tab1}

 \begin{tabular}{|c| c c c c c c c c c |}
 \hline
\textbf{Model settings} & $\boldsymbol{\sigma_1}$ & $\boldsymbol{a_1}$ & $\boldsymbol{\nu_1}$ & $\boldsymbol{\sigma_2}$ & $\boldsymbol{a_2}$ & $\boldsymbol{\nu_2}$ & $\boldsymbol{a_{12}}$ & $\boldsymbol{\nu_{12}}$ & $\boldsymbol{\rho_{12}}$  \\ [0.5ex] 
 \hline
 Model 1 & 1 & 0.5 & 1 & 1 & 0.5 & 1 & 0.5 & 1.5 & 0.05   \\ 
 \hline
 Model 2 & 1 & 1 & 2 & 1 & 1 & 3 & 1.1 & 5 & 0.1 \\
 \hline
 Model 3 & 1 & 0.6 & 3 & 1 & 1.4 & 3 & 1.5 & 4 & 0.1 \\
 \hline
 \end{tabular}
\end{table}

\begin{figure}[!t]
\centering     
\subfigure[]{\label{fig:3a}\includegraphics[width=60mm]{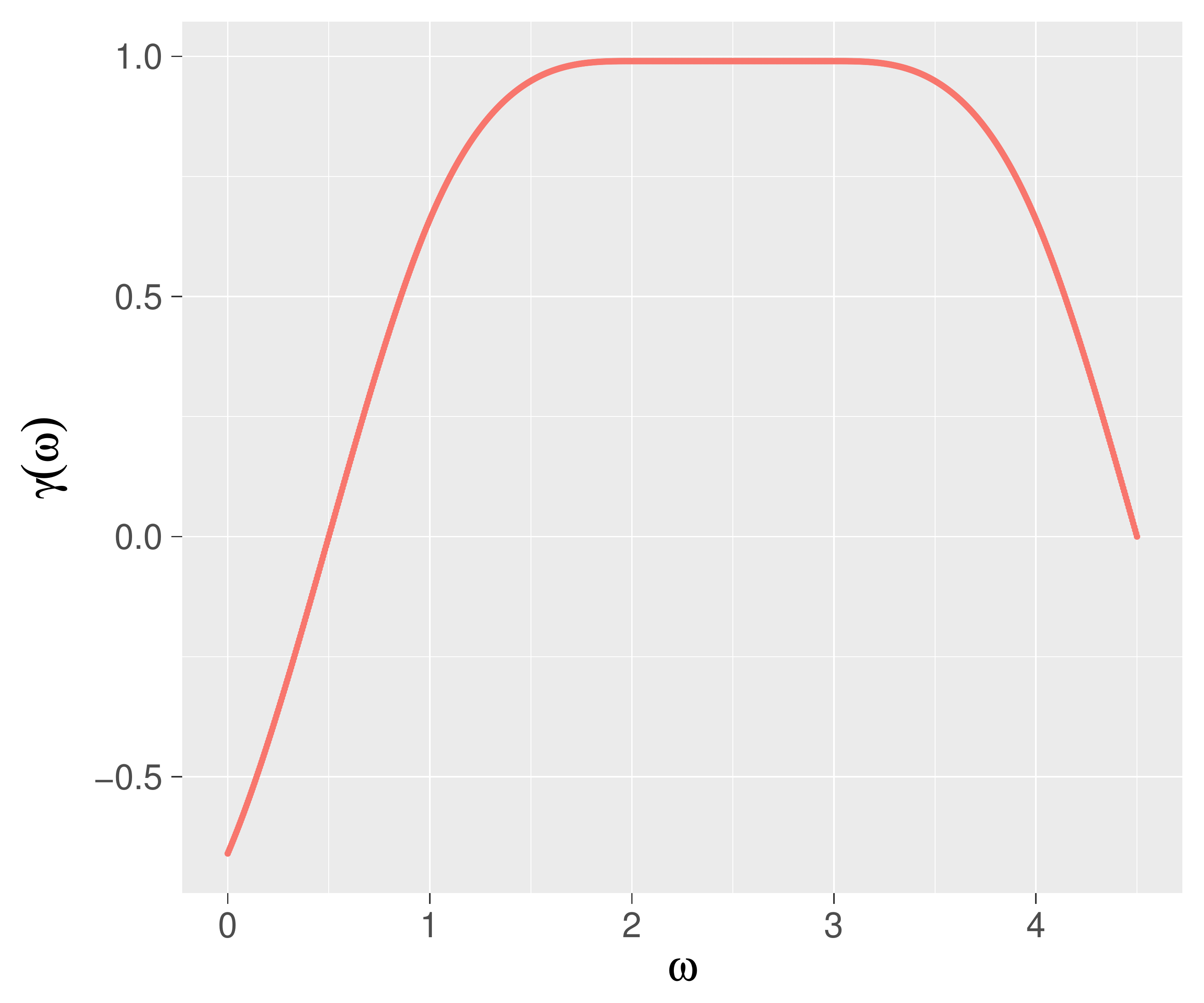}}
\subfigure[]{\label{fig:3b}\includegraphics[width=60mm]{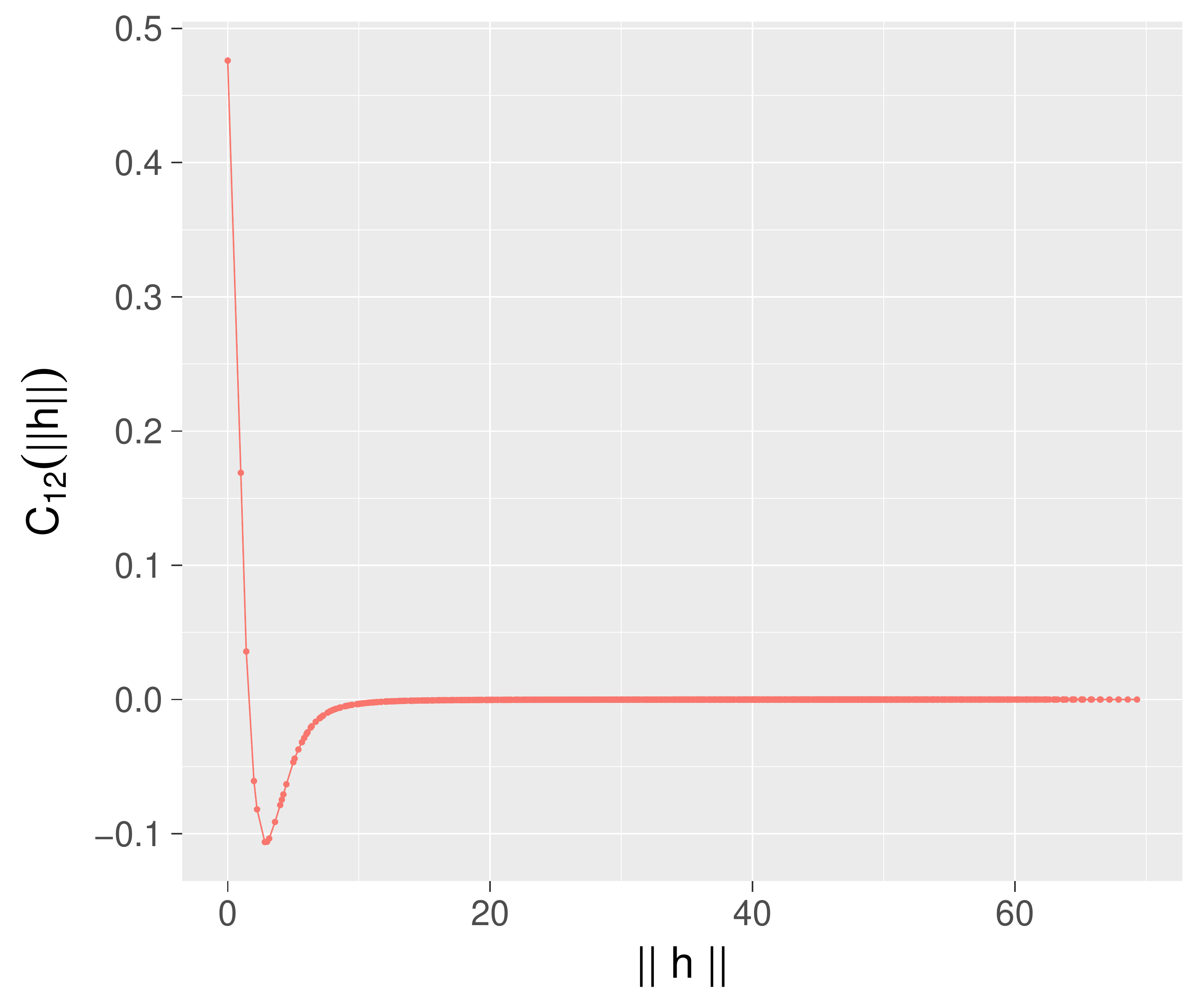}}
\subfigure[]{\label{fig:3c}\includegraphics[width=60mm]{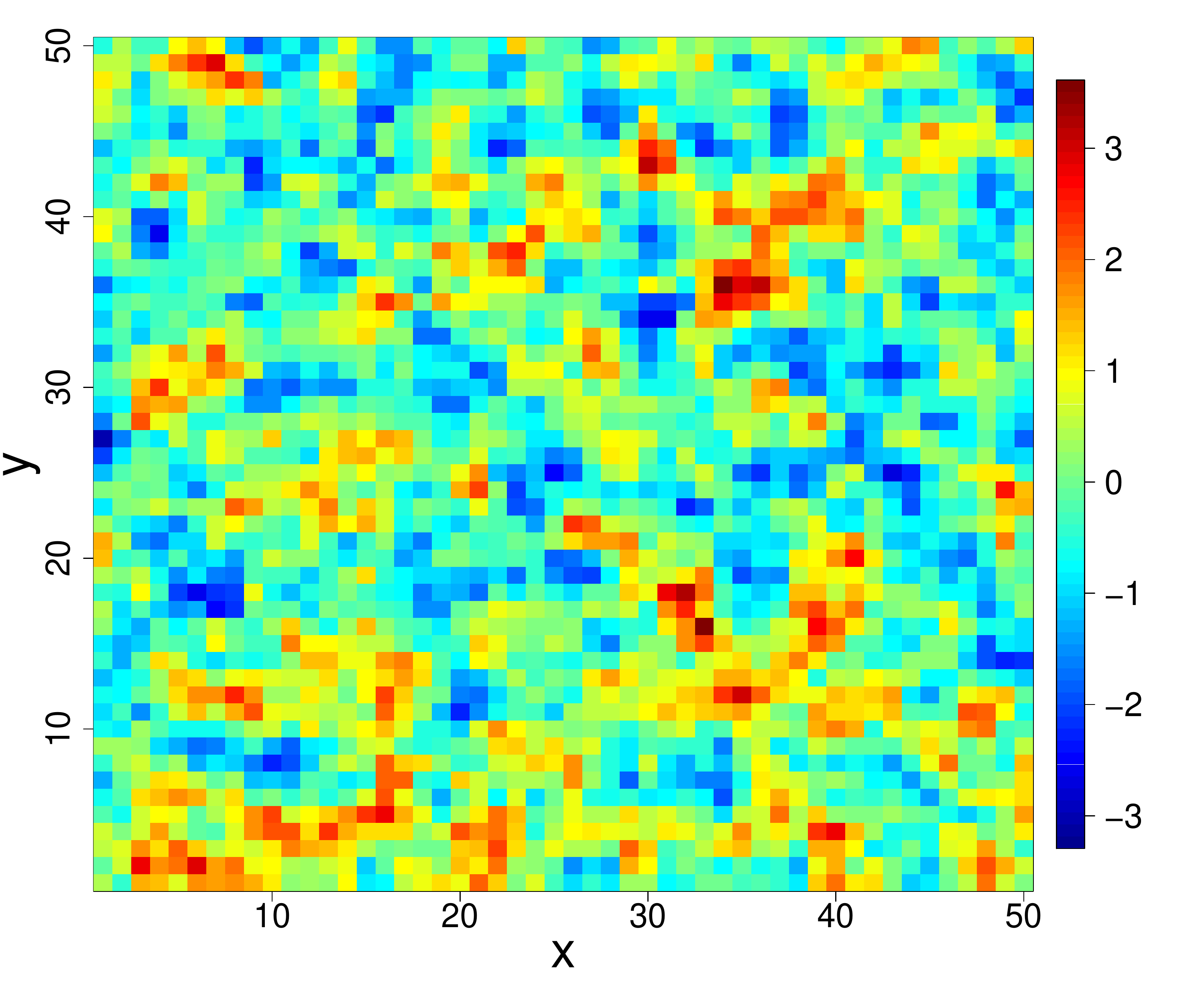}}
\subfigure[]{\label{fig:3d}\includegraphics[width=60mm]{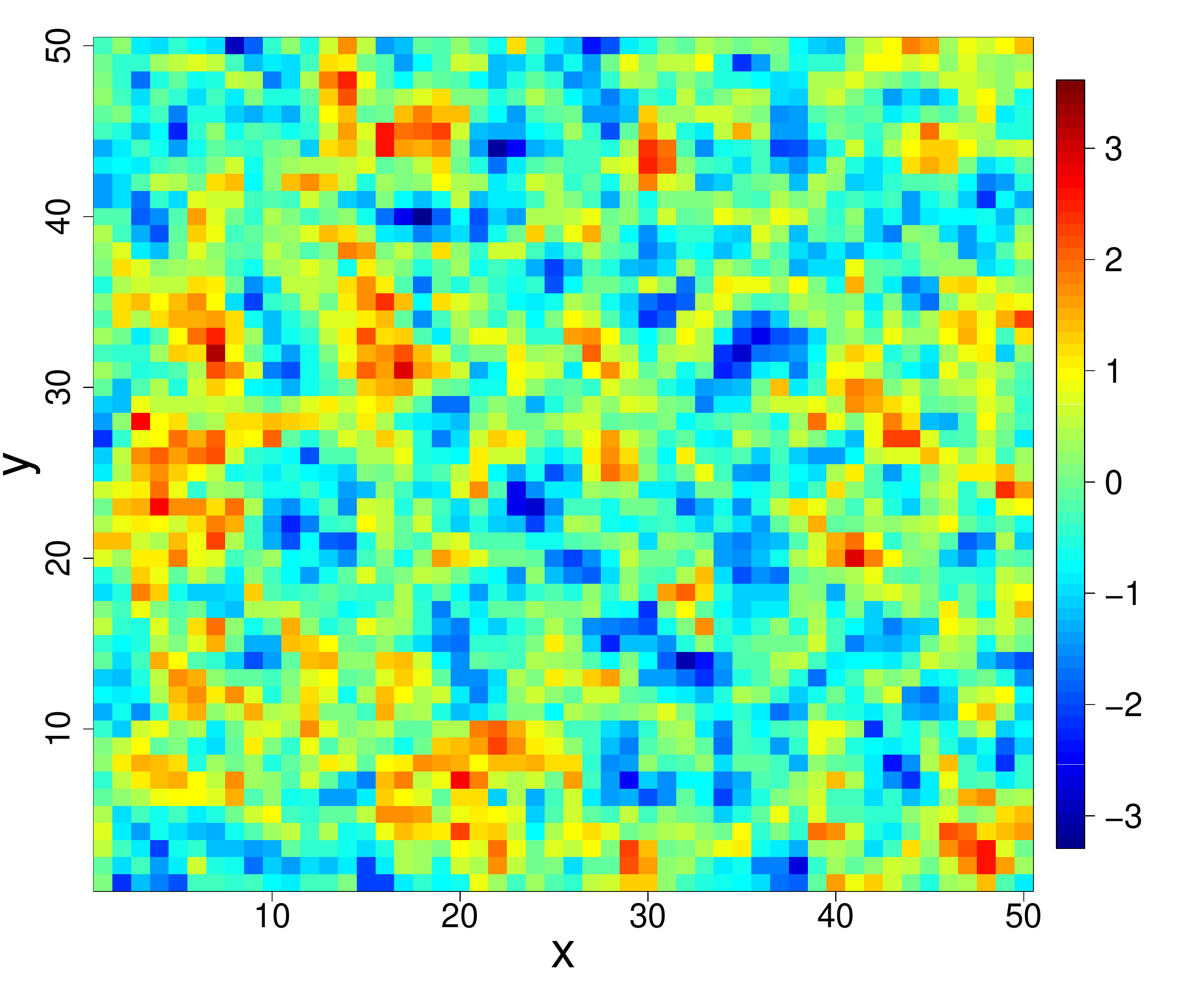}}
\caption{(a) Coherence function. (b) Cross-covariance function. (c) Simulated realization for $Y_1$ $(\sigma_1=1,\nu_1=1,a_1=1)$. (d) Simulated realization for $Y_2$ $(\sigma_2=1,\nu_2=1,a_2=1)$.} 
\label{fig:3}
\end{figure}

\begin{figure}[!t]
\centering     
\subfigure[]{\label{fig:dec2a}\includegraphics[width=60mm]{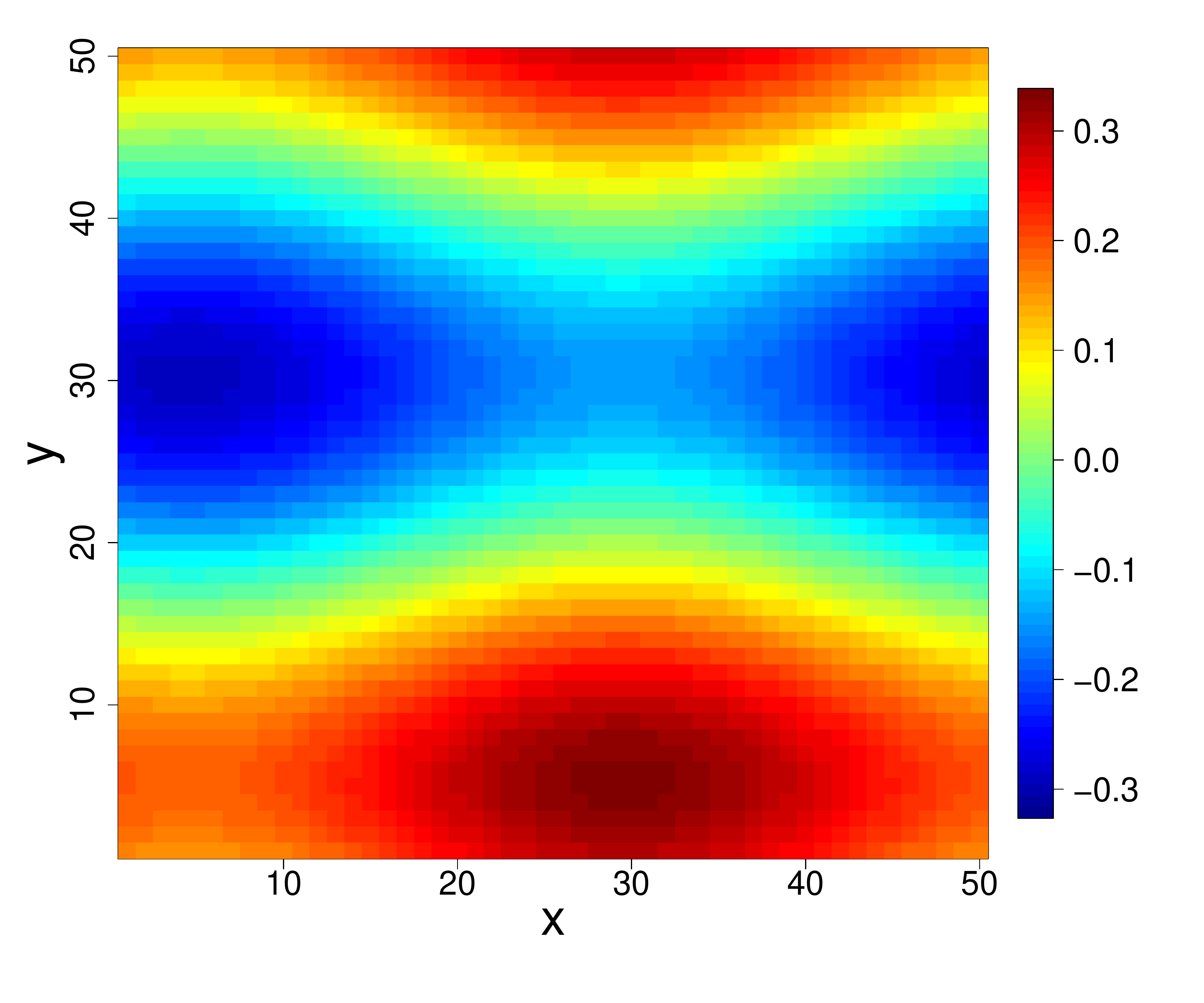}}
\subfigure[]{\label{fig:dec2b}\includegraphics[width=60mm]{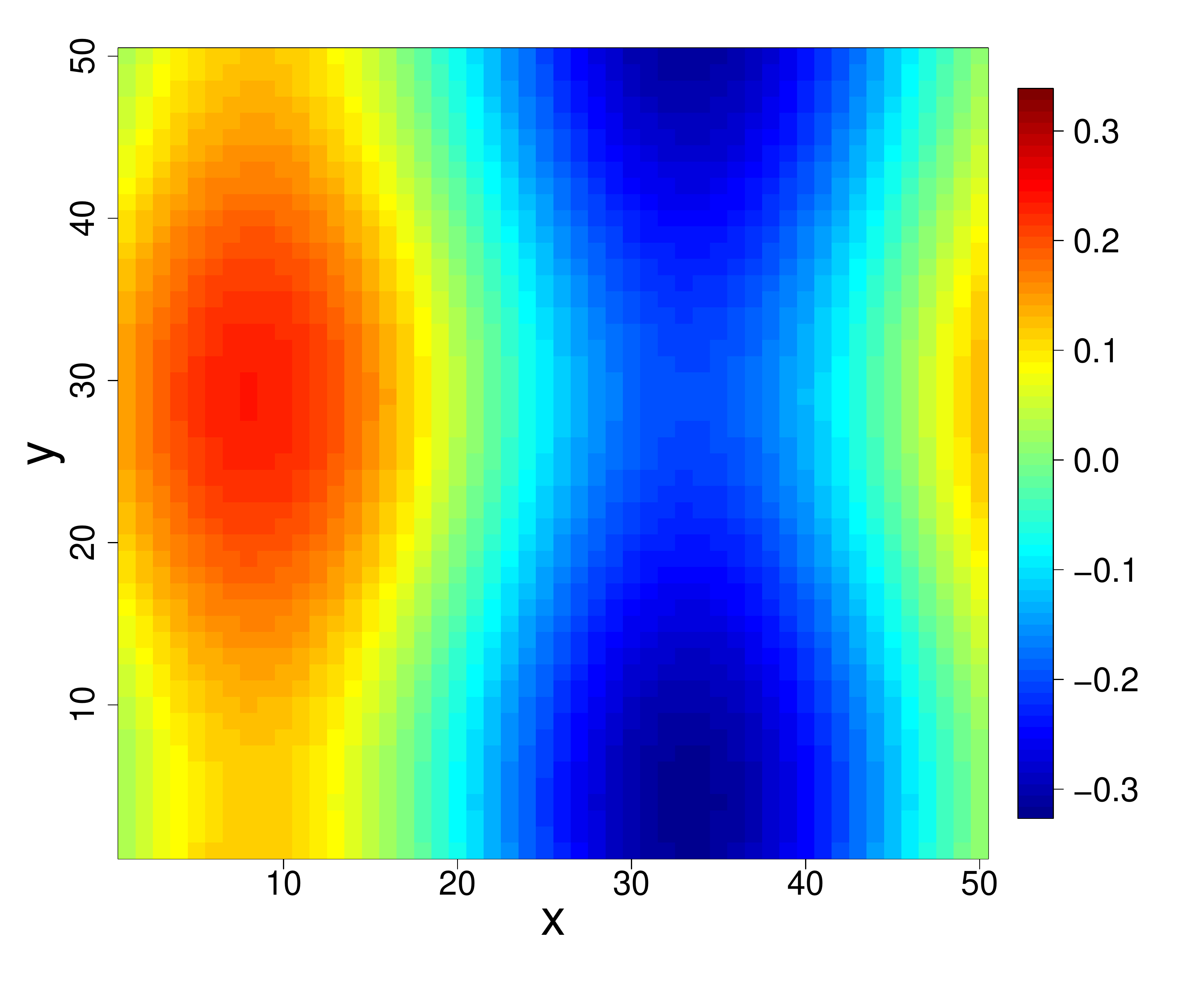}}
\subfigure[]{\label{fig:dec2c}\includegraphics[width=60mm]{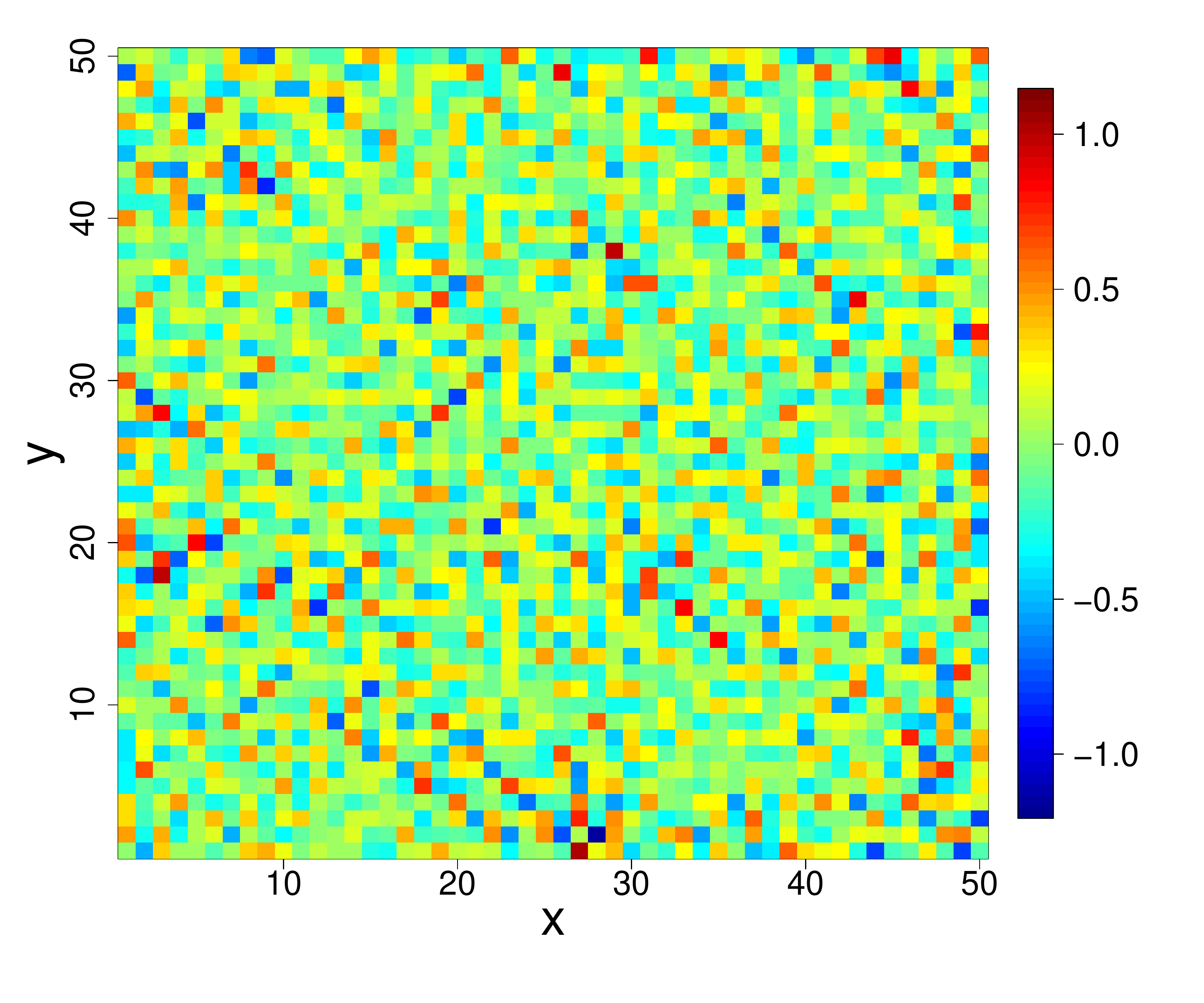}}
\subfigure[]{\label{fig:dec2d}\includegraphics[width=60mm]{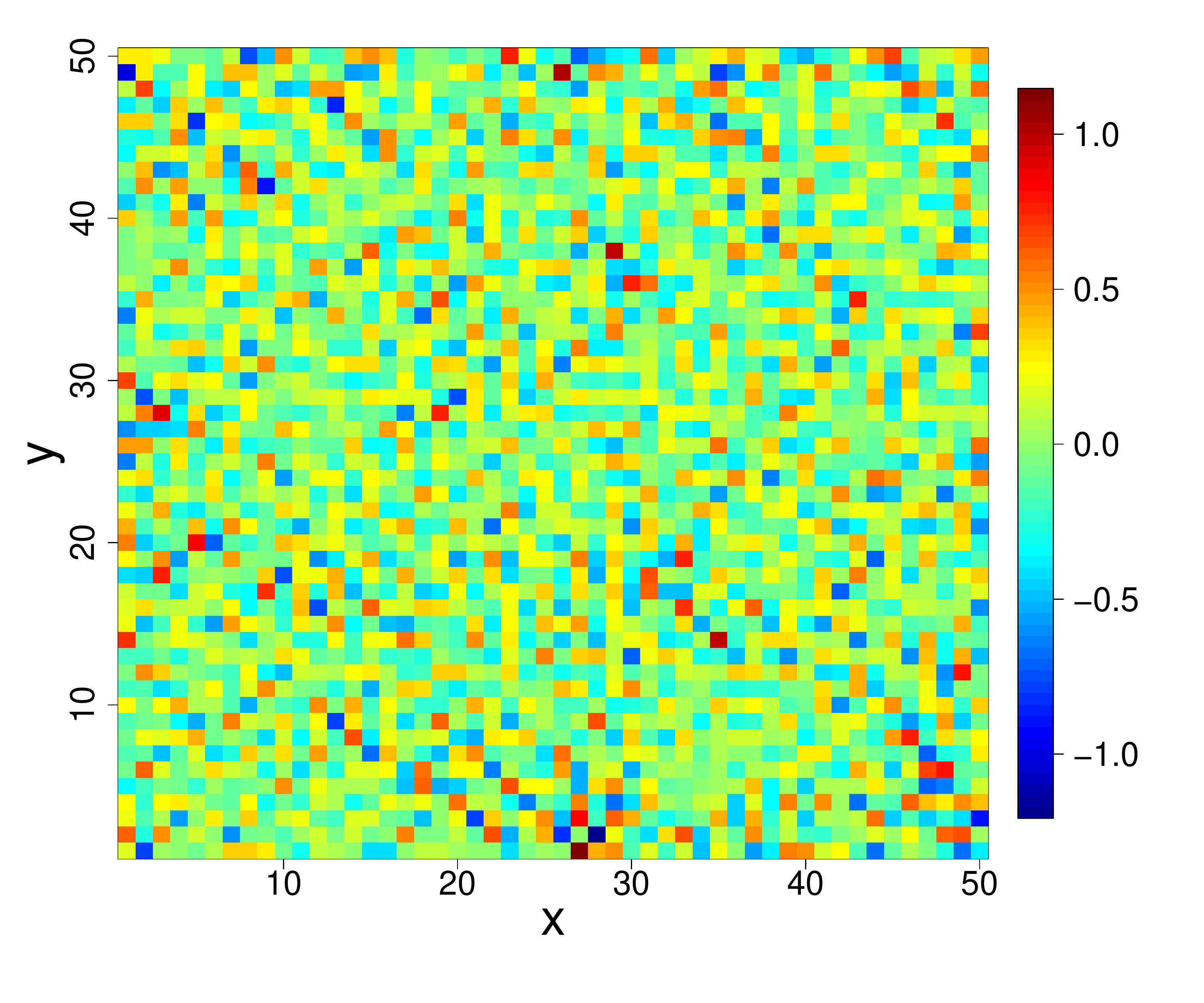}}
\caption{(a) $\check{Y_1}^{lf}$ $(Y_1$ filtered at $0\leq\omega\leq 0.2)$. (b) $\check{Y_2}^{lf}$ ($Y_2$ filtered at $0\leq\omega\leq 0.2)$. (c) $\check{Y_1}^{hf}$ $(Y_1$ filtered at $3\leq\omega\leq 4)$. (d) $\check{Y_2}^{hf}$ ($Y_2$ filtered at $3\leq\omega\leq 4)$} 
\label{fig:dec2}
\end{figure}

Various choices of spline coefficients and marginal parameters $(\sigma_i,a_i,\nu_i,\;i=1,\dots,p)$ in our model (\ref{eq6}) can imply the oscillation of coherence functions and cross-covariance functions between negative and positive values. Figure \ref{fig:3} reflects one such example where we set the marginal parameters $(\sigma_i=a_i=\nu_i=1,\;i=1,2)$, threshold frequency $\omega_t=4.5$ and $m=990$. We choose $\Delta=1$ $(K=4)$ and $\textbf{S}_{12}=\{-0.99,-0.99,0.99,0.99,0.99,0.99,-0.99,-0.99\}$ to produce negative coherence at low frequencies and positive coherence at higher frequencies (shown in Figure \ref{fig:3a}). The corresponding cross-covariance function from our model (\ref{eq6}) (shown in Figure \ref{fig:3b}) exhibits a transition from positive dependence to negative dependence with increasing distance, and eventually decays to zero at large distances. Figure \ref{fig:3c} and \ref{fig:3d} shows one realization of a zero mean bivariate Gaussian process $\textbf{Y}$ simulated with the chosen marginal and cross-covariance function. The filtered signal $\check{\textbf{Y}}^{fb}$ for the simulated dataset $\textbf{Y}$ at the low-frequency band $lf=0\leq\omega\leq 0.2$ and the high-frequency band $hf=3\leq\omega\leq4$ are shown in Figure \ref{fig:dec2}. While the empirical correlation for the filtered signal pair $(\check{Y_1}^{lf},\check{Y_2}^{lf})$ is $-0.5$, i.e., negatively correlated, the empirical correlation for the pair $(\check{Y_1}^{hf},\check{Y_2}^{hf})$ is $0.94$, i.e., positively correlated. This change of sign from negative to positive while going from  $lf$ to $hf$ is to be expected due to the oscillatory nature of the underlying coherence function. Our proposed construction provides a potential working covariance model for real multivariate datasets, which exhibits such cross-process behavior.

\subsection{Maximum Likelihood Estimation}
Let $\tilde{\textbf{X}}=\big(\textbf{X}(\textbf{s}_1)^{\text{T}},\dots,\textbf{X}(\textbf{s}_n)^{\text{T}}\big)^{\text{T}}$ be a realization from a zero mean stationary multivariate Gaussian process where $\textbf{X}(\textbf{s})=\big(X_1(\textbf{s}),\dots,X_p(\textbf{s})\big)^{\text{T}}$. Let $\Sigma_{\boldsymbol{\theta}_\mathcal{SP}}$ denote the $np\times np$ covariance matrix for $\tilde{\textbf{X}}$ where $\{\text{C}_{ij}(\textbf{s}_q-\textbf{s}_r)\}_{i,j=1}^p\in\mathbb{R}^{p\times p}$ defined in (\ref{eq6}) constitutes the $(q,r)^{th},\;q,r=1,\dots,n$ block entry of $\Sigma_{\boldsymbol{\theta}_\mathcal{SP}}$, and $\boldsymbol{\theta}_\mathcal{SP}$ denote the set of parameters in our semiparametric model (\ref{eq6}). Then $\tilde{\textbf{X}}\sim MVN_{np}(0,\Sigma_{\boldsymbol{\theta}_\mathcal{SP}})$, and the log-likelihood is given as:
\begin{equation}\label{logl}
    \ell(\boldsymbol{\theta}_\mathcal{SP}|\tilde{\textbf{X}})=-\frac{1}{2}(\text{log }\text{det}\Sigma_{\boldsymbol{\theta}_\mathcal{SP}}+\tilde{\textbf{X}}^{\text{T}}\Sigma_{\boldsymbol{\theta}_\mathcal{SP}}^{-1}\tilde{\textbf{X}}+np\;\text{log}\;2\pi)
\end{equation}
For an appropriately chosen large value of $\omega_t$ and $m$, and suitably specified uniform knot spacing $\Delta$, our semiparametric model (\ref{eq6}) entirely depends on the set of parameters $\boldsymbol{\theta}_{\mathcal{SP}}$. Here the set $\boldsymbol{\theta}_{\mathcal{SP}}$ consists of $3p$ marginal parameters $(\sigma_i,\nu_i,a_i,\;i=1,\dots,p)$ and $(K+4) {{p}\choose{2}}$ spline coefficients $\{b_k^{(ij)},\;k=-3,-2,\dots,K,\;1\leq i < j \leq p.\}$. In our implementation, we perform joint numerical maximization of the log-likelihood over the elements of the set $\boldsymbol{\theta}_{\mathcal{SP}}$, while ensuring the sufficient conditions of validity in Theorem \ref{th1} by further parameterizing the B-spline coefficients as discussed in Section \ref{subsec:spec}. In the case of $p=2$, the estimation procedure is straightforward, as restricting the values of B-spline coefficients to lie between $-1$ to 1 would suffice for the validity, and therefore does not require tricky parameterizations.  

\section{Simulation Study}\label{sec:simulation}
In this section, we explore the performance of our proposed semiparametric model (\ref{eq6}) by evaluating the maximum likelihood estimates of its marginal parameters and the underlying coherence function for bivariate processes simulated from different multivariate models. In particular, we simulate the Gaussian random field from the full bivariate Mat{\'e}rn model (see Section \ref{sim1}) and the LMC with latent Mat{\'e}rn fields (see Section \ref{sim2}), and excercise our semiparametric model to estimate the marginal and cross-process behaviour from simulated datasets.

\subsection{Simulation 1: Full Bivariate Mat{\'e}rn Model}\label{sim1}
\begin{figure}[!t]
\centering     
\subfigure[]{\label{fig:r1a}\includegraphics[width=58mm]{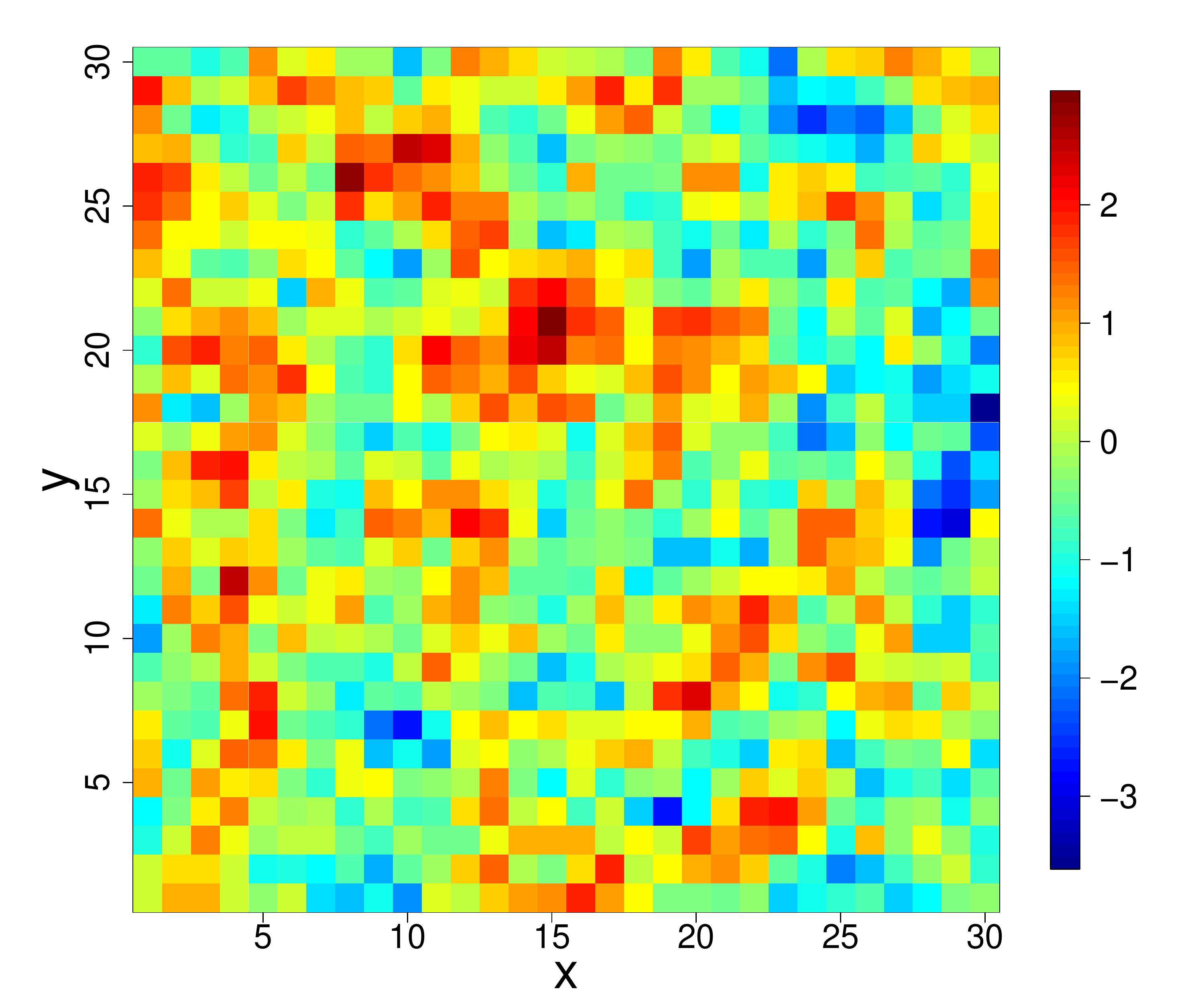}}
\subfigure[]{\label{fig:r1b}\includegraphics[width=58mm]{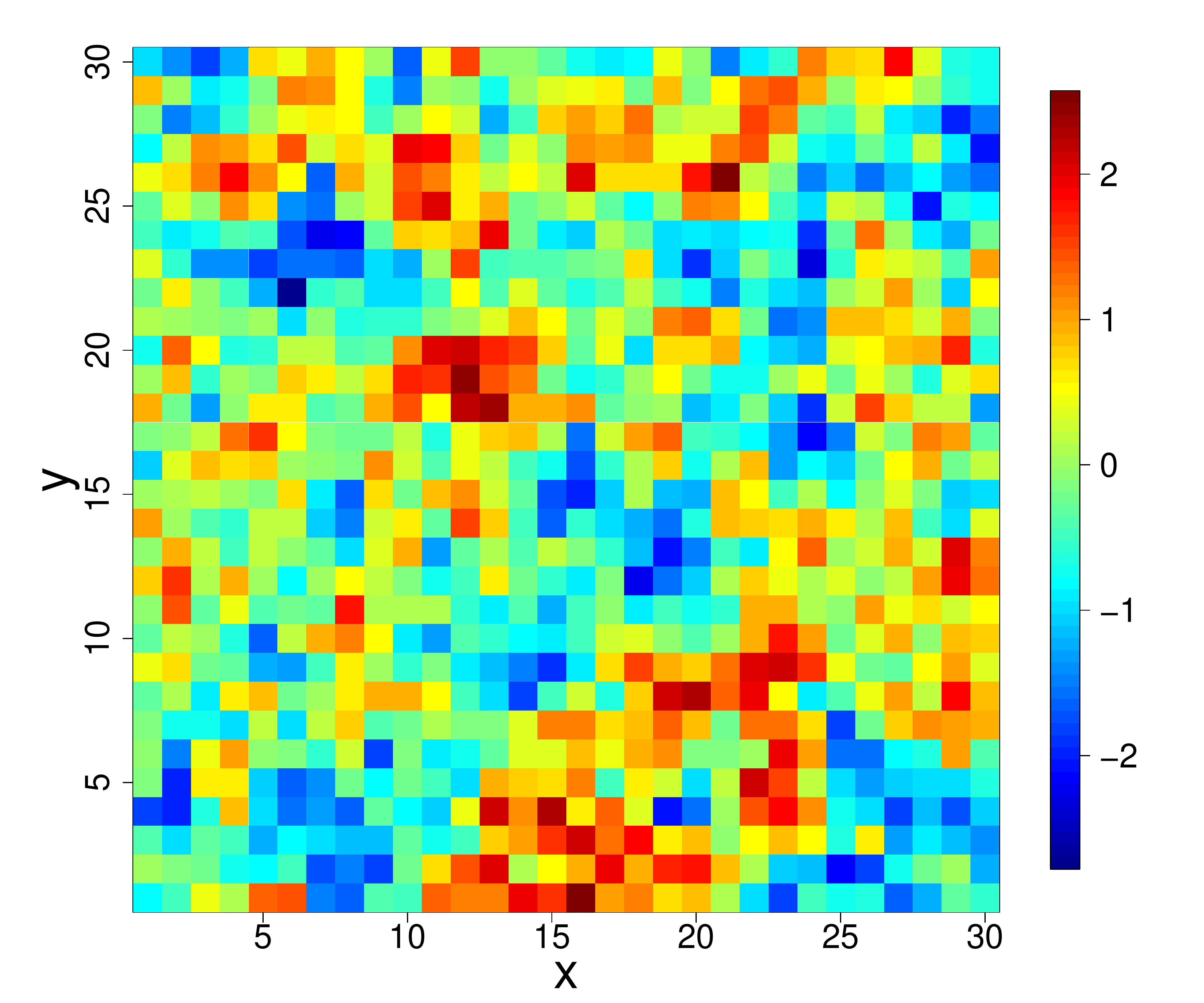}}\\
\subfigure[]{\label{fig:r1c}\includegraphics[width=58mm]{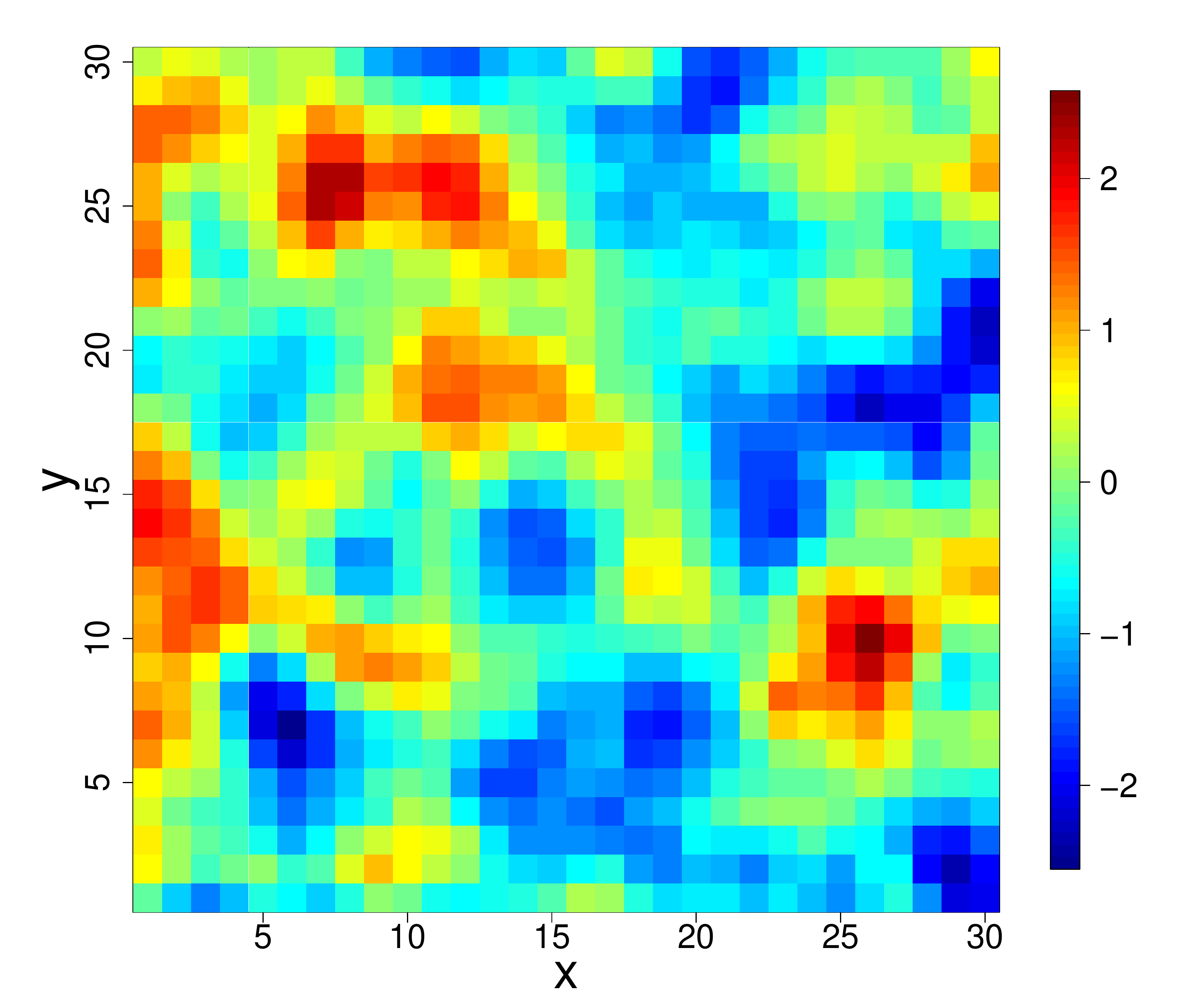}}
\subfigure[]{\label{fig:r1a}\includegraphics[width=58mm]{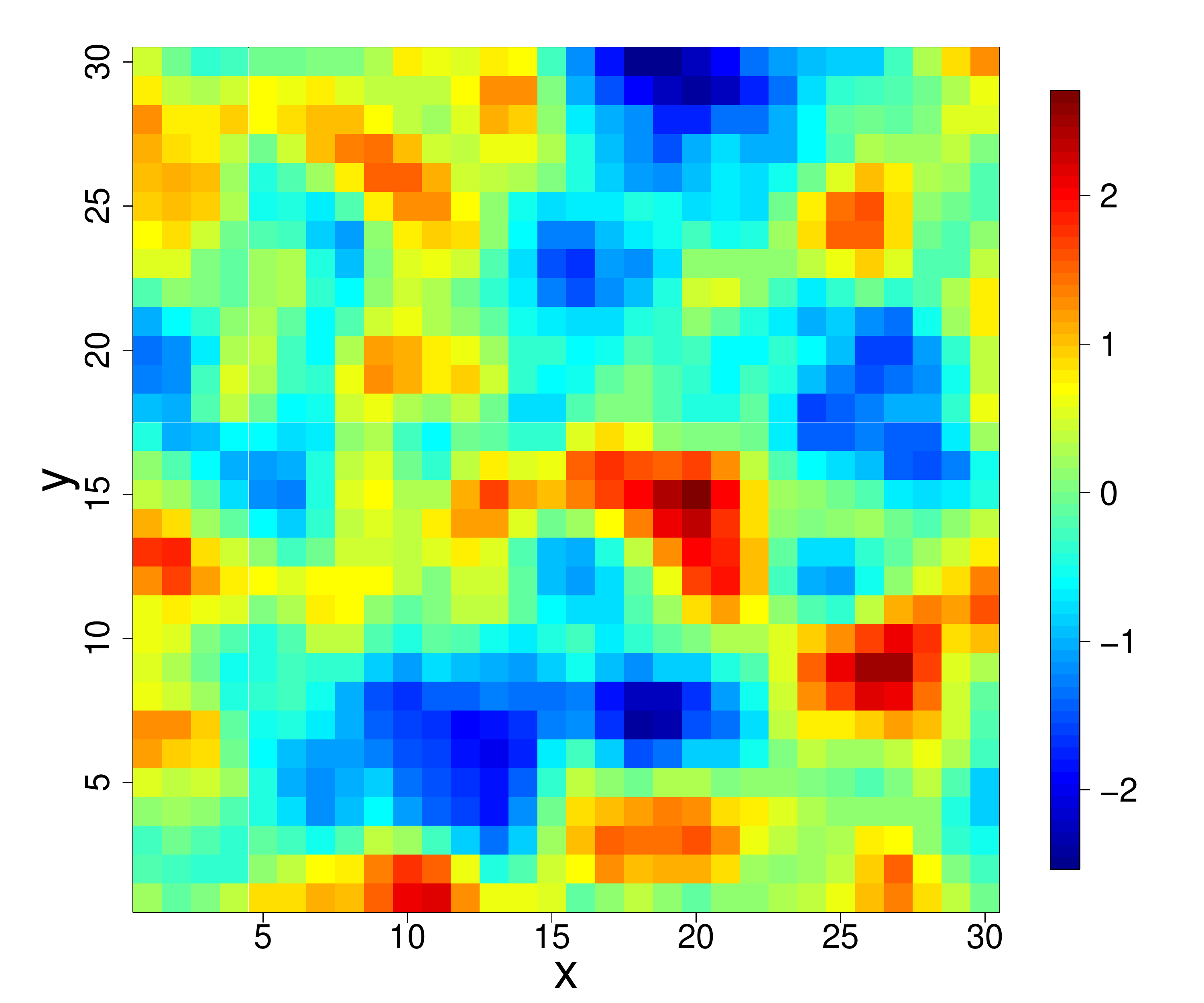}}\\
\subfigure[]{\label{fig:r1b}\includegraphics[width=58mm]{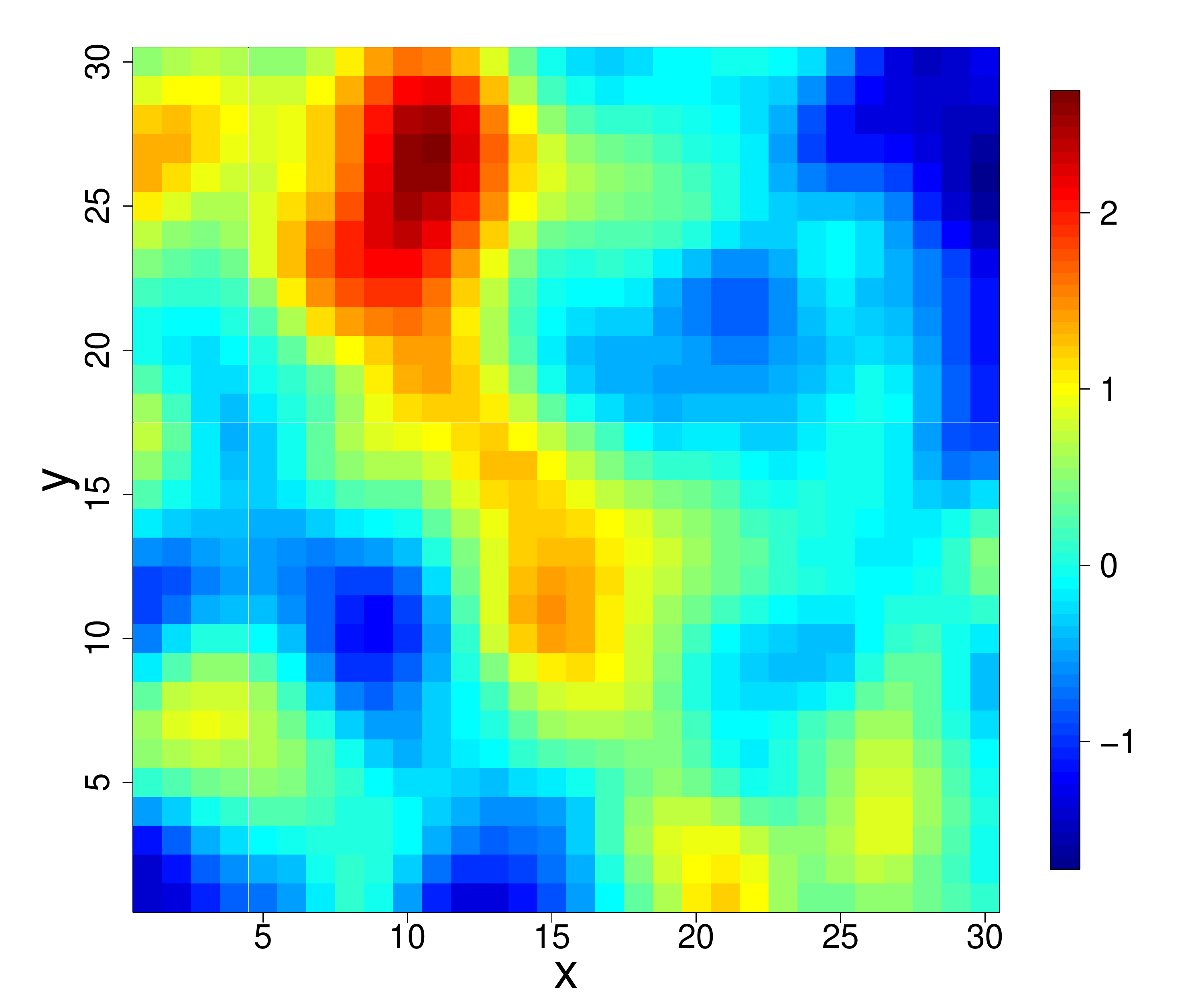}}
\subfigure[]{\label{fig:r1c}\includegraphics[width=58mm]{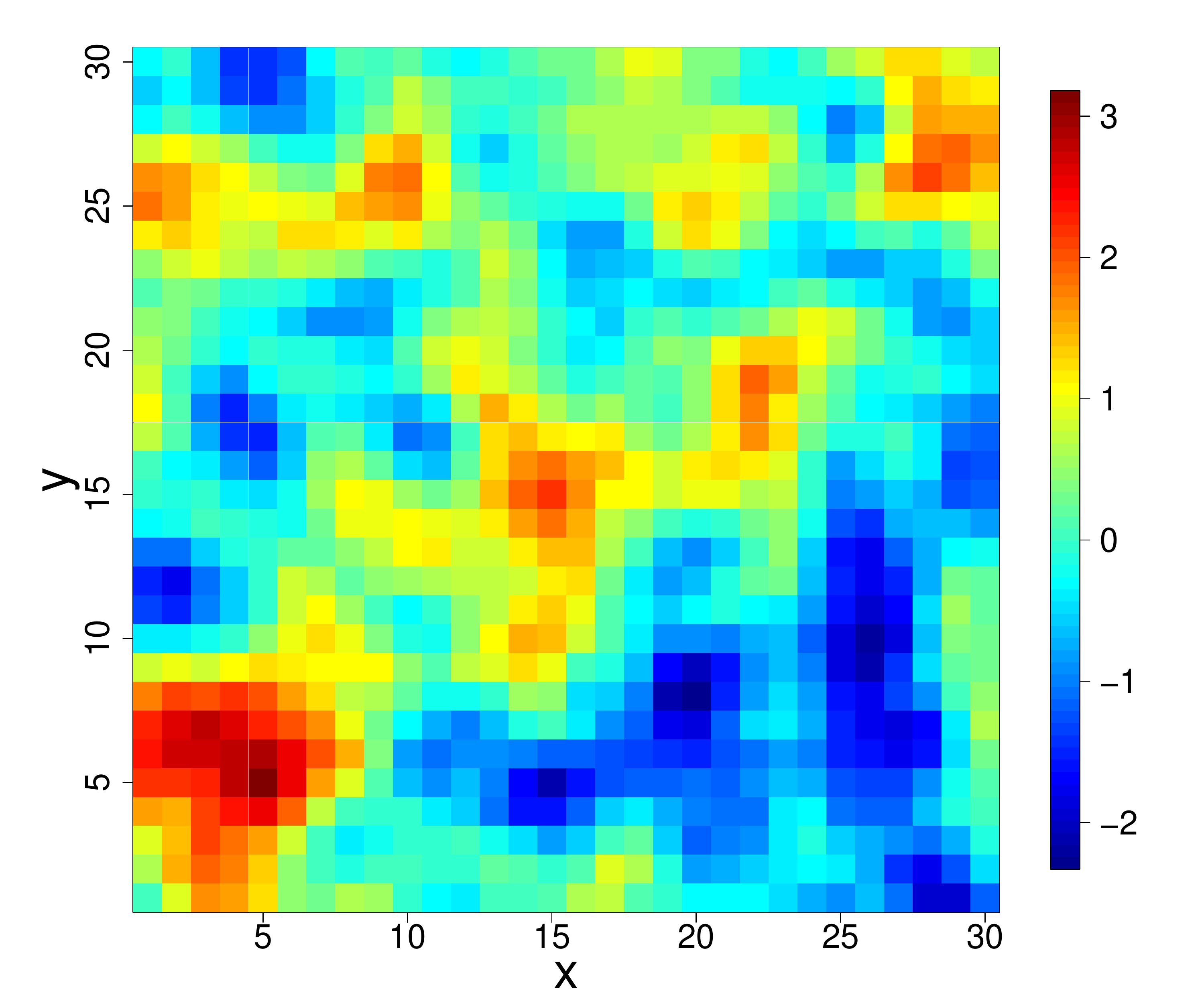}}

\caption{Example of one realization for a bivariate process from Model 1 ((a) $X_1$, (b) $X_2$), Model 2 ((c) $X_1$, (d) $X_2$) and Model 3 ((d) $X_1$, (e) $X_2$).    }
\label{fig:r1}
\end{figure}

We consider a zero mean bivariate Gaussian random field $\textbf{X}(\textbf{s})=(X_1(\textbf{s}),X_2(\textbf{s}))^{\text{T}}$ on a grid of coordinates $\{(i,j)\}_{i,j=1}^{30}$, with marginal and cross-covariances defined by the full bivariate Mat{\'e}rn model:\[\text{C}_{ii}(\textbf{h})=\text{M}(\textbf{h}|\sigma_i,\nu_i,a_i),\; i=1,2,\]
\[\text{C}_{ij}(\textbf{h})=\rho_{ij}\text{M}(\textbf{h}|\sqrt{\sigma_i\sigma_j},\nu_{ij},a_{ij}),\; 1\leq i \neq j \leq 2,\] where $\rho_{ij}$ refers to the co-located correlation coefficient that requires to satisfy the necessary and sufficient condition provided in Theorem 3 of \cite{bimat}. The full bivariate Mat{\'e}rn model implies the following isotropic coherence function in a bivariate process defined over a spatial domain $\mathcal{D} \in \mathbb{R}^d$ ($d=2$ in our case): \[{\gamma_{12}({\omega})=\rho_{12}\frac{\Gamma(\nu_{12}+d/2){\Gamma(\nu_1)}^{\frac{1}{2}}{\Gamma(\nu_2)}^{\frac{1}{2}}a_{12}^{2\nu_{12}}(a_1^2+{\omega}^2)^{\frac{\nu_1}{2}+\frac{d}{4}}(a_2^2+{\omega}^2)^{\frac{\nu_2}{2}+\frac{d}{4}}}{{\Gamma(\nu_1+d/2)}^{\frac{1}{2}}{\Gamma(\nu_2+d/2)}^{\frac{1}{2}}\Gamma(\nu_{12})a_1^{\nu_1}a_2^{\nu_2}(a_{12}^2+{\omega}^2)^{\nu_{12}+\frac{d}{2}}}}.\] We simulate 50 realizations of $\textbf{X}$, for three cases of parameter settings listed as Model 1-3 in Table \ref{tab:sim1}. An example of simulated bivariate processes from these models is shown in Figure \ref{fig:r1}. These
three models simulate bivariate processes with contrasting coherence features, broadly covering all the shapes of a coherence function that a full bivariate Mat{\'e}rn model can generate. Whereas Model 1 and 2 lead to monotonically increasing and monotonically decreasing coherence functions, respectively, Model 3 leads to a bump in the coherence function at some frequency band. 

We fit our semiparametric model (\ref{eq6}) on the simulated realizations, using the method of maximum likelihood to investigate its efficiency. For estimation in each of the three cases of simulation, we specify the threshold frequency $\omega_t=4.5$, and $m=380$ for the discretization of the frequency interval $[0,\omega_t]$. Furthermore, we set $\Delta=1$ (or equivalently $K=4$) to completely specify the coherence function, which in turn requires the estimation of eight B-spline coefficients $\{\textbf{S}_{12}= b_k^{(12)},k=-3,-2,\dots,4\}$. We also assume that the marginal smoothness parameters $\{\nu_i,i=1,2\}$ are known, and therefore are fixed to their true value in our model, to avoid possible identifiability issues \citep{zhang}. Thus, in each of the three cases, we estimate 12 parameters in total, including the 4 marginal parameters  $\{a_i,\sigma_i^2,i=1,2\}$ and a set of 8 B-spline coefficients $\textbf{S}_{12}=\{b_k^{(12)},k=-3,-2,\dots,4\}$. 

\begin{figure}[!t]
\centering     
\subfigure[]{\label{fig:4a}\includegraphics[width=52mm]{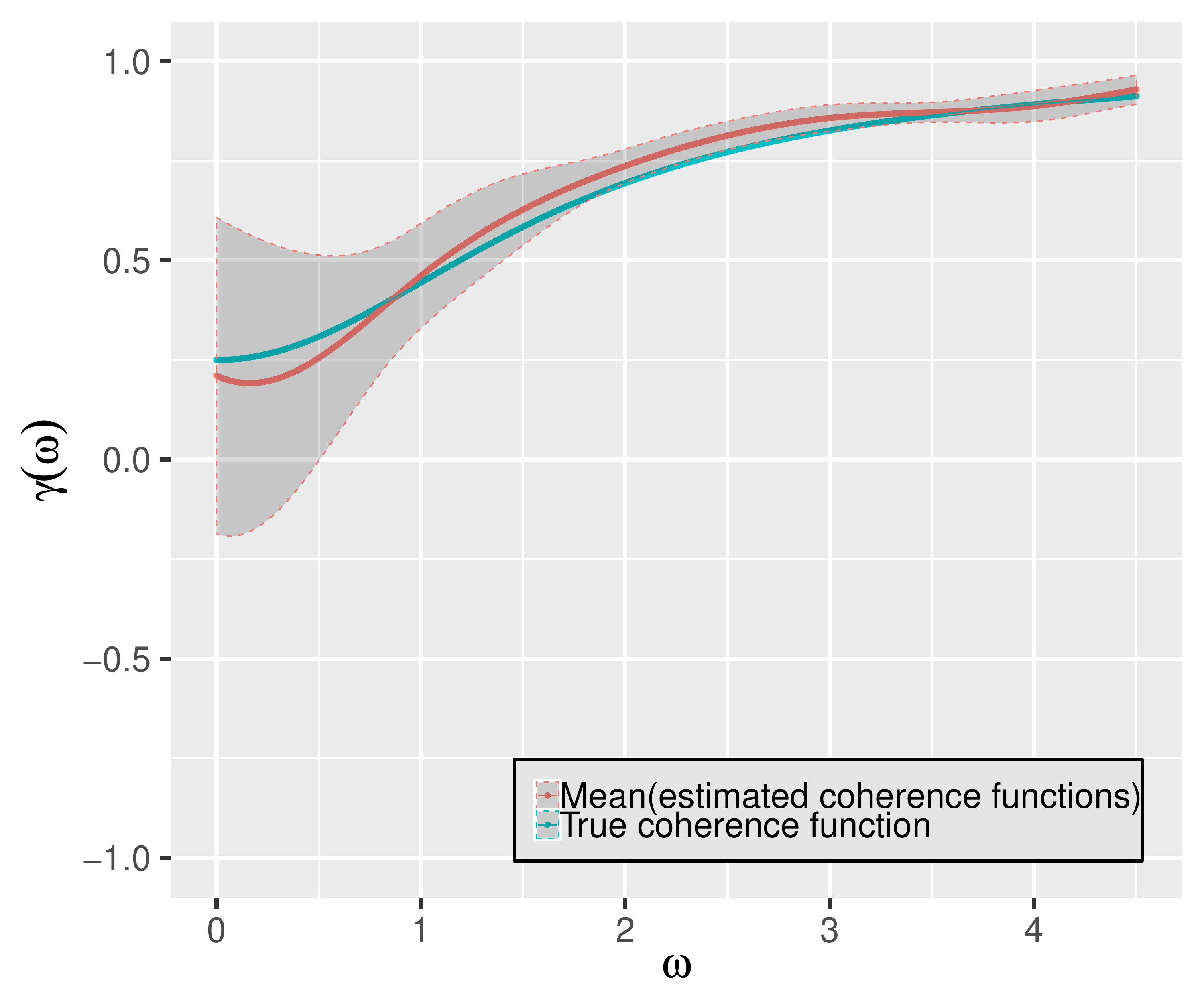}}
\subfigure[]{\label{fig:4b}\includegraphics[width=52mm]{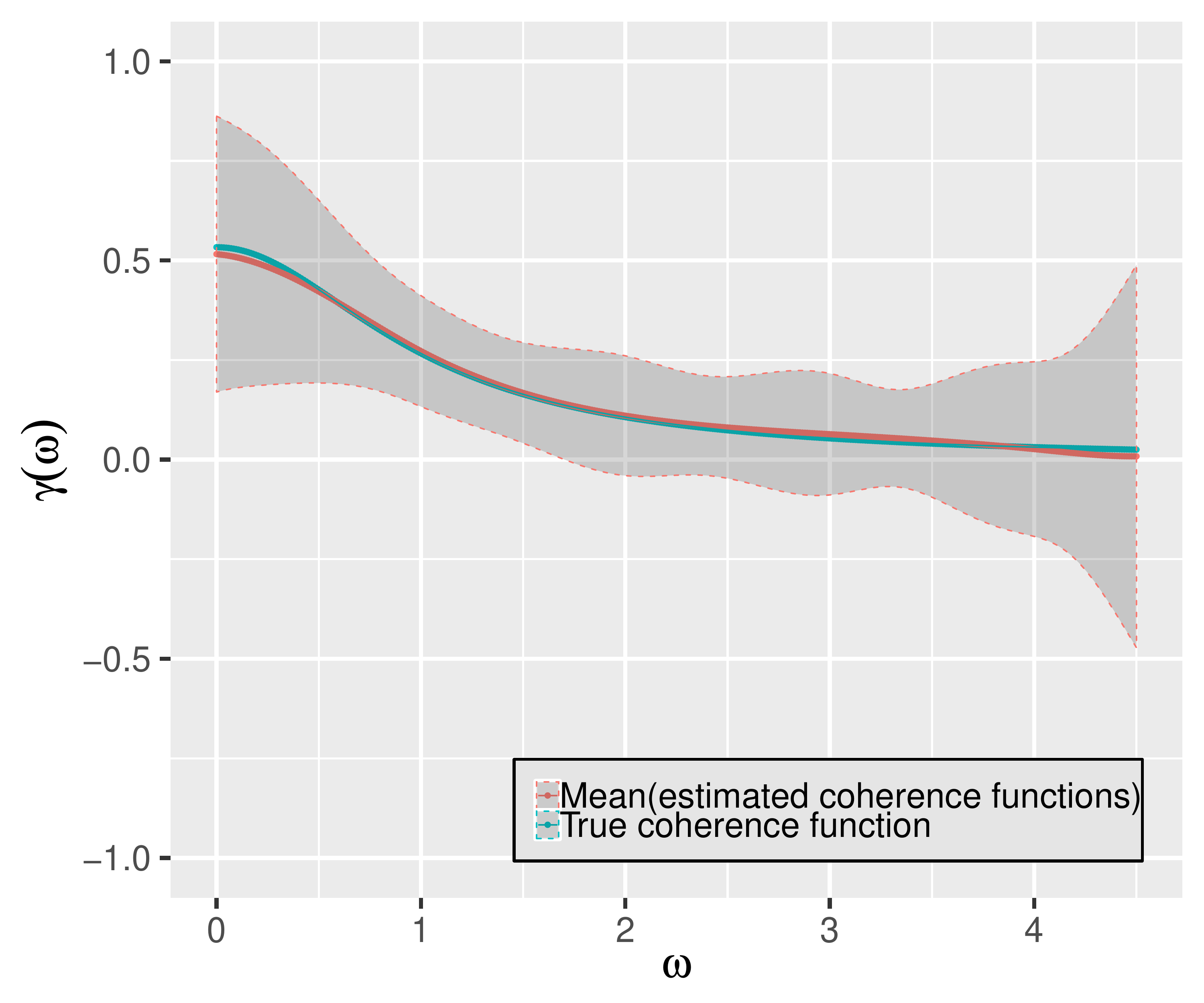}}
\subfigure[]{\label{fig:4c}\includegraphics[width=52mm]{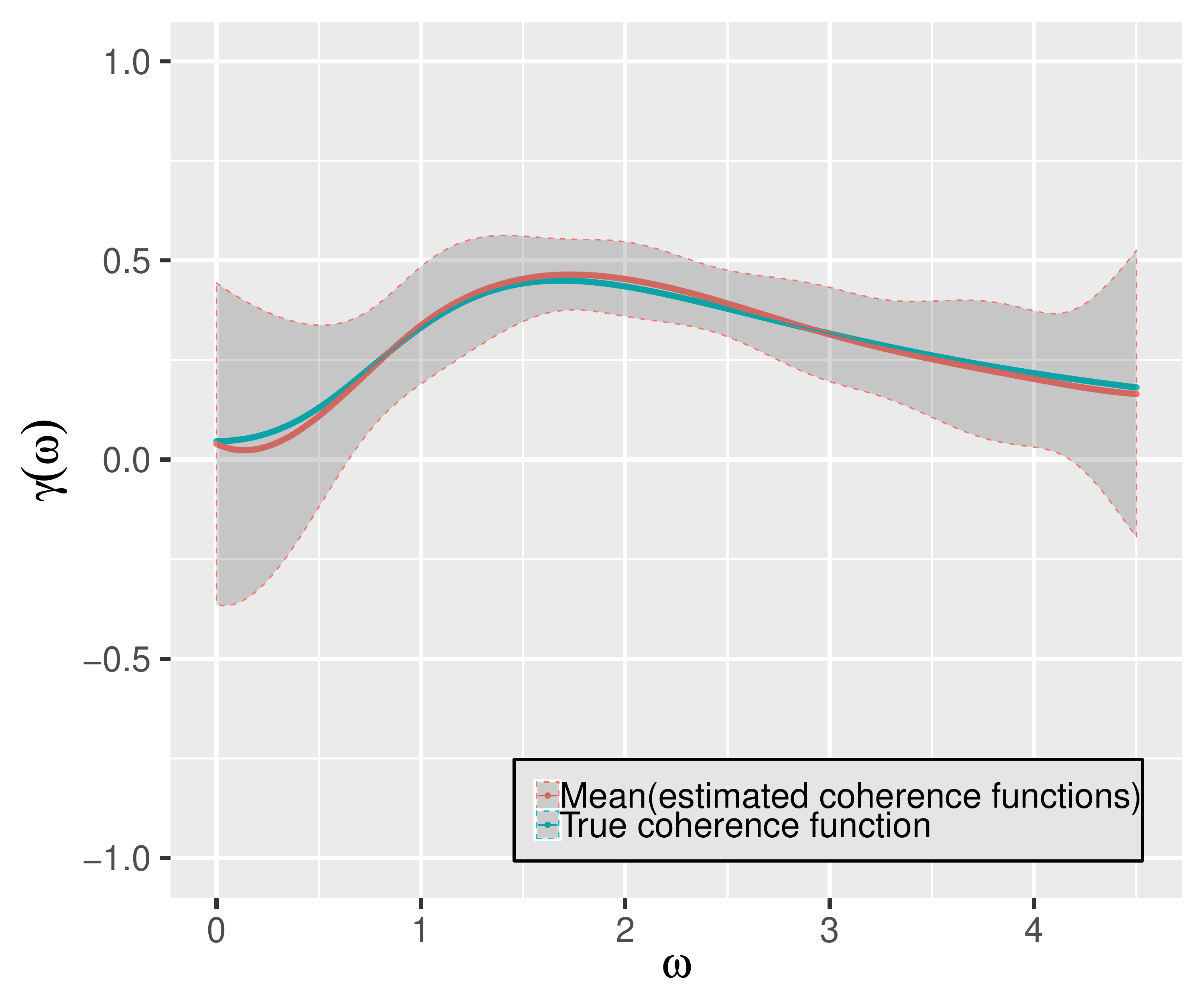}}
\caption{Comparison of the average estimate of the coherence function (95\% pointwise intervals in grey) and the true coherence functions for the processes generated from Model 1 (a), Model 2 (b) and Model 3 (c).    }
\label{fig:4}
\end{figure}

Figure \ref{fig:4} shows a comparison of the true coherence function and the averaged  estimated coherence function with 95\% pointwise intervals for the three cases of monotonically increasing coherence (Figure \ref{fig:4a}), monotonically decreasing coherence (Figure \ref{fig:4b}) and the coherence function with a bump (Figure \ref{fig:4c}). For all the three cases, the averaged estimated coherence function overlaps the true coherence function at almost all frequencies, thus indicating the efficiency of our model in adequately capturing the cross-spectral behaviour of the processes. Additionally, it also implies sufficiently reasonable fit of the cross-covariances, due to the complementary translation of coherence functions in the frequency domain to the cross-covariances in the space domain. Table \ref{tab:sim1} reports the average estimates of marginal parameters with their standard errors in parenthesis, to draw a comparison between the true parameters of the exact marginal Mat{\'e}rn and the estimated parameters from our model with approximately Mat{\'e}rn marginals. The remarkable closeness of the estimated spatial scales $\{a_i,\;i=1,2\}$  and the variances $\{\sigma_i^2,\;i=1,2\}$ of our model to the true parameter values demonstrates satisfactory marginal fits. Although our semiparametric model requires a slightly higher number of parameters as compared to the true full bivariate Mat{\'e}rn model, the validity conditions are much simpler to implement, and leads to a noticeably good fit for both the marginal and cross-process relationships.

\begin{table}[ht]
\centering
\caption{Simulation summary for marginal parameter estimates. The true values under Model 1-3 corresponds to the parameter values for the full bivariate Mat\'ern model chosen for simulations. The average estimate and standard error values under Model 1-3 corresponds the mean and standard error of the marginal parameter estimates from the semiparametric model over 50 runs. Note that average estimate and standard error entries for the last three columns are left blank since the cross-covariance part the semiparametric model is non-parametric and has been shown as comparison of coherence functions in Figure \ref{fig:4}}

\begin{tabular}{|c|c|c|c|c|c|c|c|c|c|c|}
    \hline
    \textbf{Models} & \textbf{Parameters} & $\boldsymbol{a_1}$ & $\boldsymbol{\sigma^2_1}$ & $\boldsymbol{\nu_1}$ & $\boldsymbol{a_2}$ & $\boldsymbol{\sigma^2_2}$ & $\boldsymbol{\nu_2}$ & $\boldsymbol{a_{12}}$ & $\boldsymbol{\nu_{12}}$ & $\boldsymbol{\rho_{12}}$\\\cline{1-11}
    \multirow{3}{*}{Model 1} & True value & 1 & 1 & 1 & 1 & 1 & 1 & $\sqrt{2}$ & 1 & 0.5\\ \cdashline{2-11} & Average estimate & 1.13 & 0.99 & - & 1.13  & 0.98 &- &- & - & -\\ \cdashline{2-11} & Standard error & (0.06)&  (0.08)&-& (0.06) & (0.08)& - & - & - & - \\\cline{1-11}
    \multirow{3}{*}{Model 2}&True value& 1 & 1 & 3 & 1 & 1 & 3 & 1 & 4 & 0.4\\ \cdashline{2-11}
    &Average estimate&1.01 &0.99  &-&1.02 &0.98 &-&-&-&-\\ \cdashline{2-11}&Standard error& (0.03)& (0.12) &-& (0.03)& (0.12)&-&-&-&-\\\cline{1-11}
  \multirow{3}{*}{Model 3}&True value& 0.5 & 1 & 3 & 1 & 1 & 3 & 1.2 & 4 & 0.1\\ \cdashline{2-11}
    &Average estimate&0.51&1 &-&1.01 &0.99 &-&-&-&-\\ \cdashline{2-11}
    &Standard error&(0.02)&(0.22)&-& (0.03)&(0.11)&-&-&-&-\\\hline
\end{tabular}

\label{tab:sim1}
\end{table}

\subsection{Simulation 2: Linear Model of Coregionalization}\label{sim2}
In this section, we consider a zero mean bivariate Gaussian random field $\textbf{X}(\textbf{s})=(X_1(\textbf{s}),X_2(\textbf{s}))^{\text{T}}$ on 500 irregularly spaced locations in the domain $[0,40]^2$ with cross and marginal spatial dependence described by the LMC:
\[\textbf{X}(\textbf{s})=\begin{pmatrix}X_1(\textbf{s}) \\X_2(\textbf{s})\end{pmatrix}=\begin{bmatrix}b_{11} & b_{12} \\b_{21} & b_{22}\end{bmatrix}\begin{pmatrix}Z_1(\textbf{s}) \\Z_2(\textbf{s})\end{pmatrix}=\textbf{B}\textbf{Z}(\textbf{s}),\]where \textbf{B} is the coregionalization matrix that supervises the magnitude of dependencies on the uncorrelated latent processes \textbf{Z}(\textbf{s}). We specify the independent processes $Z_1(\textbf{s})$ and $Z_2(\textbf{s})$ to marginally admit Mat{\'e}rn covariance functions  $\textbf{M}(\textbf{h}|\sigma_1,\nu_1,a_1)$ and $\textbf{M}(\textbf{h}|\sigma_2,\nu_2,a_2)$, respectively. The coherence function for the bivariate process $\textbf{X}(\textbf{s})$ is then given as:\[\gamma_{12}(\omega)=\frac{b_{11}b_{21}f_1(\omega)+b_{12}b_{22}f_2(\omega)}{\sqrt{b_{11}^2f_1(\omega)+b_{12}^2f_2(\omega)}\sqrt{b_{21}^2f_1(\omega)+b_{22}^2f_2(\omega)}},\]where $f_1(\omega)$ and $f_2(\omega)$ are the Mat{\'e}rn spectral densities corresponding to $\textbf{M}(\textbf{h}|\sigma_1,\nu_1,a_1)$ and $\textbf{M}(\textbf{h}|\sigma_2,\nu_2,a_2)$, respectively.

\begin{figure}[!t]
\centering     
\subfigure[]{\label{fig:r2a}\includegraphics[width=58mm]{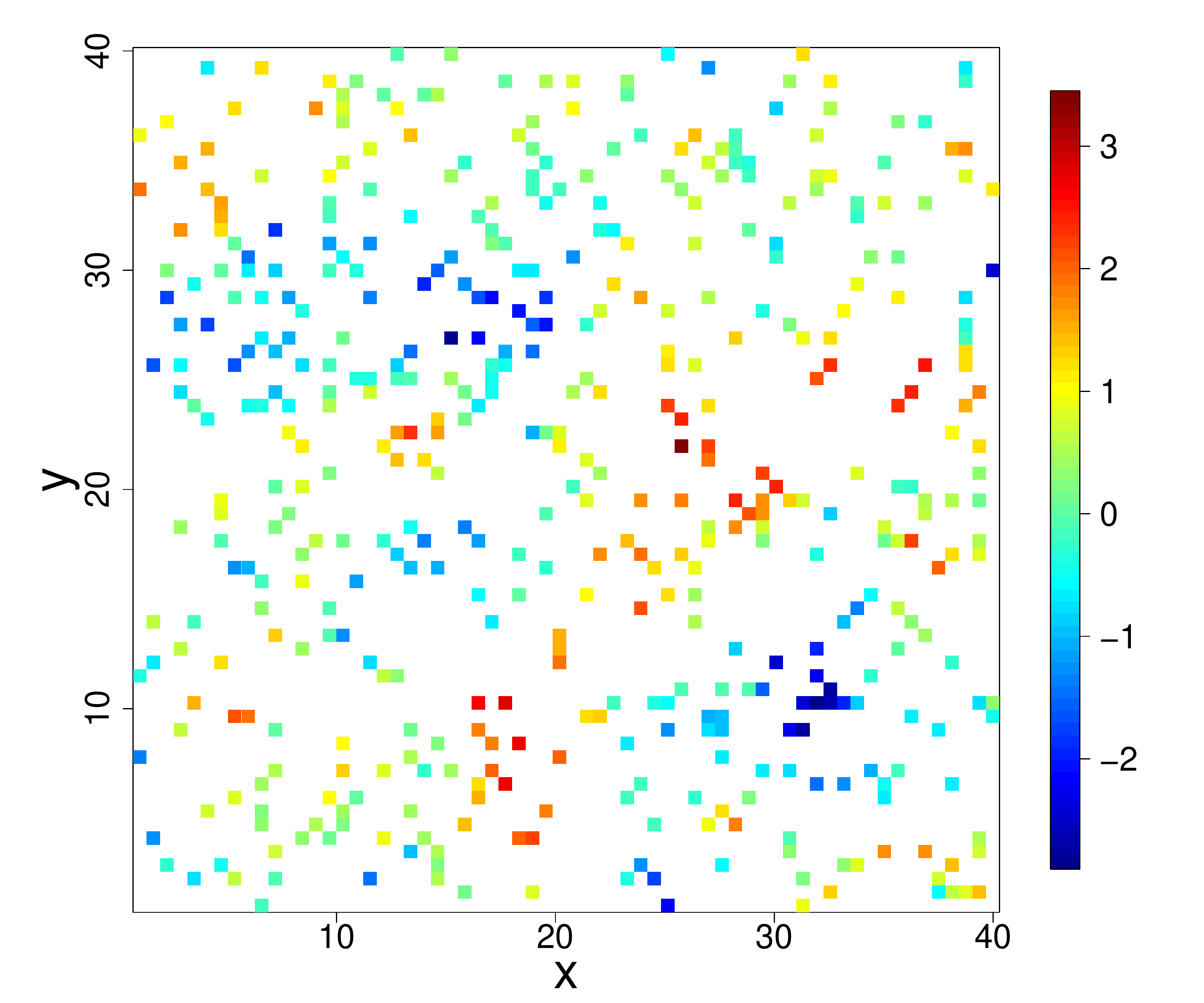}}
\subfigure[]{\label{fig:r2b}\includegraphics[width=58mm]{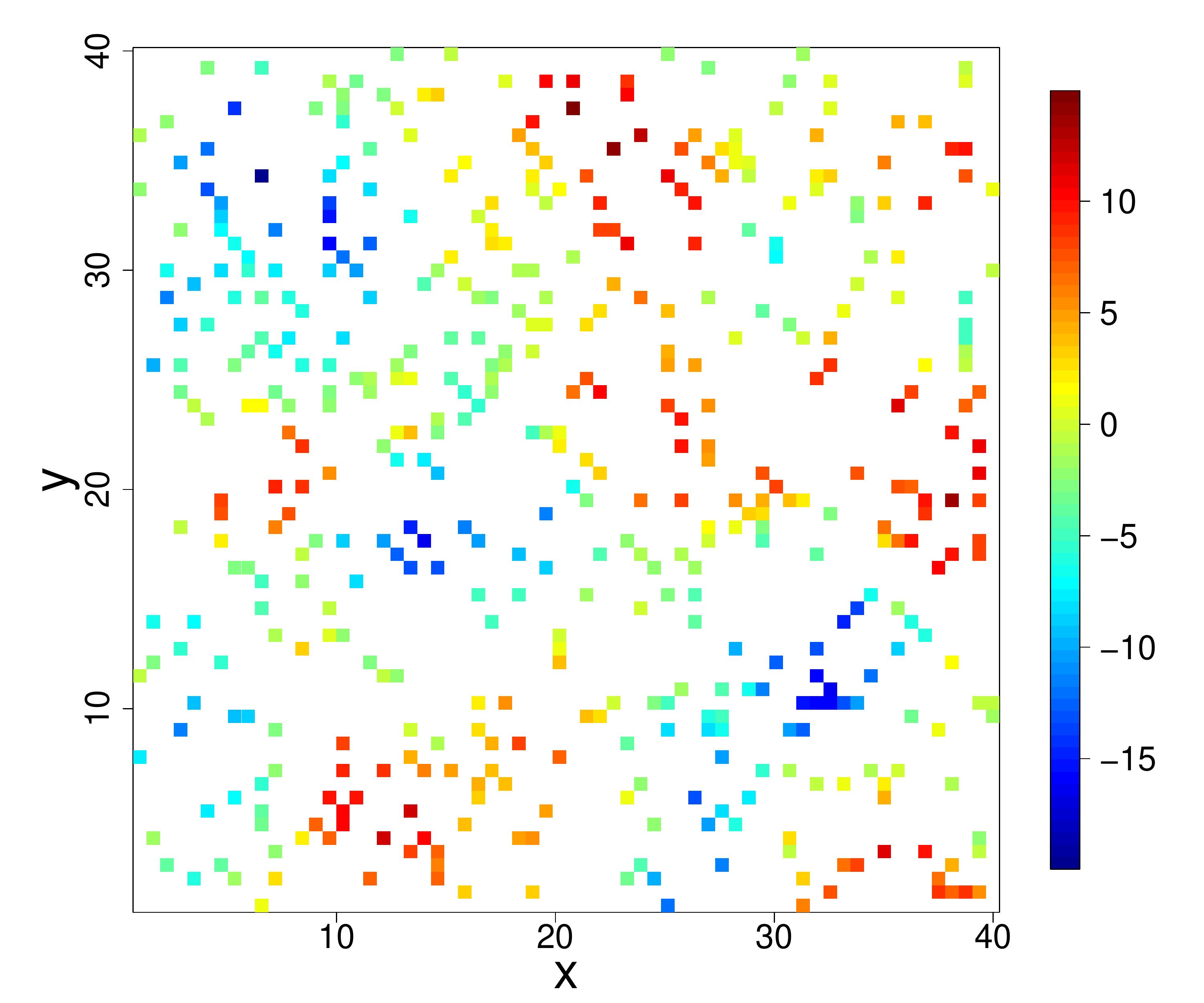}}
\caption{Example of one realization for a bivariate process from the specified LMC ((a) Variable 1, (b) Variable 2)).    }
\label{fig:r2}
\end{figure}

We consider the marginal Mat{\'e}rn parameters for $\textbf{Z}(\textbf{s})$ to be $(\sigma_1,\nu_1,a_1)=(1,1,0.5)$ and $(\sigma_2,\nu_2,a_2)=(1,2,0.5)$, and we set the entries of the   coregionalization matrix \textbf{B} as $b_{11}=1,b_{12}=0.4,b_{21}=0.9 \text{ and }b_{22}=7.5$. The coherence function for a bivariate process with this choice of parameters shows a decreasing trend at lower frequencies, followed by an increasing trend at higher frequencies. We simulate 50 realizations of the specified bivariate process $\textbf{X}(\textbf{s})$, and fit our semiparametric model (\ref{eq6}) using MLE, to model the coherence function as well as the marginal and cross-process dependence. An example realization for the simulated bivariate process from the specified LMC is shown in Figure \ref{fig:r2}.

\begin{figure}[h]
    \centering
    \includegraphics[scale=0.4]{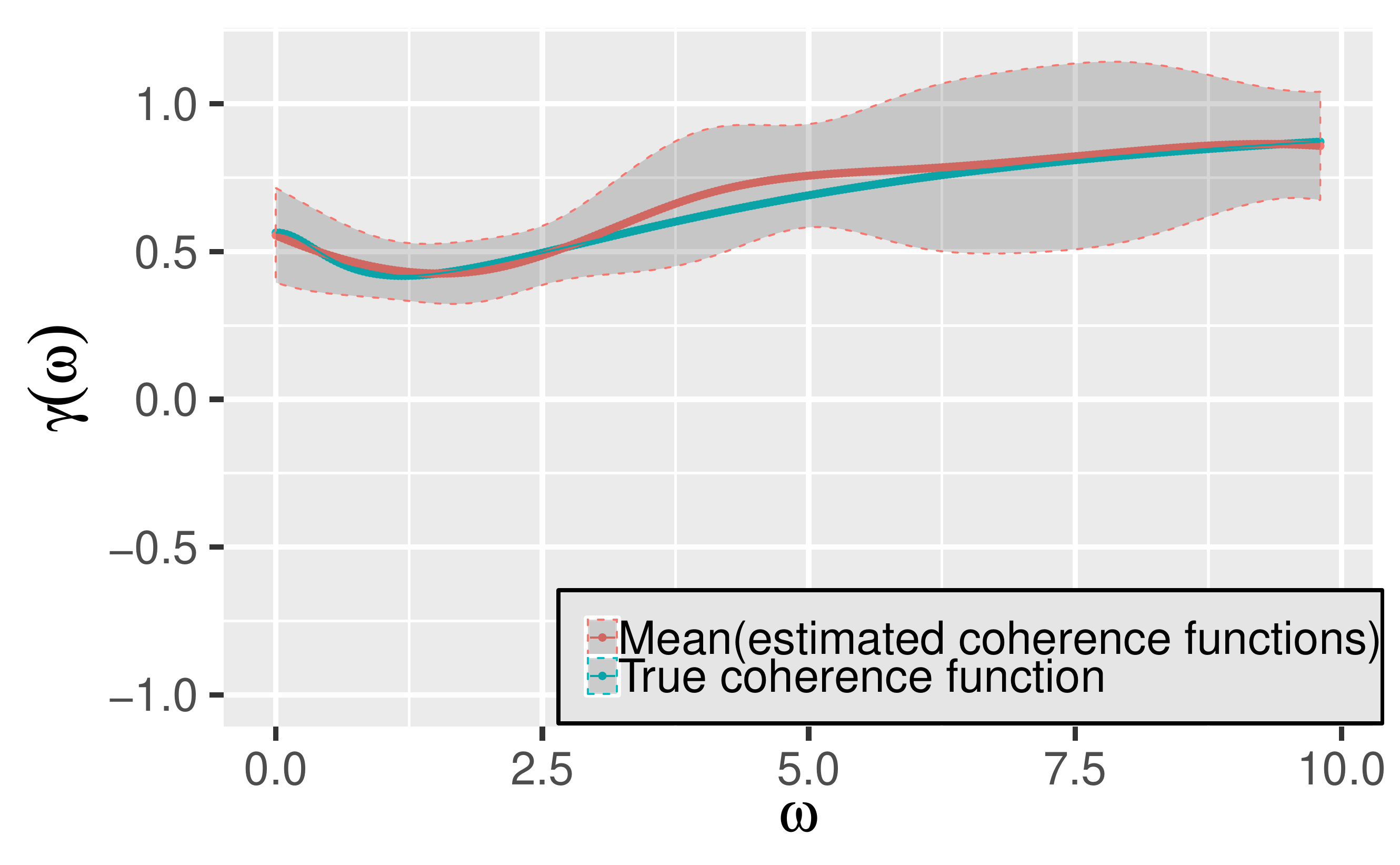}
    \caption{Comparsion of the average estimate of the coherence functions (95\% pointwise intervals in grey) and the true coherence function for the process generated from the specified LMC.}
    \label{fig:5}
\end{figure}

Prior to the estimation of our semiparametric model (\ref{eq6}) from the simulated datasets, we are required to specify the value of $m$ and the threshold frequency $\omega_t$, which we set at 299 and 9.8, respectively. For the specification of our B-spline based coherence function, we set $\Delta=2$ (or equivalently $K=4$), which in turn would require the estimation of 8 B-spline coefficients. Therefore, for this simulation study, we estimate a total of 14 parameters, that include 6 marginal Mat{\'e}rn parameters $(\sigma_i,\nu_i,a_i,\;i=1,2)$ and 8 B-spline coefficients $\textbf{S}_{12}=\{b_k^{(12)},k=-3,-2,\dots,4\}.$
\begin{table}[h!]
\centering
\caption{Average estimates and standard error of marginal parameters from the semiparametric model}
 \begin{tabular}{|c| c c c c c c  |}
 \hline
\textbf{Model Parameters} & $\boldsymbol{\sigma_1^2}$ & $\boldsymbol{a_1}$ & $\boldsymbol{\nu_1}$ & $\boldsymbol{\sigma_2^2}$ & $\boldsymbol{a_2}$ & $\boldsymbol{\nu_2}$ \\ [0.5ex] 
 \hline
 Average estimates & 1.21 & 0.45 & 0.96 & 61.63 & 0.46 & 1.84\\ 
 \hline
 Standard error & (0.19) & (0.076) & (0.11) & (1.58) & (0.04) & (0.10) \\
 \hline
  \end{tabular}
 \label{tab:sim2}
\end{table}

The averaged estimated coherence functions with 95\% pointwise interval and the true underlying coherence function shown in Figure \ref{fig:5} display conspicuous comparability. Our semiparametric model efficiently recovers the true shape of the underlying coherence function, which, moreover, signals toward decent fit of the cross-covariance function. Table \ref{tab:sim2} reports the estimates and standard errors of marginal parameters from our semiparametric model. Note that the estimates reported in Table \ref{tab:sim2} correspond to the marginal parameter estimates of our semiparametric model that describes the marginal spatial dependences of the process $\textbf{X}(\textbf{s})$, and therefore its direct comparison with the true Mat{\'e}rn parameters $(\sigma_i,\nu_i,a_i,\;i=1,2)$ of $\textbf{Z}(\textbf{s})$ is not straightforward. However, the true marginal variances for the processes $X_1(\textbf{s})$ and $X_2(\textbf{s})$ are $b_{11}^2\textbf{M}(\|0\||1,1,0.5)+b_{12}^2\textbf{M}(\|0\||1,2,0.5)=1.16$ and $b_{21}^2\textbf{M}(\|0\||1,1,0.5)+b_{22}^2\textbf{M}(\|0\||1,2,0.5)=57.06$, respectively, and are comparable with the estimated marginal variances of our semiparametric model reported in Table \ref{tab:sim2}. 

\section{Applications to $\text{PM}_{2.5}$ and Wind Speed Data}\label{sec:app}
We now illustrate the flexibility of our proposed semiparametric approach by applying our method to an atmospheric dataset consisting of a bivariate spatial field of particulate matter concentrations $(\text{PM}_{2.5})$ and wind speed. $\text{PM}_{2.5}$ is one of the principle indicators of air pollution level and represents the concentration of fine particulate matter with diameter less than $2.5\mu$m suspended in the atmosphere. Its major constituent components include nitrate, sulfate, organic carbon and elemental carbon, which in high concentrations, have hazardous effects on human health \citep{dominici,pope,samoli,chang}. While various meteorological variables such as regional stagnation, humidity, precipitation, etc., impact the concentration of $\text{PM}_{2.5}$ in polluted regions, here we focus on $\text{PM}_{2.5}$'s association with wind speed, which generally tends to be negatively correlated in nature \citep{JACOB200951}. We explore the marginal and cross-spatial dependence of $\text{PM}_{2.5}$ and wind speed by fitting various multivariate spatial models. Moreover, we perform spatial prediction to draw a comparison between the performance of our semiparametric model and other traditionally used multivariate models such as full bivariate Mat{\'e}rn and the LMC.

We study the dynamics of $\text{PM}_{2.5}$ and wind speed over the North-Eastern climatic region of the United States which comprises 11 states, namely,  Maine, New Hampshire, Vermont, New York, Massachusetts, Connecticut, Rhode Island, Pennsylvania, New jersey, Delaware and Maryland. The data for $\text{PM}_{2.5}$ is sourced from the Environmental
Protection Agency (EPA) which provides the daily average values that are generated
via Community Multiscale Air Quality Modeling System (CMAQ,
\url{https://www.epa.gov/cmaq}). The wind speed data is obtained from North
American Regional Reanalysis (NARR, \url{https://www.esrl.noaa.gov/psd}) which provides the monthly mean values of various meteorological variables. The raw datasets for our two variables differ in their spatial and temporal resolution, which we adjust by averaging the  $\text{PM}_{2.5}$ data. We average the daily $\text{PM}_{2.5}$ values over each month to comply with monthly mean wind speed data, and in addition we spatially average the monthly mean $\text{PM}_{2.5}$ data over the vicinity of 481 wind speed data locations to prepare a colocated bivariate $\text{PM}_{2.5}$/wind speed dataset. 

\begin{figure}[h]
\centering     
\subfigure[]{\label{fig:6a}\includegraphics[width=60mm]{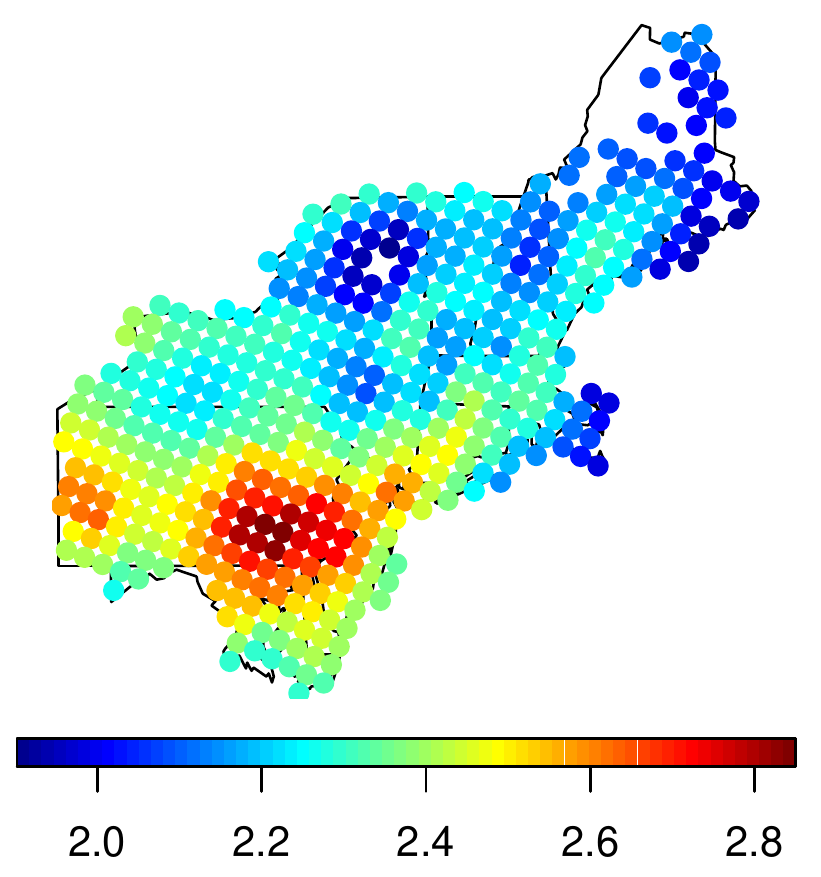}}
\subfigure[]{\label{fig:6b}\includegraphics[width=60mm]{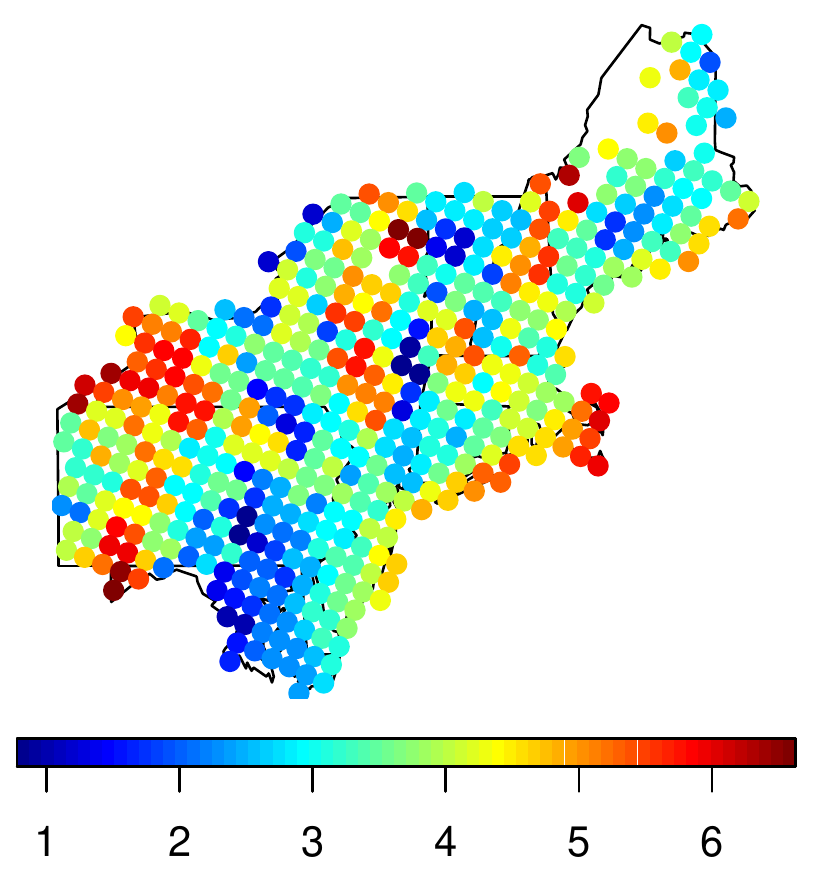}}
\caption{(a) log ($\text{PM}_{2.5}$) data over the North-Eastern climatic region of the United States. (b) Wind speed data over the North-Eastern climatic region of the United States.}
\label{fig:6}
\end{figure}

For our application, we consider the bivariate $\text{PM}_{2.5}$/wind speed data for the month of January 2013 (shown in Figure \ref{fig:6}). Whereas the wind speed exhibits approximately Gaussian distribution, the distribution of $\text{PM}_{2.5}$ shows positive skewness, which prompts us to log transform $\text{PM}_{2.5}$ to more closely satisfy the assumption of a bivariate Gaussian random field. Here, we primarily focus on modeling the second-order dependence structure of the $\text{log (PM}_{2.5})$ and wind speed; therefore, we detach the mean component by subtracting their respective empirical marginal means. Furthermore, we compute the empirical marginal variances and exercise componentwise standardization to bring (1) uniformity in the order of magnitude of process components and (2) numerical stability. Now, let us assume $\textbf{X}(\textbf{s})=\big(X_{PM_{2.5}}(\textbf{s}),X_{WS}(\textbf{s})\big)^{\text{T}}$ to be a bivariate Gaussian random field, where components $X_{PM_{2.5}}$ and $X_{WS}$ represent the standardized $\text{log (PM}_{2.5})$ and wind speed, respectively. Then, for the set of 481 observed locations $\{\textbf{s}_1,\dots,\textbf{s}_{481}\}$ (Shown in Figure \ref{fig:6}), $\textbf{X}\sim MVN_{982}(0,\Sigma_{982\times982})$, where $\Sigma_{982\times982}$ is the covariance matrix and our primary object of interest that we model using various bivariate spatial models.

 Prior to modeling the covariance matrix $\Sigma_{982\times982}$, we divide our data into a training set of 381 randomly selected locations and a validation set of the remaining 100 locations. We then proceed to fit various bivariate covariance models, augmented with nugget effects to capture the measurement errors, on 381 training locations, using the method of maximum likelihood. In particular, we consider six candidate models; an independent Mat{\'e}rn model that serves as our baseline performance standard due to its complete incomprehension of the cross-covariances between $X_{PM_{2.5}}$ and $X_{WS}$;  the commonly used full bivariate Mat{\'e}rn model; full LMC with two latent Mat{\'e}rn fields; and our proposed semiparametric model with three different choices of uniform knot spacing $\Delta$.
 
 For our semiparametric model, we specify the threshold frequency $\omega_t=9$, and set $m=499$ for the discretization of the frequency interval $[0,9]$. We consider three values of the uniform knot spacing $\Delta\in(2,4,5)$, which allows for varying degrees of flexibility in the underlying coherence function of the semiparametric model. The model with $\Delta=2$ enjoys the most flexible underlying coherence function relative to the models with $\Delta=4$ and $\Delta=5$, having a slightly tighter construct for the shape of the underlying coherence functions. The semiparametric models with $\Delta\in(2,4,5)$ require the estimation of 8,6 and 5 B-spline coefficients, respectively, in addition to 6 marginal parameters and 2 parameters representing the nugget effect of each process component.  
 
\begin{table}[h!]
\centering
\caption{Model fit summary for different candidate models. The highest log-likelihood value (shown as bold) is achieved by the semiparametric $(\Delta=2)+$Nugget model and the lowest AIC (shown as bold) is achieved by the semiparametric $(\Delta=4)+$Nugget model}
 \begin{tabular}{|l| c |c |c|}
 \hline
\textbf{Candidate Models} & \textbf{No. of  parameters} & \textbf{Log-likelihood} & \textbf{AIC} \\ [0.005ex] 
 \hline
 Independent Mat{\'e}rn + Nugget & 8 & -331.179 & 678.357 \\ 
 \hline
 Full bivariate Mat{\'e}rn + Nugget &  11 & -331.429 & 684.857 \\
 \hline
 \text{LMC} + Nugget &  12 & -312.226 & 648.452 \\
 \hline
 Semiparametric ($\Delta=2)$ + Nugget& 16 & $\mathbf{-307.989}$ & 647.977 \\
 \hline
 \text{Semiparametric ($\Delta=4)$} + Nugget  & 14 & -308.092 & $\mathbf{644.184}$  \\
 \hline
 Semiparametric ($\Delta=5)$ + Nugget  &  13 & -309.123  & 644.246\\ [0.005ex] 
 \hline
\end{tabular}

\label{tab:app1}
\end{table}

Table \ref{tab:app1} reports the maximized log-likelihood values and the  Akaike information criterion (AIC) values along with the number of parameters for the six candidate models. Strikingly, Table \ref{tab:app1} points out the comparable performance of the full bivariate Mat{\'e}rn model and the independent Mat{\'e}rn model in terms of maximized log-likelihood, and, in fact, identifes the full bivariate Mat{\'e}rn as the most inferior model in terms of the AIC values. While this result seems unrealistic and misleading at first glance due to the theoretically desired properties that the full bivariate Mat{\'e}rn model enjoys, it actually indicate towards the problems associated with its inefficient parameter estimation. We use the function \emph{\tt{RFfit}} from the R-package \emph{\tt{RANDOMFIELDS}} \citep{randf} to fit the full bivariate Mat{\'e}rn model, which in our case provides reasonably good estimates for the marginal parameters, but gives a noticeably substandard estimate for cross-covariance parameters. The estimated co-located correlation coefficient $\widehat{\rho_{12}}=-6.70\times10^{-09}$ is numerically equivalent to 0, and is indeed far from its empirical value of $-0.39$. The estimate $\widehat{\rho_{12}}=-6.70\times10^{-09}$ reduces the full bivariate Mat{\'e}rn model to almost independent Mat{\'e}rn model, thus, producing similar log-likelihood values, but a higher AIC value due to its 3 additional cross-covariance parameters. We observe a significant improvement in the log-likelihood value and the AIC value for the full LMC model as compared to the baseline independent Mat{\'e}rn case, which is not surprising because the full LMC takes into account the cross-process spatial dependence between $X_{PM_{2.5}}$ and $X_{WS}$, unlike the independent Mat{\'er}n model. Our semiparametric model in all three cases of $\Delta\in\{2,4,5\}$ outperforms all other candidate models as it achieves the highest log-likelihood values and the lowest AIC values, which is to be expected because of the flexible specification of underlying coherence function. Even the most restricted semiparametric model corresponding to $\Delta=5$ demonstrates a superior fit than all the other candidate models.  

\begin{figure}[h]
    \centering
    \includegraphics[scale=0.35]{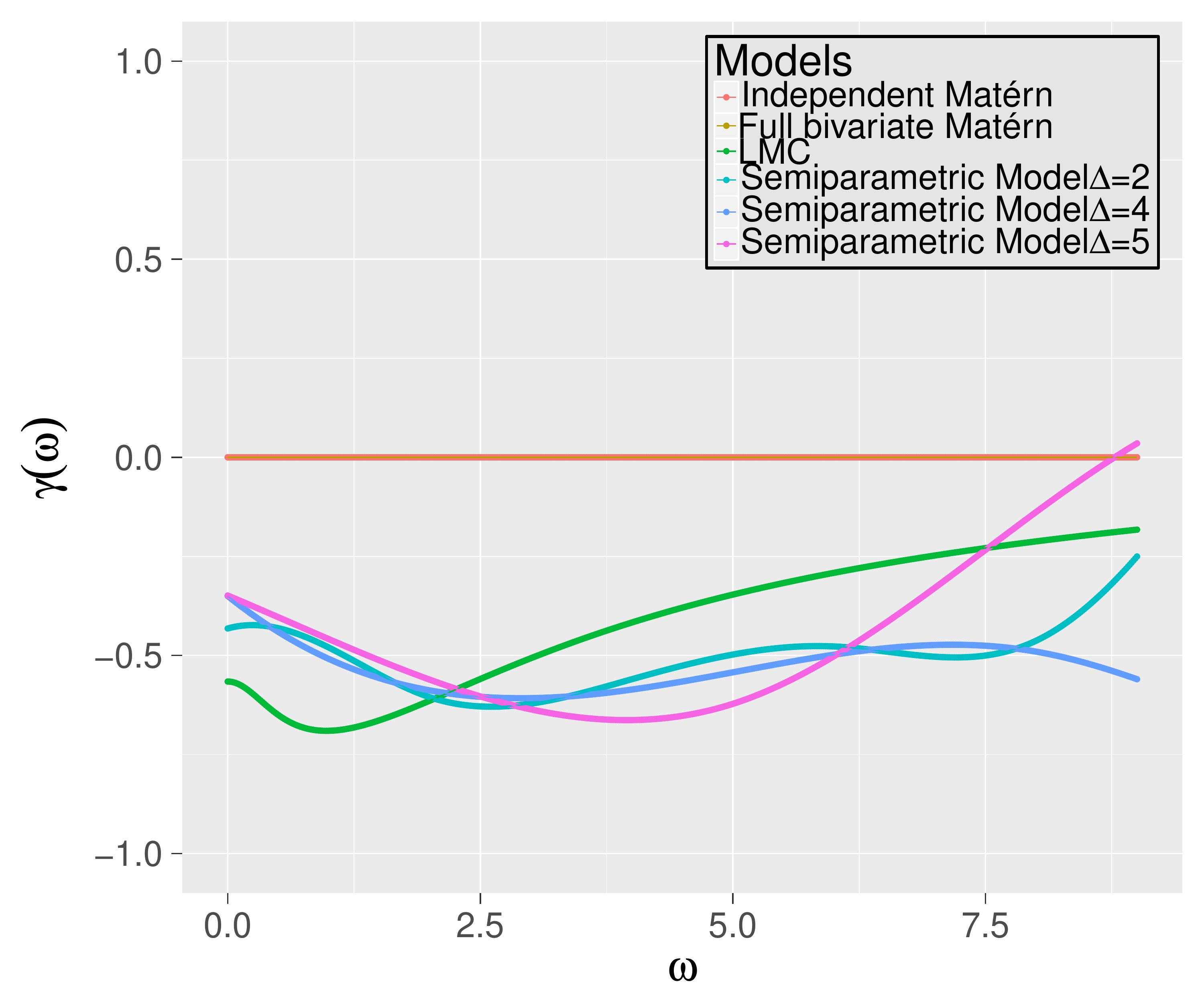}
    \caption{Coherence between log ($\text{PM}_{2.5}$) and wind speed estimated using different candidate models.}
    \label{fig:7}
\end{figure}

Figure \ref{fig:7} reveals the estimated coherence functions from all the candidate models. The independent Mat{\'e}rn model exhibits zero coherence at all frequency bands, which is obvious due to its assumed independence between $X_{PM_{2.5}}$ and $X_{WS}$. The co-located correlation coefficient $\rho_{12}$ in the full bivariate Mat{\'e}rn model acts as the scaling parameter for its coherence function, which being estimated close to zero, puts the coherence practically at 0 for all the frequency bands. The estimated coherence function for the full LMC model acquires a shape similar to the one we studied in Section \ref{sim2}, but lies in the negative axis and puts the lowest coherence (highest in magnitude) at $\omega_t\approx0.98$. The most restricted semiparametric model with $\Delta =5$ shares the common shape with the LMC; however, it puts the lowest coherence at $\omega_t\approx3.95$. The other two relatively flexible semiparametric models with $\Delta=2$ and $\Delta=4$ exhibit slightly oscillating coherence functions, and are even favoured by the log-likelihood and AIC values to represent the best fit for the true underlying coherence that cannot be captured by any existing multivariate models.

\begin{table}[h!]
\centering
\caption{Prediction scores for  different candidate models. The semiparametric $(\Delta=5)+$Nugget model shows best prediction performance in terms of RMSE, NMSE and mCRPS (shown as bold) and the semiparametric $(\Delta=4)+$Nugget model shows best prediction performance in terms of MAE and mLogS (shown as bold)}
 \begin{tabular}{|c| c |c |c| c| c|}
 \hline
\textbf{Model} & \textbf{RMSPE} & \textbf{MAE} & \textbf{NMSE} & \textbf{mCRPS} & \textbf{mLogS}\\ [0.005ex] 
 \hline
 Independent Mat{\'e}rn + Nugget & 0.533 & 0.333 & 0.746 & 0.242 & 0.232 \\ 
 \hline
 Full bivariate Mat{\'e}rn + Nugget&  0.534 & 0.333 & 0.745 & 0.243 & 0.236 \\
 \hline
 \text{LMC} + Nugget &  0.522 & 0.329 & 0.757 & 0.238 & 0.220\\
 \hline
 Semiparametric ($\Delta=2)$ + Nugget & 0.520 & 0.327 & 0.758 & 0.237 & 0.220 \\
 \hline
 \text{Semiparametric ($\Delta=4)$} + Nugget  & 0.519 & $\mathbf{0.327}$ & 0.760 & 0.236 & $\mathbf{0.218}$  \\
 \hline
 Semiparametric ($\Delta=5)$ + Nugget  &  $\mathbf{0.518}$ & 0.327  & $\mathbf{0.760}$ & $\mathbf{0.236}$ & 0.221\\ [0.005ex] 
 \hline
\end{tabular}

\label{tab:app2}
\end{table}

Here, we perform spatial predictions over the 100 left out validation locations for both the $X_{PM_{2.5}}$ and $X_{WS}$ to achieve a cross validation analysis for all the candidate models. In Table \ref{tab:app2}, we list some frequently used prediction scores combined for both the $X_{PM_{2.5}}$ and $X_{WS}$, computed over 100 validation locations. The smaller values of the root mean squared prediction error (RMSPE), mean absolute error (MAE), mean continuous ranked probability score (mCRPS) and the mean logarithmic score (mLogS) \citep{gneiting2007strictly} are suggestive of better predictions, whereas the normalised-mean-squared error (NMSE) indicates a better prediction for the value closer to unity. Here, the computed prediction scores identify the independent Mat{\'e}rn model and the full bivariate Mat{\'e}rn model as the worst among the candidate models. While this is expected for the independent Mat{\'e}rn model because the spatial predictions with the independent Mat{\'e}rn model correspond to the independent univariate kriging, which is generally inferior to the co-kriging, the poor performance of the full bivariate Mat{\'e}rn is due to its poor model estimation, and not because of its inflexibility. The LMC shows improvement in spatial prediction over the independent Mat{\'e}rn and full bivariate Mat{\'e}rn model, which is obvious as it utilizes correlations across the process components, however, due to its inflexible cross-covariance specification, its performance is not the best. Our proposed semiparametric models outperformed all the other candidate models in terms of spatial prediction, over nearly all cross-validation diagnostics combined for $X_{PM_{2.5}}$ and $X_{WS}$, which empirically substantiate the importance of flexibly modeling coherence functions for spatial predictions.

\section{Discussion}\label{sec:disc}
In this article, we introduced a semiparametric multivariate spatial covariance function via its spectral representation, that can flexibly model the coherence functions between the pair of components of a multivariate process. The B-spline based specification of the coherence function allows for more data-driven estimation of cross-covariances, relative to the available parametric models. We have presented simulation studies to demonstrate the performance of our proposed model through efficient maximum likelihood estimation of the multivariate spatial dependence, especially the underlying coherence function. The application of the proposed semiparametric model has been illustrated on a bivariate atmospheric dataset of particulate matter concentrations ($\text{PM}_{2.5}$) and wind speed over the North-Eastern region of the United States. We have shown that our semiparametric model outperformed the conventionally used full bivariate Mat{\'e}rn model and the LMC, by producing lower AIC values and prediction scores.

The choice of uniform knot spacing $(\Delta)$ is crucial, as it governs the possible shapes that the coherence function can achieve. While we tried a number of different adhoc values for $\Delta$ in our application section to choose the best model fit, the careful examination of the empirical coherence function can guide for the choice of $\Delta$ in case of complete data on a regularly spaced grid of location. However, when the spatial data is not located on grid points, we suggest to try different sensible values of $\Delta$ that maintain the trade-off between flexibility of coherence and the computational feasibility, and choose the best value based on cross-validation scores or some model selection criterion such as AIC.

In our proposed framework, we specified Mat{\'e}rn marginal, which makes our approach directly comparable with the full bivariate Mat{\'e}rn and the parsimonious multivariate Mat{\'e}rn models. However, any other choice of parametric or nonparametric spectral densities can be plugged in straightforwardly to specify marginal spatial dependence, and that would still lead to a valid multivariate model with exactly the same validity conditions provided in Theorem \ref{th1}, thus leaving the door open for any future improvements.

Our model specifies the spectral densities and coherence functions only up to a threshold frequency $\omega_t$; therefore, extending the proposed model to characterize spectral features for all frequencies $\omega\geq0$ is one potential direction for future research. This can be done by following the approach of \cite{semipzhu} to add a parametric tail part in the coherence function, which would further finding validity conditions on the tail part.


\begin{thebibliography}{}

\bibitem[Apanasovich and Genton, 2010]{doi:10.1093/biomet/asp078}
Apanasovich, T.~V. and Genton, M.~G. (2010).
\newblock Cross-covariance functions for multivariate random fields based on
  latent dimensions.
\newblock {\em Biometrika}, 97:15--30.

\bibitem[Apanasovich et~al., 2012]{multimat}
Apanasovich, T.~V., Genton, M.~G., and Sun, Y. (2012).
\newblock {A valid Mat{\'e}rn class of cross-covariance functions for
  multivariate random fields with any number of components}.
\newblock {\em Journal of the American Statistical Association}, 107:180--193.

\bibitem[Bhat et~al., 2010]{article}
Bhat, K., Haran, M., and Goes, M. (2010).
\newblock Computer model calibration with multivariate spatial output: A case
  study.
\newblock In Chen, M.-H., M{\"u}ller, P., Sun, D., Ye, K., and Dey, D.~K.,
  editors, {\em Frontiers of statistical decision making and Bayesian
  analysis}, pages 168--184. Springer, New York.

\bibitem[Chang et~al., 2011]{chang}
Chang, H.~H., Reich, B.~J., and Miranda, M.~L. (2011).
\newblock Time-to-event analysis of fine particle air pollution and preterm
  birth: Results from north carolina, 2001--2005.
\newblock {\em American Journal of Epidemiology}, 175(2):91--98.

\bibitem[Cram{\'e}r, 1940]{cramer}
Cram{\'e}r, H. (1940).
\newblock On the theory of stationary random processes.
\newblock {\em Annals of Mathematics}, 41:215--230.

\bibitem[De~Boor, 2001]{de2001}
De~Boor, C. (2001).
\newblock {\em A practical guide to splines}.
\newblock New York: Springer.

\bibitem[Dominici et~al., 2006]{dominici}
Dominici, F., Peng, R.~D., Bell, M.~L., Pham, L., McDermott, A., Zeger, S.~L.,
  and Samet, J.~M. (2006).
\newblock Fine particulate air pollution and hospital admission for
  cardiovascular and respiratory diseases.
\newblock {\em JAMA}, 295:1127--1134.

\bibitem[Gaspari and Cohn, 1999]{doi:10.1002/qj.49712555417}
Gaspari, G. and Cohn, S.~E. (1999).
\newblock Construction of correlation functions in two and three dimensions.
\newblock {\em Quarterly Journal of the Royal Meteorological Society},
  125:723--757.

\bibitem[Gaspari et~al., 2006]{doi:10.1256/qj.05.08}
Gaspari, G., Cohn, S.~E., Guo, J., and Pawson, S. (2006).
\newblock Construction and application of covariance functions with variable
  length-fields.
\newblock {\em Quarterly Journal of the Royal Meteorological Society},
  132:1815--1838.

\bibitem[Genton and Gorsich, 2002]{gor2002}
Genton, M.~G. and Gorsich, D.~J. (2002).
\newblock Nonparametric variogram and covariogram estimation with
  fourier--bessel matrices.
\newblock {\em Computational Statistics \& Data Analysis}, 41:47 -- 57.
\newblock Special issue on Matrix Computations and Statistics.

\bibitem[Genton and Kleiber, 2015]{genton2015}
Genton, M.~G. and Kleiber, W. (2015).
\newblock Cross-covariance functions for multivariate geostatistics.
\newblock {\em Statistical Science}, 30:147--163.

\bibitem[Gneiting et~al., 2010]{bimat}
Gneiting, T., Kleiber, W., and Schlather, M. (2010).
\newblock Mat{\'e}rn cross-covariance functions for multivariate random fields.
\newblock {\em Journal of the American Statistical Association},
  105:1167--1177.

\bibitem[Gneiting and Raftery, 2007]{gneiting2007strictly}
Gneiting, T. and Raftery, A.~E. (2007).
\newblock Strictly proper scoring rules, prediction, and estimation.
\newblock {\em Journal of the American Statistical Association}, 102:359--378.

\bibitem[Gorsich and Genton, 2004]{Gor2004}
Gorsich, D.~J. and Genton, M.~G. (2004).
\newblock On the discretization of nonparametric isotropic covariogram
  estimators.
\newblock {\em Statistics and Computing}, 14:99--108.

\bibitem[Goulard and Voltz, 1992]{Goulard1992}
Goulard, M. and Voltz, M. (1992).
\newblock Linear coregionalization model: Tools for estimation and choice of
  cross-variogram matrix.
\newblock {\em Mathematical Geology}, 24:269--286.

\bibitem[Greasby and Sain, 2011]{greasby2011}
Greasby, T.~A. and Sain, S.~R. (2011).
\newblock Multivariate spatial analysis of climate change projections.
\newblock {\em Journal of agricultural, biological, and environmental
  statistics}, 16:571--585.

\bibitem[Guttorp and Gneiting, 2006]{guttorp}
Guttorp, P. and Gneiting, T. (2006).
\newblock {Studies in the history of probability and statistics XLIX: On the
  Mat{\'e}rn correlation family}.
\newblock {\em Biometrika}, 93:989--995.

\bibitem[Helterbrand and Cressie, 1994]{Helterbrand1994}
Helterbrand, J.~D. and Cressie, N. (1994).
\newblock Universal cokriging under intrinsic coregionalization.
\newblock {\em Mathematical Geology}, 26:205--226.

\bibitem[Horn and Johnson, 2013]{horn2013matrix}
Horn, R.~A. and Johnson, C.~R. (2013).
\newblock {\em Matrix analysis}.
\newblock Cambridge University Press, Cambridge, 2nd edition.

\bibitem[Im et~al., 2006]{techreport}
Im, H.~K., Stein, M.~L., and Zhu, Z. (2006).
\newblock Semiparametric estimation of spectral densities with scattered data.
\newblock Technical report, University of Chicago, Center for Integrating
  Statistical and Environmental Sciences.

\bibitem[Im et~al., 2007]{semipzhu}
Im, H.~K., Stein, M.~L., and Zhu, Z. (2007).
\newblock Semiparametric estimation of spectral density with irregular
  observations.
\newblock {\em Journal of the American Statistical Association}, 102:726--735.

\bibitem[Jacob and Winner, 2009]{JACOB200951}
Jacob, D.~J. and Winner, D.~A. (2009).
\newblock Effect of climate change on air quality.
\newblock {\em Atmospheric Environment}, 43:51 -- 63.

\bibitem[Kleiber, 2017]{cohkl}
Kleiber, W. (2017).
\newblock Coherence for multivariate random fields.
\newblock {\em Statistica Sinica}, 27:1675--1697.

\bibitem[Majumdar and Gelfand, 2007]{Majumdar2007}
Majumdar, A. and Gelfand, A.~E. (2007).
\newblock Multivariate spatial modeling for geostatistical data using convolved
  covariance functions.
\newblock {\em Mathematical Geology}, 39:225--245.

\bibitem[Mardia and Goodall, 1993]{mardia1993}
Mardia, K.~V. and Goodall, C.~R. (1993).
\newblock Spatial-temporal analysis of multivariate environmental monitoring
  data.
\newblock In {\em Multivariate Environmental Statistics. North-Holland Series
  in Statistics and Probability}, volume~6, pages 347--386. North-Holland,
  Amsterdam.

\bibitem[Mat{\'e}rn, 1986]{matern}
Mat{\'e}rn, B. (1986).
\newblock {\em Spatial Variation}.
\newblock Berlin:Springer-Verlag, 2nd edition.

\bibitem[Pope~III and Dockery, 2006]{pope}
Pope~III, C.~A. and Dockery, D.~W. (2006).
\newblock Health effects of fine particulate air pollution: Lines that connect.
\newblock {\em Journal of the Air \& Waste Management Association},
  56:709--742.

\bibitem[Sain et~al., 2011]{sain2011}
Sain, S.~R., Furrer, R., and Cressie, N. (2011).
\newblock A spatial analysis of multivariate output from regional climate
  models.
\newblock {\em The Annals of Applied Statistics}, 5:150--175.

\bibitem[Samoli et~al., 2008]{samoli}
Samoli, E., Peng, R., Ramsay, T., Pipikou, M., Touloumi, G., Dominici, F.,
  Burnett, R., Cohen, A., Krewski, D., Samet, J., and Katsouyanni, K. (2008).
\newblock {Acute effects of ambient particulate matter on mortality in Europe
  and North America: Results from the APHENA study}.
\newblock {\em Environmental health perspectives}, 116:1480--1486.

\bibitem[Schlather et~al., 2015]{randf}
Schlather, M., Malinowski, A., Menck, P.~J., Oesting, M., and Strokorb, K.
  (2015).
\newblock Analysis, simulation and prediction of multivariate random fields
  with package randomfields.
\newblock {\em Journal of Statistical Software}, 63:1--25.

\bibitem[Schmidt and Gelfand, 2003]{doi:10.1029/2002JD002905}
Schmidt, A.~M. and Gelfand, A.~E. (2003).
\newblock A bayesian coregionalization approach for multivariate pollutant
  data.
\newblock {\em Journal of Geophysical Research: Atmospheres}, 108.

\bibitem[Shapiro and Botha, 1991]{botha1991}
Shapiro, A. and Botha, J. (1991).
\newblock Variogram fitting with a general class of conditionally nonnegative
  definite functions.
\newblock {\em Computational Statistics \& Data Analysis}, 11:87 -- 96.

\bibitem[Stein, 1999]{stein2012}
Stein, M.~L. (1999).
\newblock {\em Interpolation of spatial data: Some theory for kriging}.
\newblock Springer-Verlag New York.

\bibitem[Ver~Hoef and Barry, 1998]{ver1998constructing}
Ver~Hoef, J.~M. and Barry, R.~P. (1998).
\newblock Constructing and fitting models for cokriging and multivariable
  spatial prediction.
\newblock {\em Journal of Statistical Planning and Inference}, 69:275--294.

\bibitem[Ver~Hoef et~al., 2004]{doi:10.1198/1061860043498}
Ver~Hoef, J.~M., Cressie, N., and Barry, R.~P. (2004).
\newblock {Flexible spatial models for kriging and cokriging using moving
  averages and the fast Fourier transform (FFT)}.
\newblock {\em Journal of Computational and Graphical Statistics}, 13:265--282.

\bibitem[Wackernagel, 2003]{wackernagel2010multivariate}
Wackernagel, H. (2003).
\newblock {\em Multivariate geostatistics: An Introduction with Applications}.
\newblock Berlin: Springer, 3rd edition.

\bibitem[Watson, 1944]{watson1995treatise}
Watson, G.~N. (1944).
\newblock {\em A treatise on the theory of Bessel functions}.
\newblock Cambridge university press, 2nd edition.

\bibitem[Zhang, 2004]{zhang}
Zhang, H. (2004).
\newblock Inconsistent estimation and asymptotically equal interpolations in
  model-based geostatistics.
\newblock {\em Journal of the American Statistical Association}, 99:250--261.

\bibitem[Zhang, 2007]{doi:10.1002/env.807}
Zhang, H. (2007).
\newblock Maximum-likelihood estimation for multivariate spatial linear
  coregionalization models.
\newblock {\em Environmetrics}, 18:125--139.

\end{thebibliography}

\newpage
\begin{center}
\Large{Appendix}    
\end{center}
\appendix
\section{Proof of Theorem \ref{th1}}
The spectral matrix for the spectral densities in (\ref{eq4}) and (\ref{eq5}) is given as:\[\textbf{f}(\omega)=\begin{bmatrix} f_{11}(\omega) & \dots & f_{1p}(\omega) \\ \vdots & \ddots & \vdots \\ f_{p1}(\omega) & \dots & f_{pp}(\omega)
\end{bmatrix},\omega\leq\omega_t\]
\[=\begin{bmatrix}\sqrt{f_{11}(\omega)} & & \\
    & \ddots & \\
    & & \sqrt{f_{pp}(\omega)}
\end{bmatrix}\begin{bmatrix} 1 & \dots & \gamma_{1p}(\omega) \\ \vdots & \ddots & \vdots \\ \gamma_{p1}(\omega) & \dots & 1
\end{bmatrix}\begin{bmatrix} \sqrt{f_{11}(\omega)} & & \\
    & \ddots & \\
    & & \sqrt{f_{pp}(\omega)}
\end{bmatrix}\]
\[=\text{Diag}(\sqrt{f_{ii}(\omega)})_{i=1}^p\begin{bmatrix} 1 & \dots & \gamma_{1p}(\omega) \\ \vdots & \ddots & \vdots \\ \gamma_{p1}(\omega) & \dots & 1
\end{bmatrix}\text{Diag}(\sqrt{f_{ii}(\omega)})_{i=1}^p\]
The spectral matrix $\textbf{f}(\omega)$ is then nonnegative definite if the matrix \[\Gamma(\omega)=\begin{bmatrix} 1 & \dots & \gamma_{1p}(\omega) \\ \vdots & \ddots & \vdots \\ \gamma_{p1}(\omega) & \dots & 1
\end{bmatrix}\] is nonnegative definite ($\because$ if a nonnegative definite matrix $\textbf{M}$ is pre and post-multiplied by a full rank square matrix $\textbf{N}$ and its transpose $\textbf{N}^{\text{T}}$, the resulting matrix $\textbf{N}\textbf{M}\textbf{N}^{\text{T}}$ is nonnegative definite \cite[Observation~7.1.8, p.~431]{horn2013matrix} ).
\[\Gamma(\omega)=\begin{bmatrix} 1 & \dots & \sum_{k=-3}^Kb_k^{(1p)}B_k(\omega) \\ \vdots & \ddots & \vdots \\ \sum_{k=-3}^Kb_k^{(p1)}B_k(\omega) & \dots & 1
\end{bmatrix}\]
\[=\begin{bmatrix} \sum_{k=-3}^KB_k(\omega) & \dots & \sum_{k=-3}^Kb_k^{(1p)}B_k(\omega) \\ \vdots & \ddots & \vdots \\ \sum_{k=-3}^Kb_k^{(p1)}B_k(\omega) & \dots & \sum_{k=-3}^KB_k(\omega)
\end{bmatrix} (\because \sum_{k=-3}^KB_k(\omega)=1,\; \forall \omega\in [0,(K+1)\Delta) \]
\[=\sum_{k=-3}^KB_k(\omega)\boldsymbol{\beta}_k\]where $\boldsymbol{\beta}_k=\{b_k^{(ij)}\}_{i,j=1}^p$ are the $p\times p$ symmetric matrices with diagonal elements $\{b_k^{(ii)}=1$ $\forall \;i=1,2,\dots,p,\;\;k=-3,-2,\dots,K\}$. The quantity  $\sum_{k=-3}^KB_k(\omega)$ is nonnegative $\forall \omega \leq \omega_t$. Therefore the matrix $\Gamma(\omega)$ is nonnegative definite $\forall \omega \leq \omega_t$ if the matrices $\{\boldsymbol{\beta}_k,\;k=-3,\dots,K\}$ are nonnegative definite ($\because$ the linear combination of nonnegative definite matrices with nonnegative coefficients is a nonnegative definite matrix \cite[Observation~7.1.3, p.~430]{horn2013matrix}). Consequently, following the Cram{\'e}r's Theorem in its spectral density version, the matrix-valued covariance function $\textbf{C}(\textbf{h})=\{\text{C}_{ij}(\textbf{h})\}_{i,j=1}^p$ in (\ref{eq6}) is valid if the matrices $\{\boldsymbol{\beta}_k,\;k=-3,\dots,K\}$ are non-negative definite.

\section{Proof for Proposition \ref{prop1}}
For $\omega_t\rightarrow\infty$ and common spatial scale parameters $a_i=a>0,\;i=1,\dots,p$, the marginal spectral densities in (\ref{eq4}) becomes the untruncated Mat{\'e}rn spectral densities:
\[f_{ii}(\omega|\sigma_i,\nu_i,a)=\sigma_i^2\frac{\Gamma(\nu_i+d/2)a_i^{2\nu_i}}{\Gamma(\nu_i)\pi^{d/2}(a_i^2+\omega^2)^{\nu_i+d/2}}, \: \omega\geq0,\: \sigma_i,\nu_i,a_i>0,\;i=1,\dots,p\] and the corresponding marginal covariance functions are of the Mat{\'e}rn type with common spatial scales $a$, distinct smoothness $\nu_i,\;i=1,\dots,p$ and distinct variances $\sigma_i^2,\;i=1,\dots,p$:
\[\text{C}_{ii}(\textbf{h})=\int_0^\infty\|\textbf{h}\|\Bigg(\frac{2\pi\omega}{\|\textbf{h}\|}\Bigg)^{\kappa+1} J_\kappa(\omega\|\textbf{h}\|)f_{ii}(\omega|\sigma_i,\nu_i,a)\text{d}\omega=\text{M}(\textbf{h}|\sigma_i,\nu_i,a),\;i=1,\dots,p.\]

For $K\rightarrow\infty$ and common B-spline coefficients $b_k^{(ij)}=\tau_{ij},\;k=-3,\dots,K,\;1\leq i \neq j \leq p$, the coherence function for the $(i,j)^{th}$ pair of components is given as :\[\gamma_{ij}(\omega)=\tau_{ij}\sum_{k=-3}^\infty B_k(\omega)=\tau_{ij},\;\omega\geq 0,\;1\leq i \neq j \leq p.\]The cross spectral densities in (\ref{eq5}) then becomes:\[f_{ij}(\omega|f_{ii},f_{jj},\textbf{S}_{ij},K)=\tau_{ij}\mathcal{C}(\nu_i,\nu_j,d)\sigma_i\sigma_j\frac{\Gamma\big((\nu_i+\nu_j)/2+d/2\big)a^{(\nu_i+\nu_j)}}{\Gamma\big((\nu_i+\nu_j)/2\big)\pi^{d/2}(a^2+\omega^2)^{(\nu_i+\nu_j)/2+d/2}}, \:\omega\geq0,\;1\leq i \neq j \leq p,\]where \[\mathcal{C}(\nu_i,\nu_j,d)=\frac{\Gamma(\nu_i+d/2)^{\frac{1}{2}}\Gamma(\nu_j+d/2)^{\frac{1}{2}}\Gamma\big((\nu_i+\nu_j)/2\big)}{\Gamma(\nu_i)^{\frac{1}{2}}\Gamma(\nu_j)^{\frac{1}{2}}\Gamma\big((\nu_i+\nu_j)/2+d/2\big)}.\] The corresponding cross-covariances is then given as ;\[\text{C}_{ij}(\textbf{h})=\int_0^\infty\|\textbf{h}\|\Bigg(\frac{2\pi\omega}{\|\textbf{h}\|}\Bigg)^{\kappa+1} J_\kappa(\omega\|\textbf{h}\|)f_{ij}(\omega|f_{ii},f_{jj},\textbf{S}_{ij},K)\text{d}\omega\]=\[=\text{M}(\textbf{h}|\sqrt{\tau_{ij}\mathcal{C}(\nu_i,\nu_j,d)\sigma_i\sigma_j},(\nu_i+\nu_j)/2,a),\;1\leq i \neq j \leq p.\] which is a parsimonious multivariate Mat{\'e}rn cross-covariance function with the colocated correlation coefficient $\rho_{ij}=\tau_{ij}\mathcal{C}(\nu_i,\nu_j,d)$.

\end{document}